\documentclass[usenatbib,usegraphicx]{mn2e}
\usepackage{amssymb}
\usepackage{bm}
\usepackage{color}
\usepackage{graphicx}
\usepackage[normalem]{ulem}

\newcommand{\halfspace}{\hspace{1pt}}

\newcommand\dnu{{\rm d}\nu}
\newcommand\nhat{{\bm {\hat n}}}

\newcommand{\Lya}{Ly$\alpha$}

\newcommand\HI{{\hbox{H\halfspace$\rm \scriptstyle I$}}}

\newcommand\HeII{{\hbox{He\halfspace$\rm \scriptstyle II$}}}

\newcommand\lsim{~\lower.5ex\hbox{$\buildrel < \over \sim$}~}
\newcommand\gsim{~\lower.5ex\hbox{$\buildrel > \over \sim$}~}

\title[Time-dependent fluctuations in the UVBG]{Time-dependent fluctuations in the metagalactic photoionization background}
       
\author[Avery Meiksin, Matt McQuinn]{
        Avery Meiksin$^{1}$\thanks{E-mail:\ A.Meiksin@ed.ac.uk (AM)},
        Matthew McQuinn$^{2}$\\
        $^{1}$SUPA\thanks{Scottish Universities Physics Alliance},
	Institute for Astronomy, University of Edinburgh,
        Blackford Hill, Edinburgh\ EH9\ 3HJ, UK\\
        $^{2}$Astronomy Department, University of Washington,
        Seattle, WA 98195, USA}

\shortcites{}

\begin{document}

\date{Accepted . Received ; in original form }
\pagerange{\pageref{firstpage}--\pageref{lastpage}} \pubyear{2018}
\maketitle
\label{firstpage}

\begin{abstract}
  We present a formalism for computing time-dependent fluctuations in the cosmological photoionizing radiation background, extending background fluctuations models beyond the steady-state approximation. We apply this formalism to estimate fluctuations in the \HI\ \Lya\ flux redshift space power spectrum and its spatial correlation function at redshifts $2<z<4$, assuming the photoionization background is dominated by Quasi-stellar Objects (QSOs) and/or galaxies. We show the shot noise in the power spectrum due to discrete sources is strongly suppressed relative to the steady-state value at low wavenumbers by a factor proportional to the lifetime of the sources, and that this suppression may be used to constrain QSO lifetimes. The total \HI\ \Lya\ power spectrum including shot noise is affected at tens of percent on short scales, and by as much as an order of magnitude or more on scales exceeding the mean free path. The spatial correlation function is similarly found to be sensitive to the shot noise, although moderately insensitive to the effects of time-dependence on the non-shotnoise contribution. Photoionization rate fluctuations substantially modify the shape of the Baryonic Acoustic Oscillation peak in the correlation function, including a small increase in its position that must be accounted for to avoid biasing estimates of cosmological parameters based on the peak position. We briefly investigate solving the full frequency dependent equation, finding that it agrees with the frequency-independent to better than percent accuracy. Simple formulas are provided for the power spectrum of fluctuations in the photoionization rate that approximate the full computations.
\end{abstract}

\begin{keywords}
galaxies:\ formation -- intergalactic medium -- large-scale structure of Universe  -- quasars: absorption lines
\end{keywords}

\section{Introduction}
\label{sec:Intro}

The past few years have witnessed a new era in observational studies of the Intergalactic Medium (IGM) brought about by very large Quasi-Stellar Object (QSO) surveys \citep{2012ApJS..203...21A, 2017A&A...597A..79P}. The high sky coverage of QSOs has enabled 3D studies of the IGM \citep{2013AJ....145...69L, 2017arXiv171002894L, 2017Sci...356..418R}, including the discovery of the Baryonic Acoustic Oscillation (BAO) signal in the \Lya\ forest \citep{2013A&A...552A..96B}, which has continued to be measured with ever-increasing precision, providing a new means of constraining cosmological parameters \citep{2017A&A...603A..12B}.

It has long been recognised that, in addition to the contribution from matter fluctuations, fluctuations in the neutral component of the IGM will be induced by fluctuations in the metagalactic UV photoionization background as a result of source discreteness \citep{1992MNRAS.258...45Z}. Fluctuations in the UV background enhance the \Lya\ forest flux power spectrum on large scales, while suppressing it on intermediate scales \citep{2004ApJ...610..642C, 2004MNRAS.350.1107M, 2018MNRAS.473..560D}, and potentially affecting the BAO signal in the Ly$\alpha$ forest.  The magnitude of these effects depends on the sources giving rise to the metagalactic UV background at high redshifts, which are still incompletely known. Whilst the contribution of QSOs may account for the UV photoionization background up to $z\simeq2-3$, most analyses find too few QSOs at higher redshifts, suggesting an increasing contribution from galaxies with redshift at $z>3$ \citep[e.g.][]{1996ApJ...461...20H, 2005MNRAS.356..596M, 2009ApJ...703.1416F}. If galaxies source the background, knowing which type of galaxies is the dominant contributor would potentially constrain the sources of reionization. In addition to assessing the influence of the photoionization background fluctuations on the BAO signal, through the effects of their bias and shot-noise contributions to the \Lya\ forest power spectrum, it may be possible to infer the nature of the dominant sources themselves \citep{2011MNRAS.415..977M, 2014MNRAS.442..187G, 2014PhRvD..89h3010P, 2014ApJ...792L..34P}.

Estimates of the photoionization-induced fluctuations in the \Lya\ forest have previously been based on steady-state models. Because the expected lifetimes of QSOs (and possibly the durations of star formation episodes in galaxies) may be short compared with the time for a photon to travel of order a mean free path, a time-dependent formalism is required for assessing the shot noise contribution to fluctuations in the photoionizing radiation background. Evolution in the source numbers and bias (as well as in the photon mean free path) may also require time dependence to be included as photons emitted at $z\lesssim3$ typically travel for a significant fraction of the Hubble time. The purpose of this paper is to provide a time-dependent formalism of the UV background fluctuations and their impact on the \Lya\ forest flux redshift-space power spectrum and the corresponding redshift-space spatial correlation function.

Another motivation for this study is that the steady state solution formally diverges in the limit where the opacity owes entirely to a diffuse component of gas in the absence of cosmological expansion. Although the case with cosmological expansion is the one of practical interest, we show that the divergence is cured in the general case within a time-dependent framework. Another concern is the artificial distinction between a diffuse gas component optically thin at the photoelectric edge and an optically thick clumped component \citep{2014PhRvD..89h3010P}. Our model for the IGM opacity improves upon previous background fluctuation models, and is more in keeping with standard homogeneous photoionizing background calculations such as \citet{1996ApJ...461...20H}.

This paper is organised as follows. A summary of the basic formalism
is provided in the next section. Sec.3 presents our time-dependent
background model calculations under various assumptions for the
sources and absorbers, and draws contrasts with the steady-state
solutions. The observational consequences of the background fluctuations on the \HI\ \Lya\ flux redshift space power spectrum and spatial correlation function are examined in Sec.4. The reader primarily interested in these quantities may
choose to skip directly to this section. In Sec.5 we investigate the accuracy
of the frequency-integrated approximation. We end by discussing the
results and providing our conclusions. The mathematical machinery
underlying the formalism is expounded upon in an Appendix.

All numerical calculations assume {\it Planck} 2015 parameter values for a $\Lambda$CDM cosmology \citep{2016A&A...594A..13P}. Our calculations are done in a Euclidean context. Modes with $k \lesssim  H/c$ should be calculated in full General Relativity for (gauge independent) observables. We do not expect predictions for the size of effects to be altered (see \citealt{2014PhRvD..89h3010P}). Finally, we use a Fourier convention for which density fluctuations are dimensionless in real and Fourier space \citep[e.g.][]{2009RvMP...81.1405M}, and often plot the comoving power spectrum of the Fourier coefficients $\tilde \delta(k)$ as $P_\delta = V_u |\tilde \delta(k) \tilde \delta^\dag(k)|$ (defined in a comoving volume $V_u$). Dimensionless power spectra for which the integral over $d \log k$ equals the variance are given by $k^3 P_\delta/[2\pi^2]$ in our normalization convention.

\section{Basic equations}
\label{sec:basics}

\subsection{Fluctuations in the background radiation field}
\label{subsec:bgflucs}

\subsubsection{Basic results}
\label{subsubsec:tdbasics}

The radiative transfer equation for the specific intensity $I_\nu$ (in
units of energy flux per frequency per solid angle) at a position
${\bm r}$ in direction $\nhat$ at time $t$ in a cosmological setting is

\begin{eqnarray}
\frac{1}{c}\frac{\partial I_\nu({\bm r},\nhat,t)}{\partial t} &+&
\frac{1}{c}\frac{\dot a}{a}\left[3I_\nu({\bm r},\nhat,t) - \nu\frac{\partial
    I_\nu({\bm r},\nhat,t)}{\partial \nu}\right]\nonumber\\
&+& \nhat\cdot{\bm \nabla} I_\nu({\bm r},\nhat,t)\nonumber\\
&=&-\alpha_\nu({\bm r},\nhat,t)I_\nu({\bm r},\nhat,t) + j_\nu({\bm r},\nhat,t),
\label{eq:RTInu}
\end{eqnarray}
where $\alpha_\nu$ is the inverse attenuation length (also known as the absorption coefficient), $j_\nu$ is a
source term (the emission coefficient), and $a(t)$ is the expansion factor, related to redshift by
$a=1/ (1+z)$.

The following results summarised here are derived in detail in the Appendix. For most of this paper we consider solutions to the frequency-integrated form of the radiative transfer equation, but we generalize this in Sec.~\ref{sec:fullfreq}. As we are interested in fluctuations in the \HI\ photoionization rate, to derive a frequency-averaged equation, we integrate Eq.~(\ref{eq:RTInu}) over frequency, weighting by the frequency dependence of the \HI\ photoionization cross section, taken to be $\sigma_\nu\sim\nu^{-3}$:
\begin{equation}
\frac{1}{c}\dot f + \frac{1}{c}\frac{\dot 
  a}{a}\frac{I_L\sigma_L}{h_{\rm P}}+\nhat\cdot{\bm\nabla}f=-\alpha_{\rm 
  eff}f+j,
\label{eq:f}
\end{equation}

\noindent where

\begin{equation}
f\equiv\int_0^\infty\,d\nu\,\left(\frac{I_\nu}{h_{\rm P}\nu} \right)\sigma_\nu,\quad j\equiv\int_0^\infty\,d\nu\,\left(\frac{j_\nu}{h_{\rm P}\nu}\right)\sigma_\nu 
\label{eq:favg}
\end{equation}
and 
\begin{equation}
\alpha_{\rm eff}\equiv\frac{\int_0^\infty\,d\nu\,\frac{I_\nu}{h_{\rm P}\nu}\alpha_\nu\sigma_\nu}{\int_0^\infty\,d\nu\,\frac{I_\nu}{h_{\rm P}\nu}\sigma_\nu}.
\label{eq:aeff}
\end{equation}
Here $I_L$ and $\sigma_L$ are the values of the metagalactic intensity and photoelectric absorption cross-section at the photoelectric threshold energy.  Following the frequency integrated equation requires us to ignore spatial fluctuations in the spectrum of $I_\nu$; we address this assumption in Sec~\ref{sec:fullfreq}.

In terms of the dimensionless absorption coefficient and spectral redshift factor

\begin{equation}
\chi \equiv \frac{c}{H}\langle\alpha_{\rm eff}\rangle,\quad
\zeta \equiv \frac{\langle I_L\rangle\sigma_L}{h_{\rm P}\langle f\rangle},
\label{eq:chi}
\end{equation}

\noindent respectively, and the dimensionless radiation field evolution factor

\begin{equation}
\phi\equiv \frac{c\langle j\rangle/H}{(\chi+\zeta)\langle
  f\rangle},
\label{eq:phi}
\end{equation}
the spatially-averaged background radiation field obeys
\begin{equation}
\frac{\langle\dot f\rangle/H}{(\chi + \zeta)\langle f\rangle}=\phi-1.
\label{eq:fbg}
\end{equation}
\noindent Here, $\langle\dots\rangle$ denotes a spatial average, and
we have assumed that there is no preferred direction such that $\langle \nhat\cdot{\bm\nabla}f\rangle=0$. It is seen
that the dimensionless ratio $\phi$ determines whether $\langle
f\rangle$ increases or decreases with cosmic time (depending,
respectively, on whether $\phi>1$ or $\phi<1$). Assuming $\dot f=0$ is
then equivalent to asserting $\phi=1$.\footnote{Even though we are referring to this solution as the `spatially averaged' solution, note that there are complications since $\langle j \rangle$ diverges if there are point sources. More formally, the homogeneous solution (which we will perturb around) is that for a spatially uniform emission coefficient.}

We consider planewave perturbations of the mean intensity $f$ with Fourier components $\tilde \delta_f=\widetilde{\delta f}/\langle f\rangle$. Allowing for density, temperature and photoionization rate fluctuations results in fluctuations in the absorption coefficient:
\begin{equation}
  \tilde\delta\chi=\chi\left(b_{\chi, \delta}\tilde\delta
  + b_{\chi, \Gamma}\tilde\delta_\Gamma\right),
\label{eq:dchi}
\end{equation}
where $\tilde\delta$ and $\tilde\delta_\Gamma$ are the relative perturbations in the gas density and metagalactic photoionization rate, respectively.\footnote{More generally the absorption coefficient can trace temperature fluctuations with a bias parameter $b_{\chi, T}$, which Eq.~(\ref{eq:dchi}) has not included. We assume here that the temperature is a biased tracer of density alone, as would be the case if there is a one-to-one relationship between temperature and density (although helium reionization can break such a relation:\ \citealt{ 2007MNRAS.380.1369T, 2011MNRAS.415..977M,  2012MNRAS.423....7M,  2014MNRAS.442..187G}), allowing us to absorb this dependence into $b_{\chi, \delta}$.  For an IGM equation of state $T\sim\rho^{\gamma-1}$, $\tilde\delta_T=(\gamma-1)\tilde\delta$, this results in the density bias remapping to $b_{\chi, \delta} \rightarrow b_{\chi, \delta}+(\gamma-1)b_{\chi, T}$ in the limit that density fluctuations are linear.} We introduce the rescaled time variable $d\bar t'=H(t')dt'$ and wavenumber ${\bm \kappa} \equiv (c/H) {\bm k}$, where ${\bm k}$ is the comoving wavenumber of the perturbation. With these definitions, the linear-order radiative transfer equation we aim to solve is
\begin{equation}
\partial_{\bar t} {\tilde\delta}_f + \phi\zeta \tilde\delta_{I_L}-ia^{-1}{\bm \kappa}\cdot\nhat\; \tilde\delta_f=\chi\left[-\phi\tilde \delta_f +b_{\chi, \Gamma}\left( \tilde\delta_S
                       - \tilde\delta_\Gamma \right)\right].
\label{eq:df}
\end{equation}
Here, $\tilde\delta_S$ is an `effective source term' given by
\begin{equation}
  \tilde\delta_S(\bar t)=q(\bar t)\tilde\delta_j(\bar t)-\frac{b_{\chi, \delta}}{b_{\chi, \Gamma}}\tilde\delta,
\label{eq:dSnd}
\end{equation}
where $q(\bar t)\equiv[c\langle j(\bar t)\rangle/H(\bar t)]/[b_{\chi,
  \Gamma}(\bar t)\chi(\bar t)\langle f(\bar t)\rangle]=[\phi(\bar
t)/b_{\chi, \Gamma}(\bar t)[1+\zeta(\bar t)/\chi(\bar t)]$, and
$\tilde\delta_j\equiv\widetilde{\delta j}/\langle j\rangle$.

We identify $\tilde\delta_{I_L} \equiv \widetilde{\delta I_L}/\langle I_L\rangle$ with $\tilde\delta_f$ to solve the above since the fluctuations in the $\sigma_\nu$-weighted photoionizing background likely trace the fluctuations at the Lyman-limit. The general solution for $\tilde\delta_f$, with $\tilde\delta_f(t)=0$ when $t<t_i$, is
\begin{equation} 
  \tilde\delta_f(\bar t)=\int_{\bar t_i}^{\bar t}\,d\bar t'\,G(\bar t,\bar t')b_{\chi, \Gamma}(\bar t')\chi(\bar t')\left[\tilde\delta_S(\bar t')
                       - \tilde\delta_\Gamma(\bar t')\right],
\label{eq:dfnd}
\end{equation}
where $G$ is the Green's function
\begin{equation} 
  G(\bar t,\bar t') = \exp\left\{\int_{\bar t'}^{\bar t}d\bar t''\left[i\frac{{\bm \kappa}(\bar t'')\cdot{\nhat}}{a(\bar t'')}-\phi(\bar t'')\left(\chi(\bar t'') + \zeta(\bar t'')\right)\right]\right\}.  
\label{eq:Gfunc}
\end{equation}

The photoionization rate is given by $\Gamma = \int\,d^2\nhat\, f$, so that, for isotropic sources, $\tilde\delta_\Gamma(\bar t)$ is given by the implicit equation
\begin{eqnarray}
\tilde\delta_\Gamma(\bar t)&=&\frac{\langle f\rangle\int d^2\nhat\,\tilde\delta_f}{\int d^2\nhat\,\langle f\rangle}=\frac{1}{4\pi}\int d^2\nhat\,\tilde\delta_f\nonumber\\
&=&\int_{\bar t_i}^{\bar t}\,d\bar t'\,j_0[\kappa(\bar t)\bar
    \eta(\bar t,\bar t')]b_{\chi, \Gamma}(\bar t')\chi(\bar t')\left[\tilde\delta_S(\bar t')-\tilde\delta_\Gamma(\bar t')\right]\nonumber\\
&&  \times e^{-\int_{\bar  t'}^{\bar t} d\bar
   t''\,\left[\phi(\bar t'')\left(\chi(\bar t'')+\zeta(\bar t'')\right)\right]},
\label{eq:dGisofull_nondim}
\end{eqnarray}
where $j_0(x)=\sin(x)/x$ and
$\bar\eta(\bar t,\bar t')=H(t)\eta(t,t')=H(t)\int_{t'}^{t}\,dt''a^{-1}(t'')$, where $\eta(t,t')$ is the conformal time between cosmological coordinate times $t'$ and $t$. \footnote{In the Einstein-deSitter approximation,
$\eta(t,t')=3a(t)^{-1}(t-t^{2/3} t'^{1/3})=H(t)^{-1}[2/a(t)][1-(a(t')/a(t))^{1/2}]$,
and $\kappa\bar\eta=(2\kappa(a)/a)[1-(a'/a)^{1/2}]$.}  The special
case $b_{\chi, \Gamma}\chi=0$ is treated in the Appendix.

\subsubsection{Asymptotic limits}
\label{subsubsec:bgflucs:limits}
We next consider limits of the previous expressions, which are useful for intuition and for comparing with the steady state solutions. The fluctuation in the photoionization rate is exactly solvable in the limit $\kappa(\bar t)\bar\eta({\bar t}, \bar t_i) \ll 1$, valid for modes with wavelengths larger than the distances photons can travel. The solution is
\begin{eqnarray}
  \tilde\delta_\Gamma(\bar t) &=&
  \int_{\bar t_i}^{\bar t}\,d\bar t'\,b_{\chi, \Gamma}(\bar t')\chi(\bar t')\tilde\delta_S(\bar t')\nonumber\\
  &\times&e^{-\int_{\bar t'}^{\bar t} d\bar t''\,\left[(\phi(\bar t'')+b_{\chi, \Gamma})\chi(\bar t'')+\phi(\bar t'')\zeta(\bar t'')\right]}.
\label{eq:dGammakq}
\end{eqnarray}
It is seen that $\tilde\delta_S$ drives the fluctuations in
$\Gamma$. A closed form expression follows if $\chi$,
$\zeta$, $b_{\chi, \Gamma}$ and $\phi$ are taken to be temporally constant, along with the further simplifications of a negligible
contribution from $b_{\chi, \delta}\tilde\delta/b_{\chi, \Gamma}$ to
$\tilde\delta_S$, so that
$\tilde\delta_S\simeq q\tilde\delta_j$, setting
$\tilde\delta_j=b_j\tilde\delta$, and adopting the evolutionary form
$b_j=b_j(0)(1+z)^{\alpha_b}$:
\begin{eqnarray}
\delta_\Gamma(\kappa=0) &=& \frac{\chi + \zeta}{(\phi+b_{\chi, \Gamma})\chi+\phi\zeta+1-\alpha_b}\phi\tilde\delta_j(z)\nonumber\\
&\times&\left[1-x_i^{(\phi+b_{\chi, \Gamma})\chi+\phi\zeta+1-\alpha_b}\right],
\label{eq:deltaG0anal}
\end{eqnarray}
where $x_i=(1+z)/ (1+z_i)$. In the limit $(\phi+b_{\chi, \Gamma})\chi+\phi\zeta+1-\alpha_b\rightarrow0$, $\delta_\Gamma(\kappa=0)\rightarrow-(\chi+\zeta)\phi\tilde\delta_j(z)\log x_i$, avoiding a formal divergence that occurs in the steady state solution (see below).

In the limit $\kappa(a)\gg1$, the fluctuation in the photoionization
rate may be developed as an asymptotic series. With the same
simplifying approximations as above, to second order in $1/\kappa$,
\begin{eqnarray}
  \tilde\delta_\Gamma(\kappa\gg1)&\sim&\frac{\pi}{2}\phi(\chi+\zeta)\frac{a}{\kappa(a)}\tilde\delta_j(a)\nonumber\\
  &&\times\Biggl\{1 - \frac{a}{\kappa(a)}\Biggl[\frac{\pi}{2}b_{\chi,
      \Gamma}\chi\nonumber\\
    &+&\frac{2}{\pi}\left(\phi(\chi+\zeta)-\alpha_b+\frac{1}{2}\right)\nonumber\\
&+&\frac{1}{\pi}\frac{x_i^{\gamma_H}}{1-x_i^{1/2}}\cos\left[\frac{2\kappa(a)}{a}(1-x_i^{1/2})\right]\Biggr]\Biggr\},
\label{eq:dGkO2}
\end{eqnarray}
where $\gamma_H\equiv\phi(\chi+\zeta)-\alpha_b+1/2$. For finite
$x_i$, the last term describes relic memory from a flash turn on of an ionizing background, as may occur during a rapid ionization-zone overlapping phase during reionization when the mean free path in some models grows very quickly. The effect provides a BAO-like feature. However, because $\zeta$ is likely large at redshifts well after reionization, the factor $x_i^{\gamma_H}$ will be small (indicating that photons emitted at $\sim z_i$ must travel many attenuation lengths), and the
oscillations will likely contribute negligibly to $\tilde\delta_\Gamma$,
except possibly shortly after a species is reionized, or for extreme
evolution in the source bias or background radiation field.

Comparison of Eqs.~(\ref{eq:deltaG0anal}) and (\ref{eq:dGkO2}) suggests the simple Lorentzian interpolation expression
\begin{equation}
  V_u\langle\tilde\delta_\Gamma\tilde\delta^\dagger_\Gamma\rangle\simeq V_u\left\vert\frac{\chi+\zeta}{(\phi+b_{\chi, \Gamma})\chi+\phi\zeta+1-a_b}\phi\tilde\delta_j\right\vert^2\frac{1}{1+(\kappa/\kappa_*)^2},
\label{eq:aplor}
\end{equation}
where $\kappa_*=2\pi[c/H(z)]/[(1+z)\lambda_*]$ corresponds to a (proper) attenuation length in the spatial correlations of
\begin{equation}
  \lambda_*=\frac{c}{H}\frac{4}{(\phi+b_{\chi, \Gamma})\chi+\phi\zeta+1-a_b}.
\label{eq:lambdas}
\end{equation}
This is the attenuation scale of the IGM, including cosmological expansion and evolution in the bias factor of the sources. The large-scale and small-scale leading order asymptotic limits of the photoionization fluctuations are recovered for modes with wavelengths $\lambda\gg\lambda_*$ and $\lambda\ll\lambda_*$, respectively.

\subsection{Comparison between time-dependent and steady-state solutions}
\label{subsec:tdvss}

\subsubsection{Basic results}
\label{subsubsec:ssbasics}

Two distinct steady-state approximations may be made:\ (1)\ to the
unperturbed radiative transfer equation, Eq.~(\ref{eq:fbg}), and
(2)\ to the equation for the perturbations, Eq.~(\ref{eq:df}). One of
these may be approximated as in a steady state without requiring the
other also to be in a steady state, although previous studies adopted
both steady-state approximations \citep{2014MNRAS.442..187G,
  2014PhRvD..89h3010P}. Assuming the background radiation field to be
in a steady state corresponds to setting $\phi=1$ in
Eq.~(\ref{eq:fbg}). For general $\phi$, the steady-state solution to
the radiative transfer equation for the perturbed radiation field
corresponds to the steady-state ionization rate perturbation
\begin{equation}
  \tilde\delta_{\Gamma, {\rm SS}}=\frac{\phi(\chi+\zeta)\tilde\delta_j
    - b_{\chi, \delta}\chi\tilde\delta}{{\frac{\kappa}{a}\left[{\rm
    atan}{\left(\frac{\kappa}{a\phi(\chi+\zeta)}\right)}\right]^{-1}}
+b_{\chi, \Gamma}\chi}.
\label{eq:dGisofullSS}
\end{equation}

\subsubsection{Asymptotic limits}
\label{subsubsec:tdvss:limits}

With the simplifying assumption
$\tilde\delta_S=q\tilde\delta_j$, in the limit
$\kappa\rightarrow0$ Eq.~(\ref{eq:dGisofullSS}) becomes
\begin{equation}
\tilde\delta_{\Gamma, {\rm SS}}(\kappa=0) \simeq \frac{\chi+\zeta}{(\phi+b_{\chi,\Gamma})\chi + \phi\zeta}\phi\tilde\delta_j.\label{eq:deltaG0SS}
\end{equation}
Unlike the time-dependent case above, when the denominator vanishes the fluctuation formally diverges (except for specific choices of $\phi$ and $b_{\chi, \Gamma}$ to match the numerator). In particular, the steady state approximation for the choices $\phi=1$ and $b_{\chi, \Gamma}=-1$ \citep{2014PhRvD..89h3010P}
requires non-vanishing cosmological expansion ($\zeta\neq0$) for its validity. Comparison to Eq.~(\ref {eq:deltaG0anal} ) shows that the time-dependent solution approaches the steady-state result in the limit of very large $\phi(\chi+\zeta)$, corresponding to a short effective mean free path for ionizing radiation. Alternatively, the steady-state result is recovered for $\alpha_b=1$ (and taking $x_i=0$), for which the bias evolution factor cancels the evolution of the growing mode of the density fluctuations.

In the limit $\kappa\gg1$, to second order in $\kappa^{-2}$,
$\tilde\delta_{\Gamma, {\rm SS}}$ becomes

\begin{equation}
\tilde\delta_{\Gamma, {\rm
    SS}}\sim\frac{\pi}{2}(\chi+\zeta)\phi\frac{a}{\kappa}\tilde\delta_j\left[1-\frac{a}{\kappa}\left(\frac{\pi}{2}b_{\chi,\Gamma}\chi+\frac{2}{\pi}\phi(\chi+\zeta)\right)\right].
\label{eq:dGkSSO2}
\end{equation}
Comparison to Eq.~(\ref{eq:dGkO2}) shows that to order $1/\kappa$ this expression is identical to the time-dependent solution. In addition, the two expressions are identical to order
$1/\kappa^2(a)$ for $\chi\gg1$. The correspondence is exact to this
order for $x_i=0$ and $\alpha_b=1/2$, as distinct from
$\alpha_b=1$ in the limit $\kappa\rightarrow0$.  Only the time dependent solution can give rise to the reionization oscillation that is present for finite $x_i$.

\subsection{Shot noise}
\label{subsec:shotnoise}

To include shot noise, the source emissivity $\epsilon$ is perturbed,
allowing for an evolving luminosity function and evolving source
luminosity. For a periodic box of volume $V_u$, the power spectrum of
the emissivity is
\begin{eqnarray}
P_\epsilon(k,t,t^\prime)&=&V_u\langle\delta_\epsilon({\bm
                            k},t)\delta^*_\epsilon({\bm
                            k},t^\prime)\rangle\nonumber\\
&=&\frac{1}{n_{\rm eff}(t,t^\prime)}+D(t)D(t^\prime)b(t)b(t^\prime)P_{\rm init}(k),
\label{eq:powem}
\end{eqnarray}
where $D(t)$ is the linear perturbation growth factor since some
initial time when the matter spatial correlation function
$\xi_{\rm init}(\vert{\bm x}\vert)$ corresponded to an initial matter power
spectrum
$P_{\rm init}(k)=\int_{V_u}d^3x\;\xi_{\rm init}(\vert {\bm
  x}\vert)e^{i{\bm k}\cdot{\bm x}}$,
$b(t)$ is the (time-dependent) bias factor for the sources, and the effective comoving number density of sources is defined by
\begin{eqnarray}
\frac{1}{n_{\rm
  eff}(t,t^\prime)}&=&\left[{\int_0^\infty\;dL\;L\Phi(L,t)\int_0^\infty\;dL\;L\Phi(L,t^\prime)}\right]^{-1}\nonumber\\
&\times&\int_0^\infty dL\;\int_0^\infty
         dt^{\prime\prime}\;L(t-t^{\prime\prime})L(t^\prime-t^{\prime\prime})\nonumber\\
&&\times\Phi(L,t^{\prime\prime})\tau_S(L)^{-1}.
\end{eqnarray}
Here, a simple evolution model has been assumed for the spatially averaged comoving birthrate function $\bar\Psi(L,t)$ of sources of luminosity $L$ at time $t$ and lifetime $\tau_S(L)$, given by $\bar\Psi(L,t)=\Phi(L,t)/\tau_S(L)$ for a comoving source luminosity function $\Phi(L,t)$. If $\tau_S(L)=\tau_S$ for all $L$ and the luminosity function evolves slowly, so that $\tau_S\vert\dot\Phi\vert\ll\Phi$, the luminosity function may be approximated by $\bar\Phi(L,t,t^\prime)=(1/2)[\Phi(L,t)+\Phi(L,t^\prime)]$, and the expression for $n_{\rm eff}$ simplifies to
\begin{eqnarray}
\frac{1}{n_{\rm
  eff}(t,t^\prime)}&=&\frac{\int_0^\infty\;dL\;L^2\bar\Phi(L,t,t^\prime)}{[\int_0^\infty\;dL\;L\Phi(L,t)][\int_0^\infty\;dL\;L\Phi(L,t^\prime)]}\nonumber\\
&\times&{\rm Max}\left[0,\left(1-\frac{\vert t-t^\prime\vert}{\tau_S}\right)\right].
\label{eq:neffttp}
\end{eqnarray}

For a mixed population of sources, such as QSOs (\lq q\rq) and
galaxies (\lq g\rq), the shot
noise terms add:
\begin{eqnarray}
\frac{1}{n_{\rm eff}(t,t^\prime)}&=&\frac{1}{[\int_0^\infty\;dL\;L\Phi(L,t)][\int_0^\infty\;dL\;L\Phi(L,t^\prime)]}\\
&\times&\Biggl\{\int_0^\infty\;dL\;L^2\bar\Phi_q(L,t,t^\prime)
{\rm Max}\left[0,\left(1-\frac{\vert t-t^\prime\vert}{\tau_{S,
         q}}\right)\right]\nonumber\\
& +&
\int_0^\infty\;dL\;L^2\bar\Phi_g(L,t,t^\prime){\rm Max}\left[0,\left(1-\frac{\vert t-t^\prime\vert}{\tau_{S, g}}\right)\right]\Biggr\},\nonumber
\label{eq:neffttpab}
\end{eqnarray}
where $\Phi=\Phi_q+\Phi_g$. The contributions to the non-shotnoise component of the power spectrum of the background radiation field are weighted by the contribution of each population to the mean background emissivity,
\begin{equation}
\epsilon_{bg}^{(i)}(t)\simeq\int_0^\infty\;dL\; L\Phi_i(L,t).
\end{equation}
The emissivity power spectrum then becomes
\begin{eqnarray}
P_\epsilon(k,t,t^\prime)&=&\frac{1}{n_{\rm eff}(t,t^\prime)}\\
&&+\frac{1}{\epsilon_{bg}(t)\epsilon_{bg}(t^\prime)}\sum_{i,j\in q,g} \epsilon_{bg}^{(i)}(t)\epsilon_{bg}^{(j)}(t^\prime)P_{ij}(k,t,t^\prime) \nonumber
\label{eq:powemab}
\end{eqnarray}
where $P_{ij}(k,t,t^\prime)=D(t)D(t^\prime)b_i(t)b_j(t^\prime)P_{\rm
  init}(k)$ and
$\epsilon_{bg}=\epsilon_{bg}^{(q)}+\epsilon_{bg}^{(g)}$.

In constructing the power spectrum of the ionization rate fluctuations
from Eq.~(\ref{eq:dGisofull_nondim}), the shot noise contribution
arises from the $\tilde\delta_j$ contribution in $\tilde\delta_S$ and involves
double integrations over time. In the limit $\tau_S\ll t-t_i$, an
approximate form for the generic integrations entailed is given by
\begin{eqnarray}
\int_{t_i}^t\;dt^\prime\;\int_{t_i}^t\;&dt^{\prime\prime}&\;f(t^\prime)g(t^{\prime\prime})\frac{1}{n_{\rm
                                        eff}(t^\prime,t^{\prime\prime})}\nonumber\\
&\simeq&\tau_S\int_{t_i}^t\;dt^\prime\;f(t^\prime)g(t^\prime)\frac{1}{n_{\rm eff}(t',t')},
\label{eq:snapprox}
\end{eqnarray}
provided $f(t')$ and $g(t')$ are smooth functions over time intervals
$(t'-\tau_S,t')$. This shows that the shot noise scales like
$\tau_S/n_{\rm eff}$, valid for low wavenumbers $k\ll1/(c\tau_S)$.
\footnote{Since $f(t')$ and $g(t')$ involve the factor
  $j_0[\kappa\bar\eta(t,t')]$, the approximation requires
  $ck\tau_S\ll1$.} Applying Eq.~(\ref{eq:snapprox}) to
Eq.~(\ref{eq:dGammakq}), and assuming the effective comoving number density of sources evolves as $n_{\rm eff}\sim(1+z)^{-\alpha_n}$, gives
\begin{eqnarray}
V_u\langle
\tilde\delta_\Gamma\tilde\delta^\dagger_\Gamma\rangle_{\rm shot}&\sim&\frac{1}{2}\frac{\phi^2(\chi+\zeta)^2}{(\phi+b_{\chi, \Gamma})\chi+\phi\zeta+3/4-\alpha_n/2}\nonumber\\
&&\times\frac{H(a)\tau_S}{n_{\rm eff}(a,a)}\qquad\qquad\qquad\qquad\qquad;\kappa\ll1\nonumber\\
&&\times\left[1-x_i^{2\left[(\phi+b_{\chi, \Gamma})\chi+\phi\zeta+3/4-\alpha_n/2\right]}\right]
\label{eq:shotlowk}
\end{eqnarray}
(for $H\tau_S\ll1$). The corresponding limit for the steady-state solution, using Eq.~(\ref{eq:deltaG0SS}), is
\begin{equation}
V_u\langle
\tilde\delta_\Gamma\tilde\delta^\dagger_\Gamma\rangle_{\rm shot}^{\rm SS}\sim\left[\frac{\phi(\chi+\zeta)}{(\phi+b_{\chi, \Gamma})\chi+\phi\zeta}\right]^2\frac{1}{n_{\rm eff}}\qquad;\kappa\ll1.
\label{eq:shotsslowk}
\end{equation}
A comparison between Eqs.~(\ref{eq:shotlowk}) and (\ref{eq:shotsslowk}) shows that the finite lifetime of the sources reduces the shot noise at low wavenumbers compared with the steady-state limit approximately by the factor $H\tau_S[(\phi+b_{\chi, \Gamma})\chi+\phi\zeta]$. This corresponds to the ratio of the distance light travels during the lifetime of the sources to the effective total mean free path, including both attenuation by intervening gas and redshifting.

For high wavenumbers, the asymptotic expansion of
$\tilde\delta_\Gamma$ may be used. The leading order noise
contribution then scales as $\kappa^{-2}$, with the same leading order behaviour as for
the steady-state solution,
\begin{equation}
V_u\langle
\tilde\delta_\Gamma\tilde\delta^\dagger_\Gamma\rangle_{\rm shot}\sim\left[\frac{\pi}{2}\phi\left(\chi+\zeta\right)\frac{a}{\kappa}\right]^2\frac{1}{n_{\rm eff}}\qquad;\kappa\gg1.
\label{eq:shothighk}
\end{equation}
Comparison of Eqs.~(\ref{eq:shotlowk}) and (\ref{eq:shothighk}) suggests an approximately Lorentzian interpolation expression. Whilst a fixed scale factor may be chosen, we find it is more accurate to allow for a sliding scale factor dependent on $\kappa$. Then
\begin{equation}
V_u\langle
\tilde\delta_\Gamma\tilde\delta^\dagger_\Gamma\rangle_{\rm shot}\simeq
\frac{1}{A}\frac{\left[\phi(\chi+\zeta)\right]^2}{1 + [\kappa/\kappa_*^{\rm SN}(\kappa)]^2}\frac{1}{n_{\rm eff}},
\label{eq:apshotLor}
\end{equation}
where
\begin{eqnarray}
  A &=& \frac{2}{H\tau_S}\left[(\phi+b_{\chi, \Gamma})\chi+\phi\zeta+\frac{3}{4}-\frac{1}{2}\alpha_n\right]\\
  &&-\left[\frac{4}{\pi^2}\left(\phi(\chi+\zeta)-1/2\right)+b_{\chi, \Gamma}\chi\right]^2,\nonumber
  \label{eq:defAlor}
\end{eqnarray}
  and $\kappa_*^{\rm SN}(\kappa) = 2\pi[c/H(z)]/[(1+z) \lambda_*^{\rm SN}(\kappa)]$ with
\begin{eqnarray}
  \lambda_*^{\rm SN}&=&4\frac{c}{H}A^{-1/2}\\
  &&\times\left\{1+\frac{\pi}{2}\left[\frac{4}{\pi^2}\left(\phi(\chi+\zeta)-1/2\right)+b_{\chi, \Gamma}\chi\right]\frac{a}{\kappa(a)}\right\}\nonumber,
  \label{eq:lambdas_shot}
\end{eqnarray}
(for $H\tau_S\ll1$). This recovers both the leading order behaviour and first order correction to Eq.~(\ref{eq:shothighk}), which follows from using Eq.~(\ref{eq:dGkO2}) with $\alpha_b=1$ (and assuming for simplicity $x_i=0$). For short wavelength modes, the shot noise scalelength is proportional to the geometric mean between the distance light travels during the lifetime of the sources and the total mean free path; for long wavelength modes this scale is lengthened by the ratio of the wavelength to the total mean free path.

In the Appendix, we demonstrate the Lorentzian approximations provide accurate predictions for the redshift-space spatial correlations in the \HI\ \Lya\ flux fluctuations. We find that the Lorentzian form recovers the spatial correlation function with an accuracy of better then 10 percent for the non-shotnoise contribution, but may be off by as much as $\sim40-70$ percent for the shot noise contribution.

\subsection{Method of solution}
\label{subsec:meth}

Eq.~(\ref{eq:dGisofull_nondim}) is readily solved as a matrix equation. By subdividing the time into intervals $d\bar t_j$ (not necessarily equal), the perturbation in the radiation field may be expressed at time $\bar t_i$ as $\tilde\delta_\Gamma(\bar t_i)=\tilde\delta_{\Gamma,i}$, where
\begin{equation}
\tilde\delta_{\Gamma,i}=\sum_j{\rm M}_{ij}\left[\tilde\delta_{S,j}-\tilde\delta_{\Gamma,j}\right]
\end{equation}
and
\begin{equation}
{\rm M}_{ij}=\sum_{j}\; d\bar t_j w_j j_0(\kappa_i\bar\eta_{ij})(b_{\chi,\Gamma})_j\chi_j e^{-\int_{\bar t_j}^{\bar t_i}\;d\bar t^\prime\;\phi(\bar t')\left[\chi(\bar t')+\zeta(\bar t')\right]}.
\end{equation}
Here, integration weights $w_j$ have been allowed for. In matrix notation,
\begin{equation}
{\bm{ \tilde\delta}_\Gamma}={\mathbf M}\left({\bm{ \tilde\delta}_S}-{\bm{ \tilde\delta}_\Gamma}\right),
\label{eq:dgammat}
\end{equation}
which has the solution
\begin{equation}
{\bm{ \tilde\delta}_\Gamma} = ({\mathbf M} + {\bm 1})^{-1}{\mathbf M}{\bm{ \tilde\delta}_S}.
\end{equation}

Starting from $({\mathbf M}+{\bm 1}){\bm{\tilde\delta}_\Gamma}={\mathbf M}{\bm {\tilde\delta}_S}$, the radiation fluctuation power spectrum is given through

\begin{equation}
  \left({\mathbf M}+{\bm 1}\right)V_u\langle{\bm{\tilde\delta}_\Gamma}{\bm{\tilde\delta}_\Gamma^\dagger}\rangle({\mathbf M}+{\bm 1})^\dagger={\mathbf M}V_u\langle{\bm{\tilde\delta}_S}{\bm{\tilde\delta}_S^\dagger}\rangle{\mathbf M}^\dagger.
\end{equation}
The product of source terms is given by the power spectrum of the emissivity terms in Eq.~(\ref{eq:powem}) ($\delta_\epsilon=\delta_j$), including the shot noise term. In the low $\kappa$ limit, Eq.~(\ref{eq:dgammat}) becomes ${\bm{\tilde\delta}}_\Gamma={\mathbf M}{\bm{\tilde\delta}}_S$, so that
\begin{equation}
V_u\langle{\bm{\tilde\delta}_\Gamma}{\bm{\tilde\delta}_\Gamma^\dagger}\rangle={\mathbf M}V_u\langle{\bm{\tilde\delta}_S}{\bm{\tilde\delta}_S^\dagger}\rangle{\mathbf M}^\dagger;\qquad\kappa(\bar t)\bar\eta(\bar t,\bar t_i)\ll1.
\end{equation}

\section{Background radiation fluctuations}
\label{sec:bgflucs}

\subsection{Model formulation}
\label{subsec:models}

Now we discuss the details of our models for the sources and absorbers needed to evaluate the source models. Our calculations concentrate on $z=2-5$ since this redshift range is most relevant for Ly$\alpha$ forest observations. We show in Sec.~\ref{subsubsec:uvbg} that our simplified model provides a good match to the \citet{2012ApJ...746..125H} model over this redshift range.

The comoving emissivity of the sources is modelled as
\begin{equation}
\epsilon_\nu(z)=\epsilon_L\left(\frac{\nu}{\nu_L}\right)^{-\alpha_j}(1+z)^{-\alpha_S},
\label{eq:emiss}
\end{equation}
where $\epsilon_L$ is a normalization factor formally corresponding to the comoving emissivity at the photoelectric edge at $z=0$ (although the model need not extend to $z=0$), and $\nu_L$ denotes the threshold frequency for photoelectric absorption.
 
The angle-averaged intensity, which sets the mean ionization rate and neutral hydrogen fraction, is given by
\begin{equation}
4\pi J_\nu(z) =
\int_z^\infty\,dz'\frac{dl_p}{dz'}\epsilon_{\nu'}(z')(1+z)^3e^{-\tau_\nu(z,z')},
\label{eq:Jnu}
\end{equation}
where $dl_p/dz = c/[H(z)(1+z)]$, $\epsilon_{\nu'}$ is the comoving
emissivity, $\nu'=\nu(1+z')/(1+z)$, and $\tau_\nu(z,z')$ is the optical depth due to IGM attenuation along a path from $z'$ to $z$. The hydrogen ionization rate is
\begin{equation}
\Gamma(z) = \int_{\nu_L}^\infty d\nu\,\frac{4\pi J_\nu}{h_P\nu}\sigma_\nu,
\label{eq:Gamma}
\end{equation}
where $\sigma_\nu$ is the photoelectric cross-section.

Our model for absorption builds off the largely empirical formalism for attenuation used by homogeneous uniform background models \citep[e.g.][]{2012ApJ...746..125H} (HM12). Under the approximation that the clumped component follows a power law in \HI\ column density, $\partial^2N/\partial N_{\rm HI}\partial z=A N_{\rm LLS}^{-1}(N_{\rm HI}/N_{\rm LLS})^{-\beta} (1+z)^{\gamma_a}$, where $N_{\rm LLS}\sigma_L=1$, the attenuation coefficient is given by
\begin{eqnarray}
\alpha_\nu&=&A(1+z)^{\gamma_a}\left(\frac{dl_p}{dz}\right)^{-1}\int_0^\infty\,dyy^{-\beta}\left[1-e^{-y(\nu/\nu_L)^{-3}}\right]\nonumber\\
&=&A(1+z)^{\gamma_a}\left(\frac{\nu}{\nu_L}\right)^{-3(\beta-1)}\nonumber\\
&&\times\left(\frac{dl_p}{dz}\right)^{-1}\int_0^\infty\,dx x^{-\beta}\left(1-e^{-x}\right),
\label{eq:attLLScolden}
\end{eqnarray}
where $dl_p/dz = 1/[H(z)(1+z)]$. This formulation approaches the `diffuse' limit for attenuation by optically thin absorbers in the IGM for $\beta\rightarrow2$, although with a logarithmic divergence at the lower end point (and so not very sensitive to it). Since a clumped component will alway be present, we expect in practical situations $1<\beta<2$. We note this is a simplification, as observations indicate both $\beta$ and $\gamma_a$ depend on $N_{\rm HI}$, with the values not well established over some column density ranges \citep[e.g.][]{2012ApJ...746..125H, 2014MNRAS.438..476P}.

We adopt an empirical normalization of the average opacity, similar to homogeneous ionizing background models. Observations suggest a (proper) mean free path at the Lyman edge of
\begin{equation}
\lambda_L\simeq (26\pm1)h^{-1}\,{\rm Mpc}\left[0.2(1+z)\right]^{-5.4\pm0.4}
\label{eq:mfpL}
\end{equation}
\citep{2014MNRAS.445.1745W}. As a comparison, a good match to the
frequency and redshift dependence of the net attenuation in the
Ly$\alpha$ absorber model of \citet{2012ApJ...746..125H} is given by
$\beta=1.2$ for $\nu_L<\nu<3.9\nu_L$, accurate to 15 percent, and
usually better than 10 percent, with $\gamma_a=5.2$. Using the frequency dependence of Eq.~(\ref{eq:attLLScolden}) then yields
\begin{equation}
\alpha_\nu=\lambda_L^{-1}\left(\frac{\nu}{\nu_L}\right)^{-3(\beta-1)}.
\label{eq:attLLS}
\end{equation}

The first integral in Eq.~(\ref{eq:attLLScolden}) is dominated by
values $y\sim1$ for fixed $\nu$. The functional form of the integrand
suggests a dependence on the \HI\ density for $y\sim1$ of $\alpha_\nu
\sim n_{\rm HI}^{(\beta-1)}$. In photoionization equilibrium, $n_{\rm
  HI}=n_en_{\rm H}\alpha_A(T)/\Gamma$ for electron density $n_e$,
hydrogen density $n_{\rm H}$ and (Case A) radiative recombination
coefficient $\alpha_A(T)$. The perturbation of the attenuation
coefficient is then $\delta\alpha_\nu=b'_{\chi,
  \delta}\alpha_\nu\delta+(\beta-1)(\delta_{\alpha_A}-\delta_\Gamma)\alpha_\nu$,
where $b'_{\chi,
  \delta}$ is the bias ignoring the contribution from temperature fluctuations and $\delta_{\alpha_A}=\delta\alpha_A/\alpha_A=-0.75\delta_T$ for
$\alpha_A(T)\sim T^{-0.75}$. For an approximate IGM equation of state $T\sim\rho^{\gamma-1}$ (and supposing $\gamma$ is scale-independent),
\begin{equation}
  \delta_{\alpha_\nu}\equiv\frac{\delta\alpha_\nu}{\alpha_\nu}=b'_{\chi, \delta}\delta-(\beta-1)\left[\delta_\Gamma+0.75(\gamma-1)\delta\right].
   \label{eq:danu}
  \end{equation}
This corresponds to $b_{\chi, \Gamma}=1-\beta$. Assuming the
absorption systems trace fluctuations in the density field to linear order for
simplicity, we take $b'_{\chi, \delta}=1$. For a nearly isothermal gas
($\gamma\gsim1$) \citep{2012MNRAS.424.1723G}, $b_{\chi, \delta}\simeq
b'_{\chi, \delta}$; we accordingly set $b_{\chi, \delta}=1$ in Eq.~(\ref{eq:dSnd}).\footnote{Discrepancies persist in the observational determination of $\gamma$ depending on analysis method, which may possibly result from an inadequacy of a simple polytropic equation of state as a description of the thermal state of the IGM. For $2<z<3$, values ranging over $1.0<\gamma<1.6$ have been reported \citep{2014MNRAS.438.2499B, 2018MNRAS.474.2871R}. These all give small corrections to our fiducial $b_{\chi, \delta}=1$.} Because the source bias term dominates over the absorption bias term in Eq.~(\ref{eq:dSnd}), the results are not very sensitive to the absorption bias.

We note that the response of the absorption coefficient to the ionizing background has been studied in simulations and with analytic models finding the absorption coefficient changes by a factor proportional to $(1-\beta') \delta \Gamma$, where $\beta'$ is the spectral index of the column density distribution if there were no self-shielding. As discussed in HM12, the observed column density of \HI\ absorption systems is best approximated as a series of broken power laws, with the power-law index ranging over $\beta = 0.5-0.8$ for Lyman Limit Systems (self shielding flattens the index of Lyman Limits Systems over the intrinsic $\beta'$), and varying over $\beta = 1.5-2$ for the more diffuse \Lya\ forest in good agreement with simulations  \citep{2011ApJ...743...82M, 2014MNRAS.442..187G}.  Our fiducial model with $\beta=1.2$ provides a good match to the HM12 average intensity and \HI\ photoionization rate, although since our model approximates the absorption coefficient response as $(1-\beta) \delta \Gamma$ rather than $(1-\beta') \delta \Gamma$ it may underestimate this response. We therefore also examine a model with $\beta=1.5$.

\subsection{Scaled attenuation model}
\label{subsec:scaledatten}

To examine the roles of source bias and attenuation on the magnitude
of the ionization rate fluctuations, it is instructive to consider a toy
model with the attenuation coefficient scaled to the horizon size (so $\chi$ is
constant).   We compute the power spectrum of fluctuations in the UV
ionization rate for a model with emissivity parameters $a_j=1.8$,
$a_S=0$. A few evolutionary models for the source bias are considered,
all corresponding to $b_j=4$ at $z=3$:\ $b_j=4$, $b_j=(1+z)$ and
$b_j=0.25(1+z)^2$. A matter power spectrum computed with the concordance $\Lambda$CDM parameters is
adopted. Rather than using Eq.~(\ref{eq:mfpL}), we set the attenuation
proportional to the Hubble constant:\ either $\alpha_{\rm eff}=H(z)/c$
or $\alpha_{\rm eff}=10H(z)/c$. A power-law column density
distribution with $\beta=1.2$ is assumed, giving $b_{\chi,
  \Gamma}=1-\beta=-0.2$.  (This simple model does not consider shot noise, which future sections show potentially lead to even larger differences with the steady state.)

\begin{figure}
\scalebox{0.86}{\includegraphics{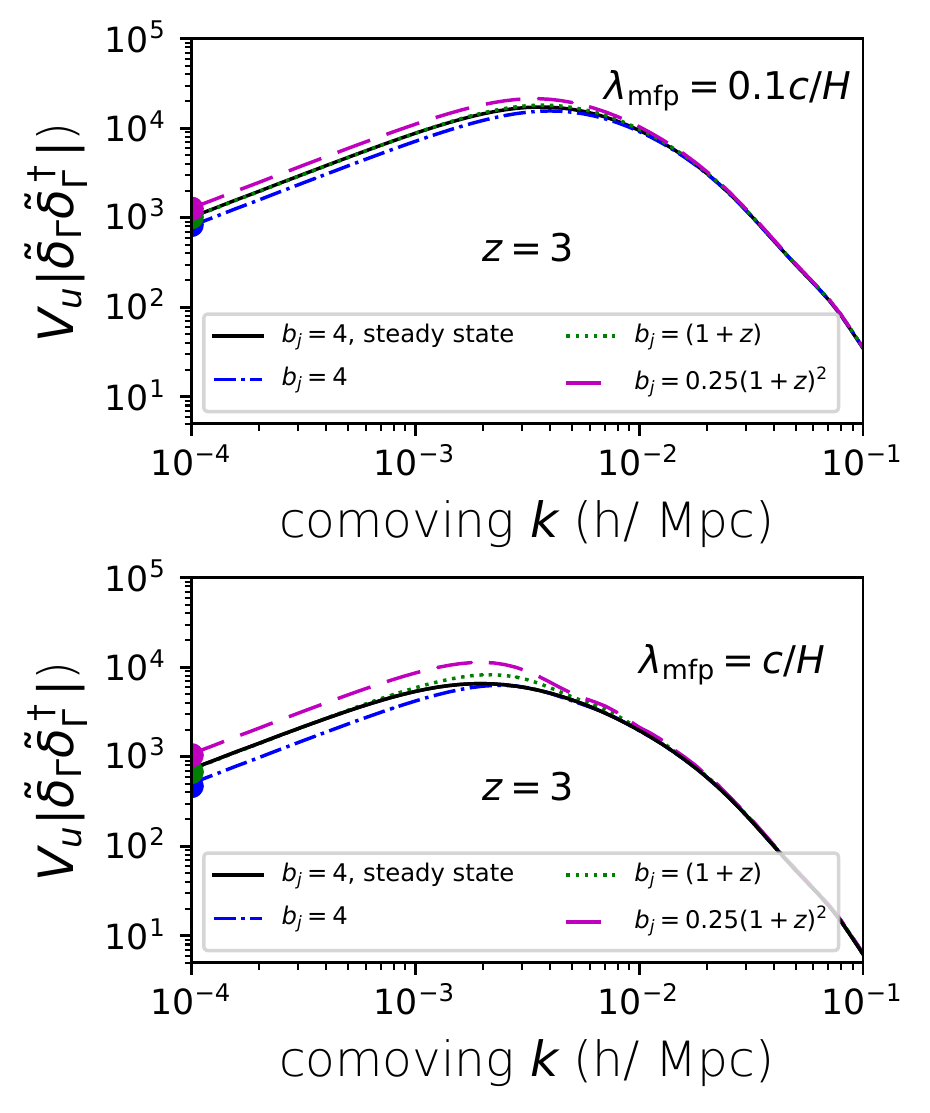}}
\caption{Comoving power spectrum (in units $h^{-3}\,{\rm Mpc}^3$),  of fluctuations in the UV ionization rate as a
  function of comoving wavenumber at $z=3$ with $b_{\chi,
    \Gamma}=-0.2$. (No shot noise is included.) The source bias
  factor is normalized to $b_j=4$ at $z=3$, with results for different
  evolutionary trends shown. The steady-state solutions are shown for
  comparison. The emissivity
  parameters are $\alpha_j=1.8$ and $\alpha_S=0$. (Top panel):\ The
  attenuation coefficient is taken as $\alpha_{\rm
    eff}(z)=10H(z)/c$. (Bottom panel):\ The attenuation coefficient is
  taken as $\alpha_{\rm eff}(z)=H(z)/c$. The points in the lower panel
  on the axis at $k=10^{-4}h{\rm Mpc}^{-1}$ are our analytic $k\rightarrow$0 limits. 
\label{fig:SolvedGammaCorrTDvsSS}
}
\end{figure}

\begin{figure}
\scalebox{0.55}{\includegraphics{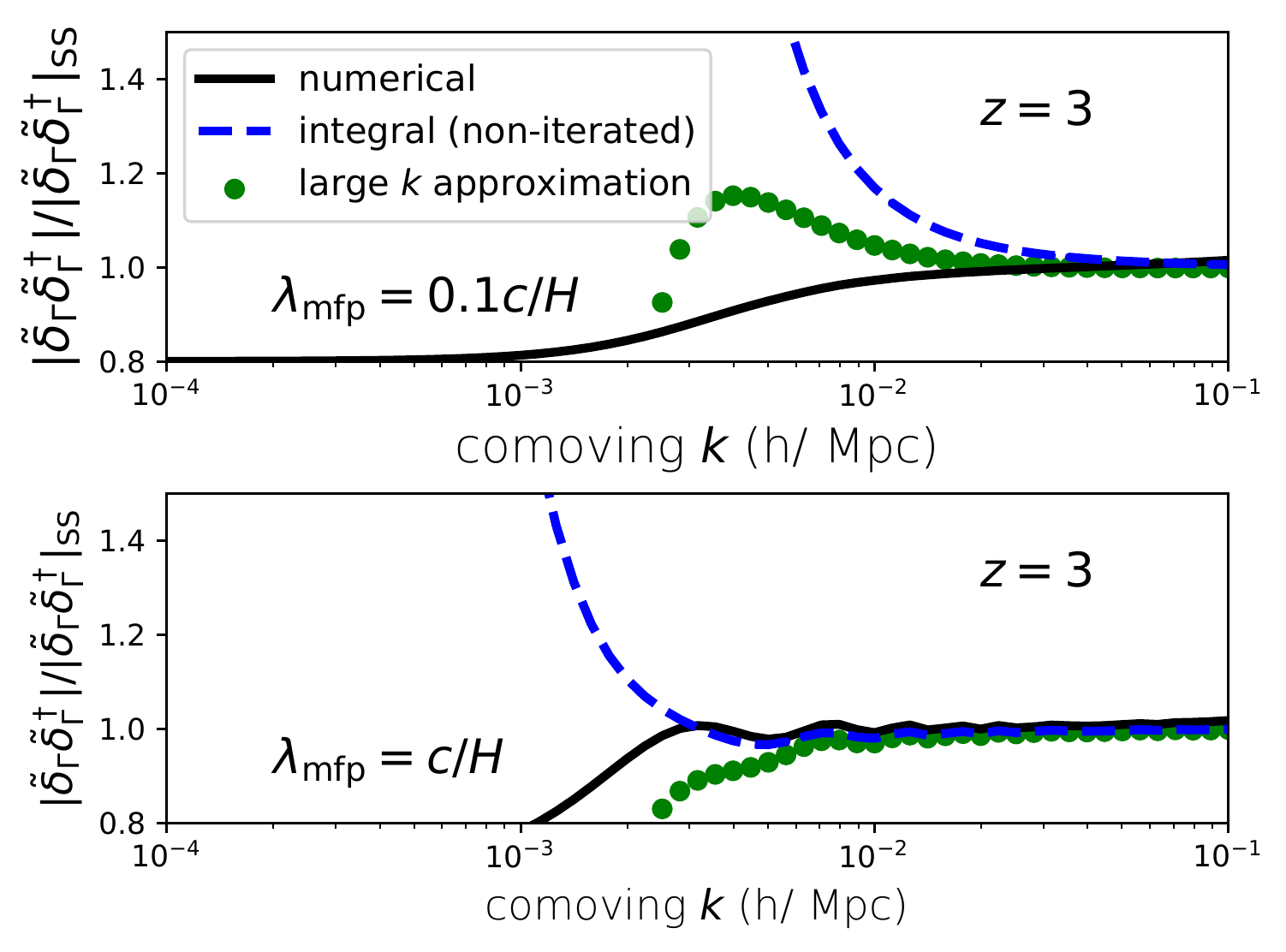}}
\caption{Ratio of time-dependent to steady-state power spectra of
  fluctuations in the UV ionization rate as a function of comoving
  wavenumber at $z=3$ for $b_{\chi, \Gamma}=-0.2$. The source bias
  factor is fixed at $b_Q=4$. The emissivity parameters are
  $\alpha_j=1.8$ and $\alpha_S=0$. (Top panel):\ The attenuation
  coefficient is taken as $\alpha_{\rm eff}(z)=10H(z)/c$. (Bottom
  panel):\ The attenuation coefficient is taken as $\alpha_{\rm
    eff}(z)=H(z)/c$. The black solid curves show the results from the full
  numerical solution of the time-dependent fluctuations.  We also compare with our analytic expressions. The blue dashed
  curves show the results using the integral expression for the
  time-dependent case without iteration (see text), while the green
  points show the asymptotic ratio for large $k$. The reionization
  oscillations are visible in the lower panel.
}
\label{fig:SolvedGammaCorrTDvsSSratio}
\end{figure}

In Fig.\ref{fig:SolvedGammaCorrTDvsSS}, the time-dependent and
steady-state solutions are compared. For $\alpha_{\rm eff}=10H/c$
(corresponding to an effective absorption mean free path $\lambda_{\rm mfp}\equiv\alpha_{\rm eff}^{-1}=0.1c/H$) (top panel), the solution to the
unperturbed time-dependent radiative transfer equation gives
$\phi=0.75$ at $z=3$. Analytic estimates for $k\rightarrow0$ are shown
as points on the $k=0.0001h{\rm Mpc}^{-1}$ axis. The steady-state and
time-dependent solutions agree well at low $k$, as expected in the
large $\chi$ limit.

Choosing instead $\alpha_{\rm eff}=H/c$ ($\lambda_{\rm mfp}=c/H$)
(bottom panel of Fig.\ref{fig:SolvedGammaCorrTDvsSS}), shows good agreement between the steady-state and
time-dependent solutions at low $k$ for $b_j$ held fixed, but
evolution in $b_j$ produces increases by factors of a few for the
time-dependent solutions compared with the steady-state. Here, the
solution to the unperturbed time-dependent radiative transfer equation
gives $\phi=0.82$ at $z=3$, used for the analytic
estimates for $k\rightarrow0$ shown as points on the $k=0.0001h{\rm Mpc}^{-1}$ axis.

The ratios of the time-dependent to steady-state solutions are shown in Fig.\ref{fig:SolvedGammaCorrTDvsSSratio}, adopting the same value of $\phi$ for the steady-state solution as for the time-dependent. For $k\rightarrow0$, the ratio reaches a constant offset as given by Eqs.~(\ref{eq:deltaG0anal}) and (\ref{eq:deltaG0SS}). For large $k$, the ratio approaches 1, as required by Eqs.~(\ref{eq:dGkO2}) and (\ref{eq:dGkSSO2}). The lower panel displays the reionization oscillation predicted by Eq.~(\ref{eq:dGkO2}). Also shown, following discussion in the Appendix, is the non-iterated solution using the full integral Eq.~(\ref{eq:Z00_ap}), except the leading order of $Z_0^{(1)}$, Eq.~(\ref{eq:Z01O2_ap}), has also been added to maintain at least order $1/k^2$ accuracy for large $k$. The large $k$ asymptotic form well recovers the full integral. The approximation breaks down for $k<k_{\rm mfp}=2\pi a/\lambda_{\rm mfp}\simeq0.023\,(0.0023)\,h\,{\rm Mpc}^{-1}$ at $z=3$ for $\lambda_{\rm mfp}=0.1c/H$ ($\lambda_{\rm mfp}=c/H$).

\subsection{Haardt-Madau (2012) QSOs$+$galaxies model}
\label{subsec:HM12tdvsSS}

\subsubsection{Metagalactic UV background model}
\label{subsubsec:uvbg}

\begin{figure}
\scalebox{0.55}{\includegraphics{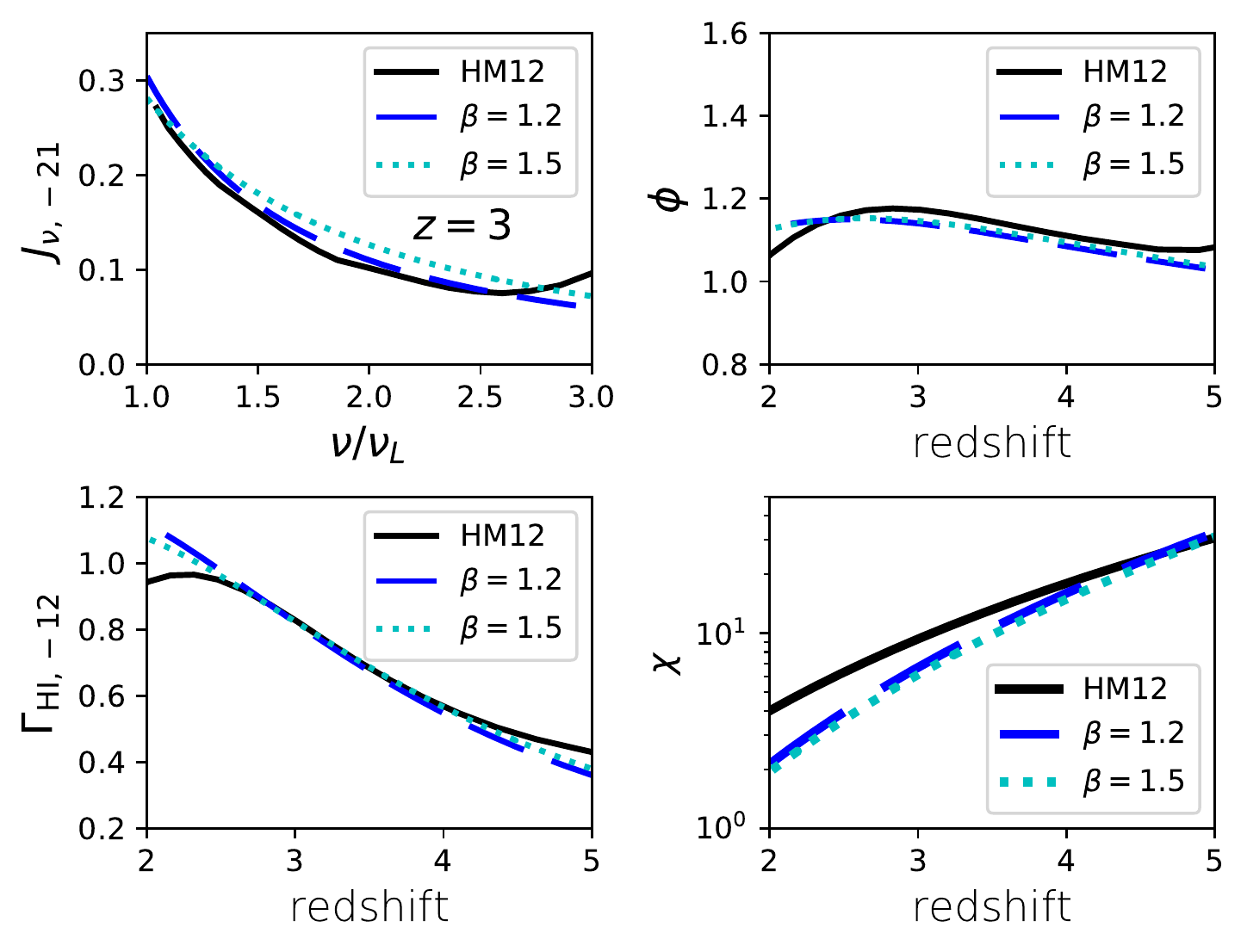}}
\caption{Model metagalactic UV background, compared with the HM12 model. Two models are considered, with $\beta=1.2$ and 1.5. (Top left panel):\ The angle-averaged specific intensity at $z=3$ ($10^{-21}{\rm erg\,s^{-1}\,cm^{-2}\,Hz^{-1}\,sr^{-1}}$). (Bottom left panel):\ The \HI\ photoionization rate ($10^{-12}\,{\rm s}^{-1}$), (Top right panel):\ Dimensionless ratio $\phi$ indicating deviation from a steady-state UV background (see text). (Bottom right panel):\ The flux-weighted mean intergalactic \HI\ attenuation coefficient. See \S~\ref{subsubsec:tdbasics} for definitions.\label{fig:UVbg}}
\end{figure}

The \citet{2012ApJ...746..125H} spectrum for QSO and galaxy
sources is well approximated by an emissivity of the form
Eq.~(\ref{eq:emiss}), with $\alpha_j=1.8$ and $\alpha_S=0.8$, as shown
in Fig.~\ref{fig:UVbg}. Rather than using the HM12 attenuation model,
we adopt the more recent estimate based on Eq.~(\ref{eq:mfpL}). Two
attenuation models are considered, with $\beta=1.2$ and 1.5. The
photoionization rate at $z=3$ is normalized to
$\Gamma_{\rm HI}=0.82\times10^{-12}\,{\rm s}^{-1}$ to match the HM12
model, designed to agree with observational constraints, with
emissivity coefficient
$\epsilon_L=2.4\times10^{25}\,{\rm erg\,s^{-1}\,Hz^{-1}\,Mpc^{-3}}$ in
Eq.~(\ref{eq:emiss}) for $\beta=1.2$. For $\beta=1.5$, the emissivity
spectrum is tilted slightly to $\alpha_j=1.9$, with
$\epsilon_L=2.1\times10^{25}\,{\rm
  erg\,s^{-1}\,Hz^{-1}\,Mpc^{-3}}$.
The corresponding equivalent HM12 value is
$\epsilon_L\sim2\times10^{25}\,{\rm erg\,s^{-1}\,Hz^{-1}\,Mpc^{-3}}.$
We do not include radiative recombination or line emission from the
IGM itself. Whilst IGM emission would dilute fluctuations in the
photoionization background, the effect is expected to be small on the
\HI\ photoionization rate because of the moderate contribution of IGM
emission to the total photoionization rate (\citealt{2009ApJ...703.1416F}, HM12).

The evolution in the \HI\ photoionization rate we find agrees well
with the HM12 estimate. The HM12 spectrum shows a small amount of
time-dependence, with $\phi\simeq1.17$ at $z=3.0$ and decreasing
mildly towards 1 at higher redshifts (but rising again towards
$z=6$)\footnote{Much more rapid evolution in $\phi$ is found for $z<2$, which likely results in larger differences between the time-dependent and steady-state solutions.} Deviations from steady-state evolution in the radiation field,
as quantified through $\phi$, found for the models presented here are
similar to those in the HM12 model. In the comparisons below between
the time-dependent and steady-state UV background fluctuation power
spectra, the same values of $\phi$ are adopted for both.

\subsubsection{Sources}
\label{subsubsec:sources}

The QSO contribution in the HM12 model is based on a QSO emissivity
fit that closely matches the results of \citet{2007ApJ...654..731H}, who
provide several fits to the QSO luminosity function. Here, we
consider a few of the \citet{2007ApJ...654..731H} fits:\ the full redshift
evolution fit to a double-power law
luminosity function (full $z$-fit), the pure luminosity evolution fit
(PLE), the modified Schechter function fit (mS) and the redshift
evolution fit to the high luminosity end (HL). Unless stated
otherwise, we adopt the full $z$-fit model, as this provides the best-fitting and most complete description of the QSO data \citep{2007ApJ...654..731H}. We allow for an evolving QSO
bias factor $b_Q=0.278(1+z)^2+0.57$, based on the extended-BOSS QSO
survey \citep{2017JCAP...07..017L}.

For the galaxy contribution, we adopt the galaxy luminosity function
of \citet{2015ApJ...803...34B}, along with the Lyman Limit escape
fraction $f_{\rm esc}=1.8\times10^{-4}(1+z)^{3.4}$ from
\citet{2012ApJ...746..125H}. The main effect of the galaxies is to
dilute the UV background power spectrum compared with the QSO-only
case. At $z=3$, the contributions of the galaxies and QSOs are
comparable and are similar at all redshifts to the HM12 model.

The galaxy bias factor to use depends on the halo masses of the
galaxies dominating the photoionization background. In general some
weighted average over galaxy halo mass or luminosity should probably
be used. Measurements of high redshift star-forming galaxies suggest a
bias factor at $z=2-3$ of $b_G=2-3$
\citep[e.g.][]{2013MNRAS.430..425B}.

\subsubsection{Shot noise estimates}
\label{subsubsec:shotnoise}

Whilst the mean emissivities predicted by the various QSO luminosity functions are similar, the fluctuations in the emissivity are not. The fluctuations depend little on the lower end of the bolometric luminosity function but, depending on model, may be sensitive to the upper end, particularly if the upper end extends above $10^{16}\,L_\odot$. The effect of changing the upper luminosity is discussed in the Appendix for various QSO luminosity function models. For the results in this paper, we limit the QSO bolometric luminosities to the range $10^{10}-10^{15}\,L_\odot$, corresponding approximately to the range supported by the data \citep{2007ApJ...654..731H}. The effective number density of galaxies is computed for $L>0.01L^*$. Their contribution to the shot noise is negligible compared with that of the QSOs.

\begin{figure}
\scalebox{0.8}{\includegraphics{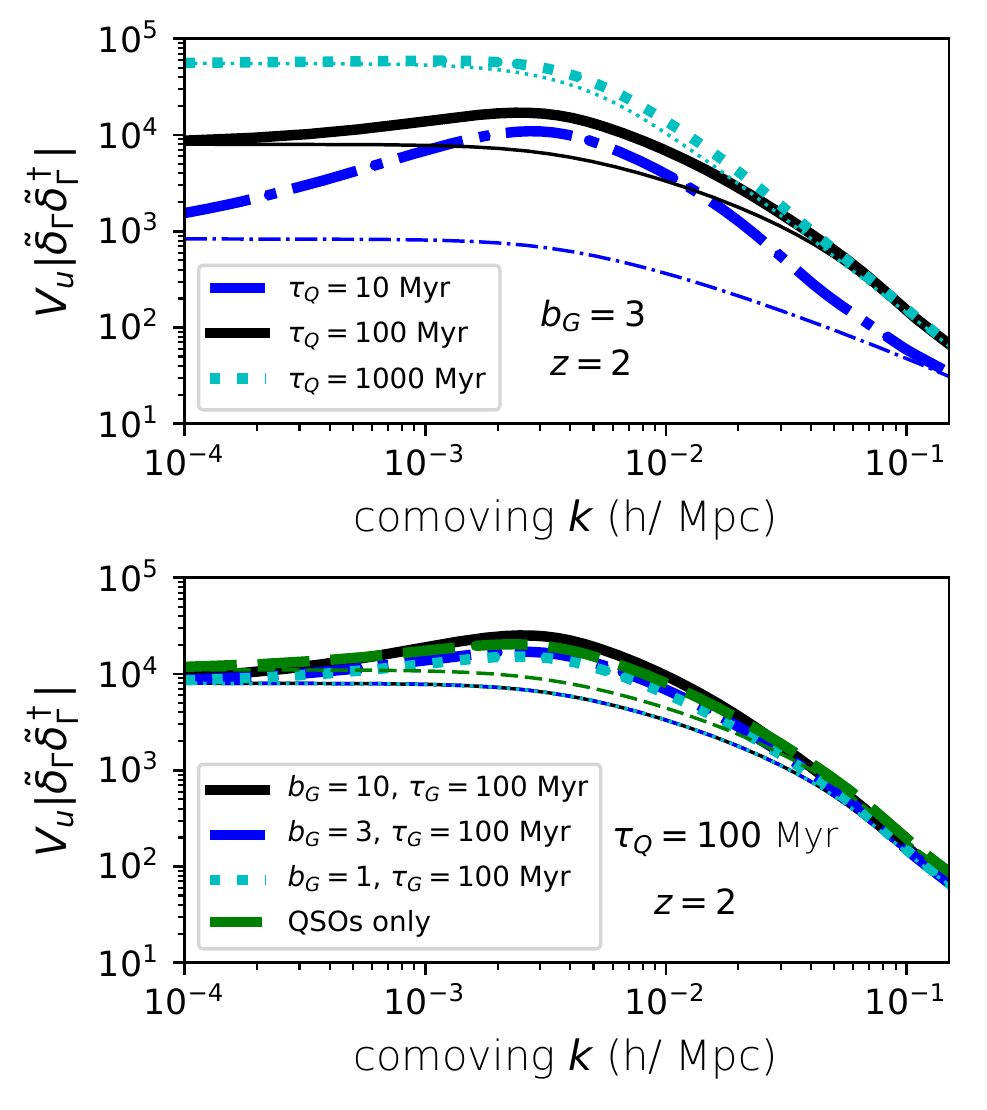}}
\caption{Comoving power spectrum (in units $h^{-3}\,{\rm Mpc}^3$) of
  fluctuations in the photoionization rate as a function of comoving
  wavenumber at $z=2$. The BOSS estimate for QSO bias is used. The
  emissivity parameters are $\alpha_j=1.8$ and $\alpha_S=0.8$, and
  $\beta=1.2$ is adopted for the attenuation coefficient. The heavy
  curves show the total power spectrum while the light curves show the
  shot noise contribution. (Upper panel):\ The QSO shot noise term
  adopts a redshift dependent luminosity density ($z$-evol) and source
  lifetimes of $\tau_Q=10$, 100 and 1000~Myr. A contribution from
  galaxies is added with the indicated bias parameter $b_G=3$ and
  lifetime $\tau_G=100$~Myr.  (Bottom panel):\ The QSO and galaxy
  lifetimes are $\tau_Q=\tau_G=100$~Myr, and the galaxy bias factors
  are $b_G=1$, 3 and 10. Also shown is a case using only QSOs as the
  source of emissivity (green short-dashed curves).\label{fig:SolvedGammaCorrQSOGal}}
\end{figure}

The effect of varying the QSO lifetime on the shot noise is shown in
Fig.~\ref{fig:SolvedGammaCorrQSOGal}. For a lifetime $\tau_Q=10$~Myr,
the shot noise contribution to the total power in the radiation field
fluctuations is subdominant over comoving wavenumbers
$0.001<k<0.04\,h{\rm Mpc}^{-1}$ for $2<z<3$. For $\tau_Q=100$~Myr, the
range reduces to $0.002\lsim k\lsim0.01\,h{\rm Mpc}^{-1}$ for
$2<z<2.5$, and disappears altogether by $z=3$. For $\tau_Q=1000$~Myr,
shot noise dominates everywhere for $2<z<3$. Note that at high
wavenumbers, the shot noise converges to a $k^{-2}$
dependence, given by Eq.~(\ref{eq:shothighk}), with the convergence
extending to lower $k$ values for larger $\tau_Q$. A case with $b_G=3$
and $\tau_G=500$~Myr was also computed; the results agree closely with
the case $b_G=3$ and $\tau_G=100$~Myr, showing the contribution from
galaxies to the total shot noise is negligible, except for diluting
the power spectrum through its contribution to the total emissivity.

We have not considered beaming. Beaming can be included in our formalism by including a beam profile correlation function, ${\cal B}(\nhat -\nhat')$, into our model for the source power-spectrum and then by averaging the differential equation for $\delta_f$ over angle at a later stage. Using a different formalism, \citet{2017MNRAS.472.2643S} showed that beamed quasar emissions will modify the shot noise term.

\subsubsection{Effect of galaxy bias}
\label{subsubsec:galaxybias}

While the shot noise from galaxies is small compared with that of the
QSOs under the model assumptions adopted here, galaxies are an
important contributor to the UV background and its power spectrum and
will dilute the shot noise. The contributions from galaxies for
various bias factors are shown in the lower panel of
Fig.~\ref{fig:SolvedGammaCorrQSOGal} using Eq.~(\ref{eq:powemab}).
Results are shown for unevolving galaxy bias factors of $b_G=1$, 3 and
10, for $\tau_G=100$~Myr. Galaxy bias boosts the power spectrum for comoving wavenumbers $k<0.1\,h{\rm Mpc}^{-1}$. This may be useful as a means of
constraining the rarity (halo masses) of the galaxies dominating the
photoionization background, although the effects are subtle for
$b_G<3$.

As a comparison, a case with only QSOs as the emissivity source
is also shown (green short-dashed curves). Now undiluted by the galaxy contribution to the mean background, the shot noise
rises by about 30 percent (increasing to 70 percent at $z=3$), with the full power spectrum nearly overlying the
power spectrum from QSOs $+$ galaxies for $b_G=3$.

\subsubsection{Time-dependent vs steady-state solutions}
\label{subsubsec:tdvss}

\begin{figure}
\scalebox{0.55}{\includegraphics{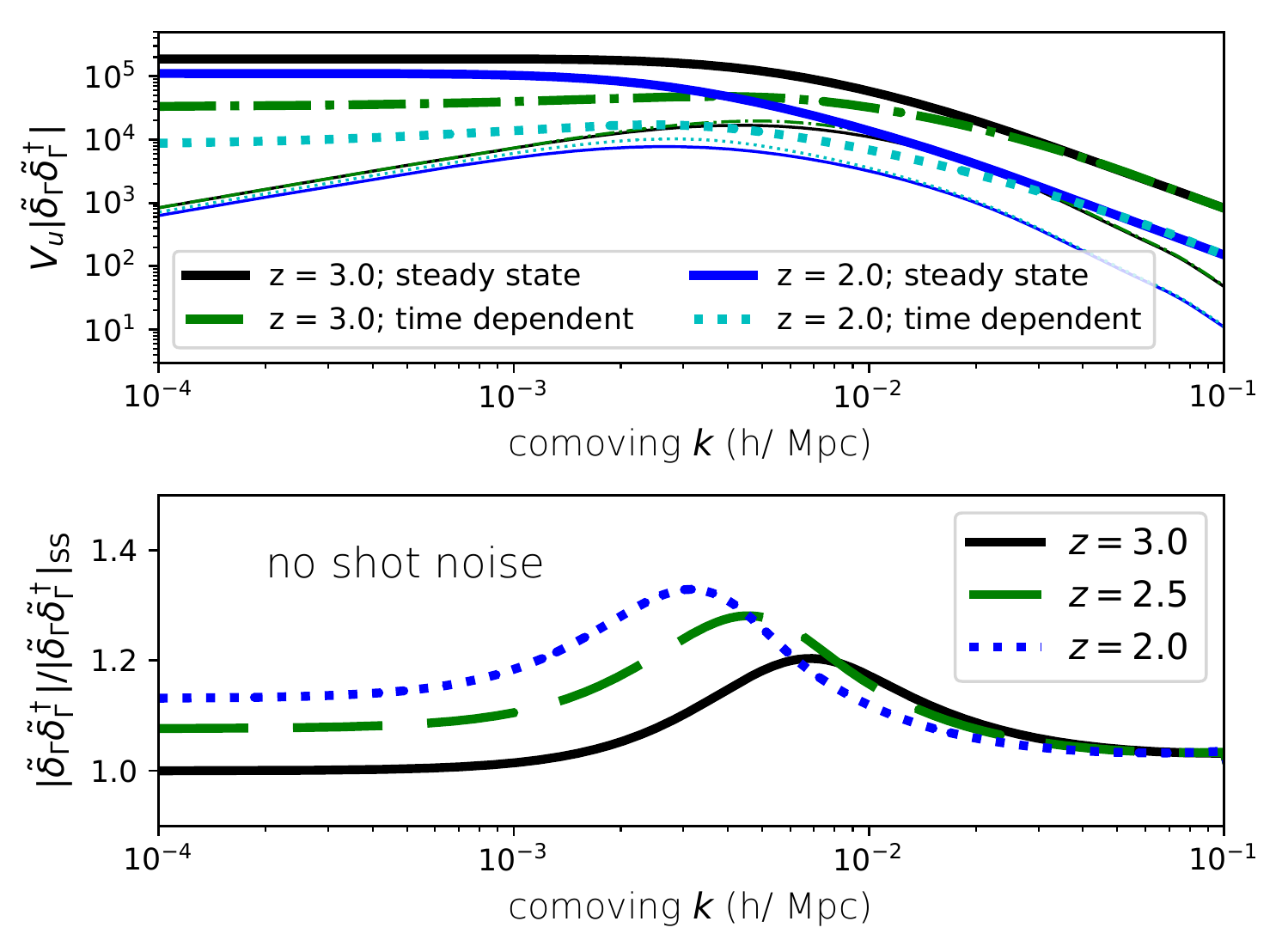}}
\caption{Comparison between time-dependent and steady-state estimates
  of the power spectrum (in units $h^{-3}\,{\rm Mpc}^3$) of
  fluctuations in the photoionization rate over $2<z<3$, as a function
  of comoving wavenumber. The BOSS estimate for QSO bias is used, and
  the galaxy bias is set at $b_G=3$. The emissivity parameters are
  $\alpha_j=1.8$ and $\alpha_S=0.8$, and $\beta=1.2$ is adopted for
  the attenuation coefficient. For the time-dependent model,
  $\tau_Q=\tau_G=100$~Myr is used. The steady-state model adopts the
  same value for $\phi$ as found in the time-dependent model (see
  text). Heavy lines show the power spectra including the shot noise
  contribution, and the light lines without. (Upper panel):\ The QSO
  shot noise term dominates the power in the steady-state model for
  all wavenumbers over $2<z<3$. By contrast, QSO shot noise is
  subdominant over central ranges of wavenumber, depending on redshift
  (see text). (Bottom panel):\ The ratio of time-dependent to
  steady-state power in the radiation fluctuations are shown, with the
  shot noise term excluded. While the estimates converge at high $k$,
  they differ by a few tens of percent at low $k$. A peak in the ratio
  migrates towards lower wavenumbers for decreasing redshift, with the steady-state estimate
  20--30 percent low.
  \label{fig:SolvedGammaCorrTDVSS}
}

\end{figure}

The predictions for the time-dependent and steady-state estimates of the comoving photoionization rate power spectrum are shown in Fig.~\ref{fig:SolvedGammaCorrTDVSS} over the redshift range $2<z<3$ for $\tau_Q=\tau_G=100$~Myr. The steady-state solutions have been normalized using the same value for $\phi$ as found for the mean UV radiation background, shown in Fig.~\ref{fig:UVbg}, thus allowing for a small amount of evolution in the mean background. Without this correction, the steady-state predictions would be $2-3$ percent lower at low $k$ and by $\sim30$ percent at high $k$. The QSO shot noise term dominates the power in the steady-state model over all wavenumbers for $2<z<3$. By contrast, as discussed above, QSO shot noise may be subdominant for a range of wavenumbers in the time-dependent calculation over $2<z<2.5$ , but by $z=3$ shot noise dominates everywhere.

The non-shotnoise contributions agree more closely. For $k>0.1\,h{\rm
  Mpc}^{-1}$, the time-dependent and steady-state models predict
comparable power in the photoionization rate fluctuations. The
asymptotic values at low $k$ disagree by typically 10--20 percent. The
largest level of disagreement is found at intermediate wavenumbers,
with the steady-state estimate 20--30 percent low and with a peak that
grows and migrates towards lower wavenumbers at lower redshifts.

\section{\HI\ fluctuation observational signatures}
\label{sec:HIsignatures}

\subsection{\Lya\ flux redshift-space power spectrum}
\label{subsec:LyaPk}

Fluctuations in the UV background radiation field are detectable
through their effect on intergalactic \HI, in particular as measured
through \Lya\ absorption towards QSOs or bright galaxies. We estimate
the effect using linear perturbations, valid for wide spectral
regions. We follow the approach of \citet{2014MNRAS.442..187G}.

The fluctuations in the measured \Lya\ flux $f_\alpha$ in a bright
background QSO or galaxy are characterised as
\begin{equation}
\delta_\alpha = f_\alpha/\bar f_\alpha-1,
\end{equation}
where $\bar f_\alpha$ is the mean \Lya\ flux across the spectrum (or a
sufficiently broad section of the spectrum to be representative of a
narrow redshift range). The measured flux depends on the gas density,
temperature and velocity and on the photoionization rate. For a wide
range of densities, the temperature is correlated with the
density.\footnote{The correlation becomes noisy during the
  \HeII\ reionization epoch $3\lsim z\lsim 4$, but is expected to
  resume by $z\lsim 2.5$ \citep{2007MNRAS.380.1369T,
    2012MNRAS.423....7M}.} Similarly, the velocity field is correlated
with the density field through mass continuity. Consequently
$\delta_\alpha$ dependences may be approximately confined to gas
density perturbations $\delta$ and ionization rate perturbations
$\delta_\Gamma$.

The dependence of the velocity may be quantified in terms of the peculiar velocity gradient  $\theta=n_in_j(\partial v_i/\partial x_j)$, where $\bm v$ is the peculiar velocity at comoving position $\bm x$, and $\bm n$ is a unit vector specifying direction. The dependence of $\theta$ on the density may be estimated using linear theory for the Fourier modes for wavevector $\bm k$:
\begin{equation}
\tilde\theta=f(\Omega_m)\mu_k^2\tilde\delta,
 \end{equation}
\citep{1987MNRAS.227....1K}, where $\mu_k\equiv {\bm n}\cdot
{\bm k}/ k$ and $f(\Omega_m)$ is a peculiar velocity growth factor. Allowing for bias factors
\begin{equation}
b_\delta=\partial \tilde\delta_\alpha/\partial \tilde\delta, \quad b_\theta=\partial \tilde\delta_\alpha/\partial \tilde\theta\quad {\rm and}\ b_\Gamma=\partial \tilde\delta_\alpha/\partial \tilde\delta_\Gamma,
\end{equation}
the redshift-space \Lya\ flux power spectrum in a comoving volume $V_u$ may be expressed as
\begin{eqnarray}
  P_\alpha(k,\mu_k,z)&\equiv& V_u\langle \tilde\delta_\alpha\tilde\delta^\dagger_\alpha\rangle\nonumber\\
  &=&b_\delta^2P_{\rm L}(k,z)+V_ub_\theta^2\langle \tilde\theta \tilde\theta^\dagger\rangle+V_u b_\Gamma^2\langle \tilde\delta_\Gamma \tilde\delta^\dagger_\Gamma\rangle\nonumber\\
  &&+V_ub_\delta b_\theta\langle \tilde\delta \tilde\theta^\dagger\rangle+V_ub_\delta b_\theta\langle \tilde\theta \tilde\delta^\dagger\rangle\nonumber\\
  &&+V_ub_\delta b_\Gamma\langle \tilde\delta \tilde\delta_\Gamma^\dagger\rangle+V_ub_\delta b_\Gamma\langle \tilde\delta_\Gamma \tilde\delta^\dagger\rangle\nonumber\\
  &&+V_ub_\theta b_\Gamma\langle \tilde\theta \tilde\delta_\Gamma^\dagger\rangle+V_ub_\theta b_\Gamma\langle \tilde\delta_\Gamma \tilde\theta^\dagger\rangle\nonumber\\
  &=&b_\delta^2\left(1+\beta_v\mu_k^2\right)^2P_{\rm L}(k,z)+V_ub_\Gamma^2\langle \tilde\delta_\Gamma \tilde\delta^\dagger_\Gamma\rangle\nonumber\\
  &&+V_ub_\delta b_\Gamma\left(1+\beta_v\mu_k^2\right)\langle \tilde\delta \tilde\delta^\dagger_\Gamma+\tilde\delta_\Gamma\tilde\delta^\dagger\rangle,
  \label{eq:Pkaa}
\end{eqnarray}
where $P_{\rm L}(k,z)$ is the linear matter comoving power spectrum at redshift
$z$ and $\beta_v\equiv f(\Omega_m)b_\theta/b_\delta$. Following
\citet{2014MNRAS.442..187G}, we shall adopt $\beta_v=1$,
$b_\delta=-0.17$ and $b_\Gamma=0.13$, noting that these values are applicable at $z\simeq2$.
  
\begin{figure}
\scalebox{0.55}{\includegraphics{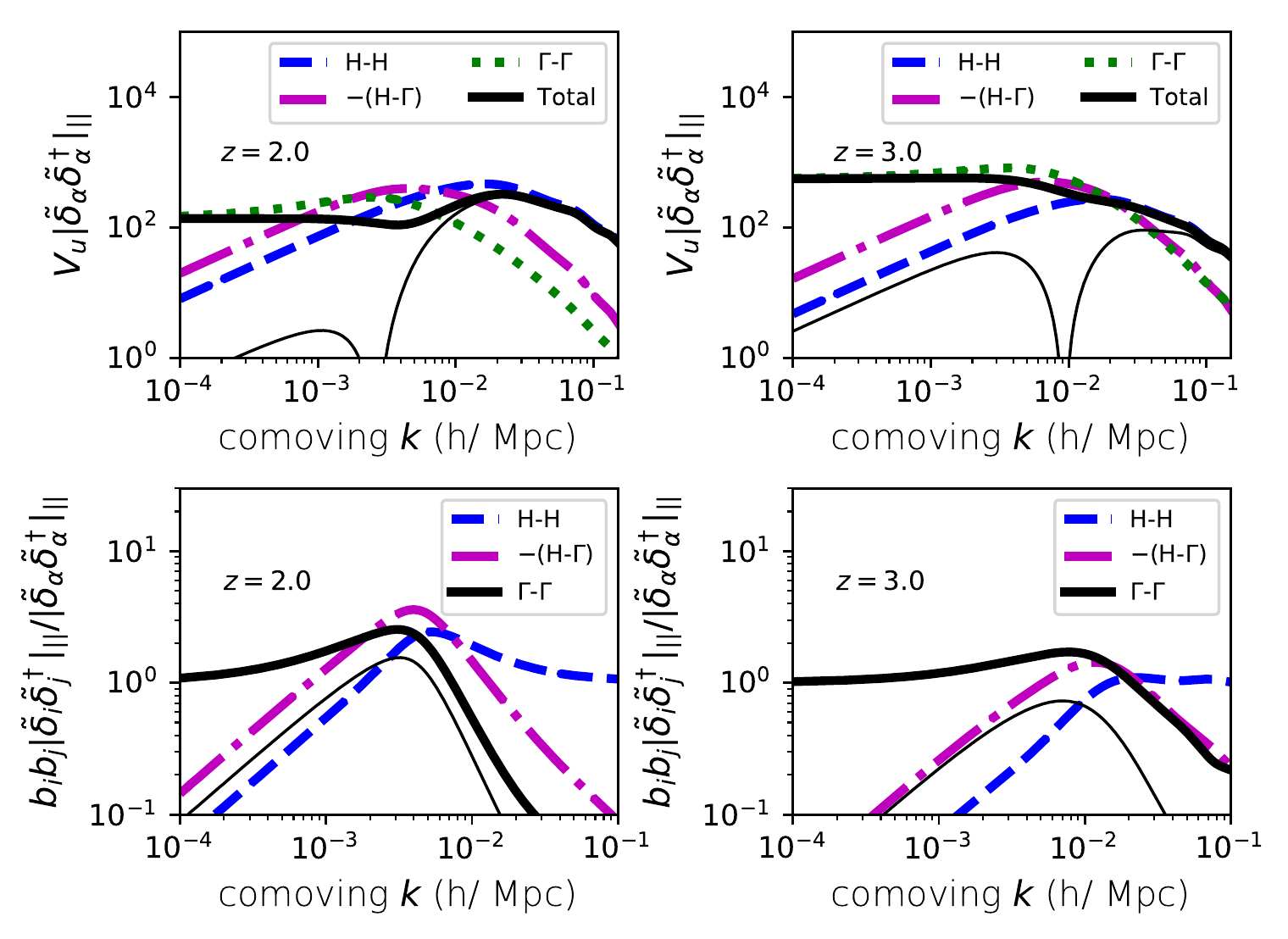}}
\caption{Break-down of the contributions to the line-of-sight component of the
  \HI\ \Lya\ flux comoving redshift-space 3D power spectrum as a function of comoving
  wavenumber. The BOSS estimate for QSO bias is used, and the galaxy
  bias is set at $b_G=3$. The emissivity parameters are $\alpha_j=1.8$
  and $\alpha_S=0.8$, and $\beta=1.2$ is used for the attenuation
  coefficient. Shot noise is included. Results are shown for QSO and
  galaxy lifetimes $\tau_Q=\tau_G=100$~Myr and at redshifts $z=2$ and 3. The gas density fluctuations (blue dashed lines) dominate at high wavenumber while the photoionization rate fluctuations (green dotted lines), including shot noise from the sources, dominate at low wavenumbers. The cross term between the gas density and photoionization rate fluctuations (magenta dot-dashed lines) contributes a large negative component. In the top panels, the thin solid lines show the total power spectrum without the shot noise contribution. In the bottom panels, the thin solid lines show the fractional contribution of the photoionization rate fluctuations without shot noise.
}
\label{fig:Pkmu_tauQ_zred_TDV_components}
\end{figure}

The contributions to the line-of-sight ($\mu_k=1$) component of the comoving \HI\ \Lya\ redshift-space power spectrum from Eq.~(\ref{eq:Pkaa}) are shown in Fig.~\ref{fig:Pkmu_tauQ_zred_TDV_components}. While the gas density fluctuations ($H$-$H$) dominate at high wavenumbers, the fluctuations in the photoionization rate ($\Gamma$-$\Gamma$), including the shot noise from the sources, become increasingly important towards lower wavenumbers and dominate at the lowest. The cross term between the hydrogen density and photoionization rate ($H$-$\Gamma$) reduces the power, producing an inflection in the total power at intermediate wavelengths at comoving $k=0.001-0.01\,h\,{\rm Mpc^{-1}}$, where $\tilde\delta_\alpha$ vanishes, as indicated by the curves without shot noise in the upper panels.

\begin{figure*}
\includegraphics{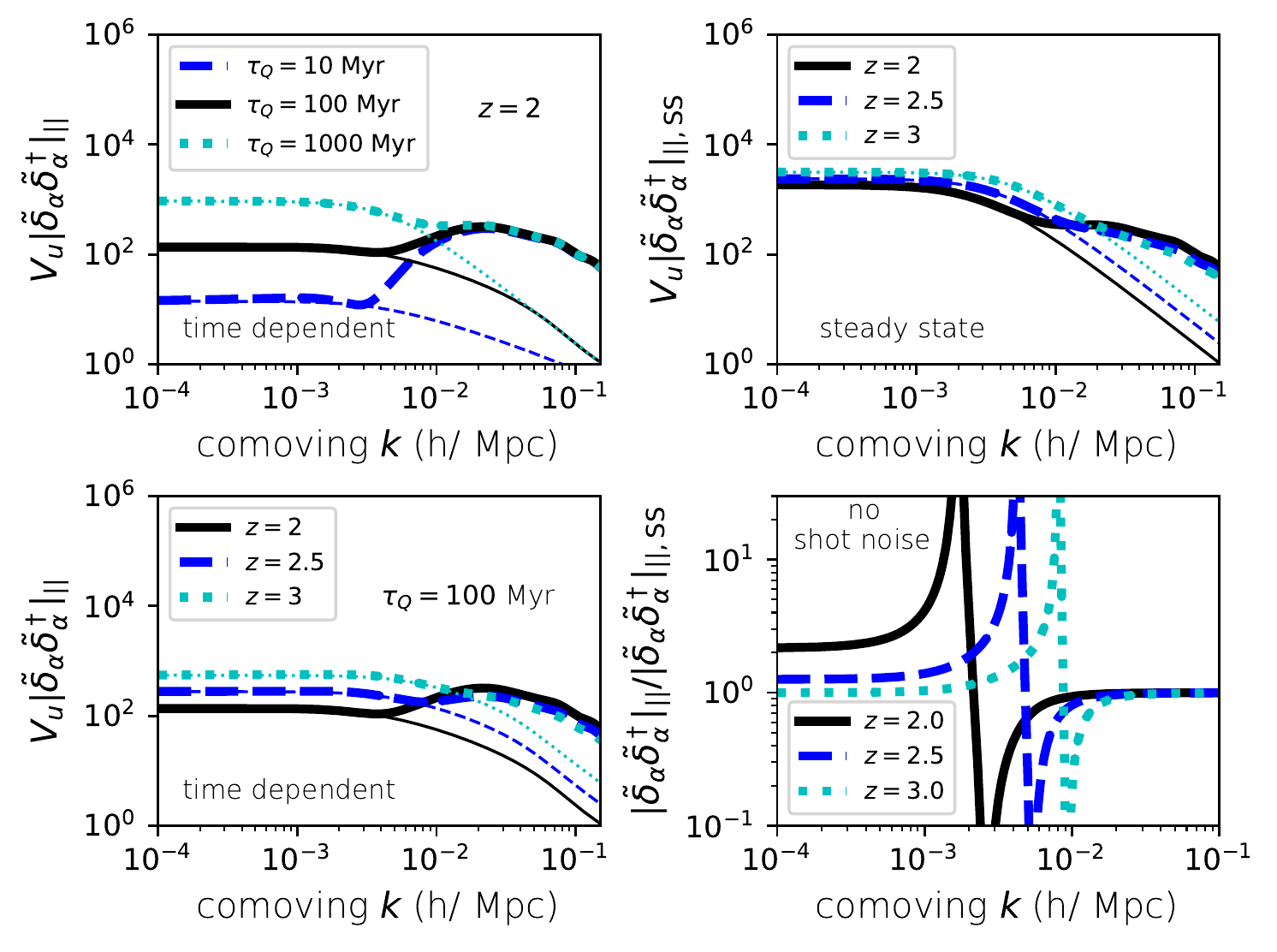}
\caption{Line-of-sight ($\mu_k = 1$) component of the redshift-space comoving 3D power
  spectrum (in units $h^{-3}\,{\rm Mpc}^3$) of \HI\ \Lya\ flux
  fluctuations as a function of comoving wavenumber. The BOSS estimate
  for QSO bias is used, and the galaxy bias is set at $b_G=3$. The
  emissivity parameters are $\alpha_j=1.8$ and $\alpha_S=0.8$, and
  $\beta=1.2$ is used for the attenuation coefficient. Heavy lines
  show the power spectra including the shot noise contribution; light
  lines show the shot noise contribution. (Top left
  panel):\ Prediction for the time-dependent calculation at $z=2$ for
  $\tau_G=100$~Myr and $\tau_Q=10$, 100 and 1000~Myr. Near $k=0.005\,h{\rm
    Mpc^{-1}}$, the dip in power is limited by shot noise for all
  values of $\tau_Q$ shown. (Bottom left panel):\ Evolution of the
  power spectrum for $\tau_Q=100$~Myr. The dip in power migrates to
  lower wavenumbers with decreasing redshift. (Top right
  panel):\ Predictions for the steady-state calculation over
  $2<z<3$. Shot noise dominates the power at low wavenumbers. (Bottom
  right panel):\ The ratio of the time-dependent to steady-state
  calculations without the shot-noise contributions. Whilst the
  predictions agree well at high and low wavenumbers, they disagree
  for a range of intermediate wavenumbers.
\label{fig:Pkmu_tauQ_zred_TDVSS}}
\end{figure*}

\begin{figure}
\scalebox{0.55}{\includegraphics{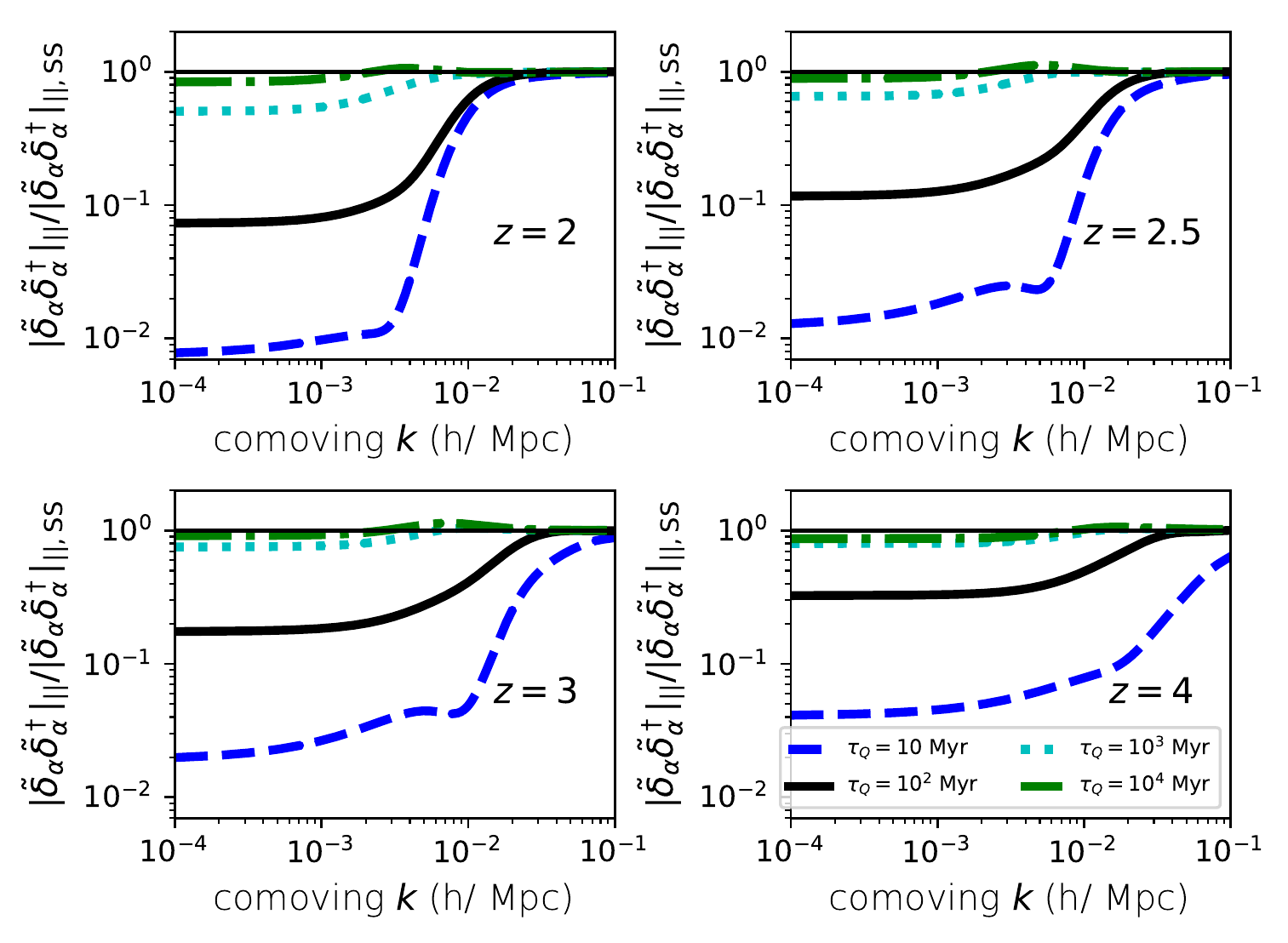}}
\caption{Ratio of time-dependent to steady-state line-of-sight
  component of the redshift-space 3D power spectrum of \HI\ \Lya\ flux
  fluctuations as a function of comoving wavenumber. The BOSS estimate
  for QSO bias is used, and the galaxy bias is set at $b_G=3$. The
  emissivity parameters are $\alpha_j=1.8$ and $\alpha_S=0.8$, and
  $\beta=1.2$ is used for the attenuation coefficient. Shot noise is
  included. Results are shown for galaxy and QSO lifetimes
  $\tau_G=100$~Myr and $\tau_Q=10$, $10^2$, $10^3$ and $10^4$~Myr at
  redshifts $z=2$, 2.5, 3 and 4. The QSO lifetimes curtail the power
  spectra at the low wavenumbers ($k\lsim0.01\,h{\rm Mpc^{-1}}$), for
  which fluctuations in the photoionizing background dominate the
  signal.\label{fig:Pkmu_tauQ_zred_TDVSS_ratio}
}
\end{figure}

The predictions for the total line-of-sight component of the redshift-space comoving 3D \Lya\ flux power spectrum are shown in Fig.~\ref{fig:Pkmu_tauQ_zred_TDVSS} for a range of QSO lifetimes and redshifts for the time-dependent calculation. As shown in the top left panel, for $\tau_Q=10$~Myr the shot noise dominates for $k<0.004\,h\,{\rm Mpc^{-1}}$, extending up to $k<0.006\,h\,{\rm Mpc^{-1}}$ by $\tau_Q=1000$~Myr. At higher wavenumbers, the power spectrum is dominated by density fluctuations rather than photoionization rate fluctuations, but at $k\lesssim 0.01\,h\,{\rm Mpc^{-1}}$ the impact of intensity fluctuation is substantial. Shallower dips to low wavenumbers, limited by shot noise, are apparent for the longer QSO lifetime cases.

Shot noise dominates at $k<0.02\,h\,{\rm Mpc^{-1}}$ in the
steady-state calculation at $z=3$, moving to $k<0.008\,h\,{\rm
  Mpc^{-1}}$ at $z=2$, as shown in the top right panel of
Fig.~\ref{fig:Pkmu_tauQ_zred_TDVSS}. The steady-state and time-dependent predictions for the power agree at high wavenumbers, as shown in the lower right panel. The values converge towards approximate agreement asymptotically at low wavenumbers, although with an offset, as discussed in Sec.~\ref{subsubsec:tdvss:limits}. For a range of intermediate wavenumbers near the dip in power in the time-dependent computation there is substantial disagreement.

The ratios of the time-dependent to steady-state solutions are shown in Fig.~\ref{fig:Pkmu_tauQ_zred_TDVSS_ratio} for a range of QSO lifetimes and redshifts. The region of the dip moves towards higher wavenumbers for higher redshifts. Measurements of the power spectrum at $k<0.01-0.1\,h\,{\rm Mpc^{-1}}$ may provide a means of constraining the lifetime of QSO sources.

\begin{figure}
\scalebox{0.55}{\includegraphics{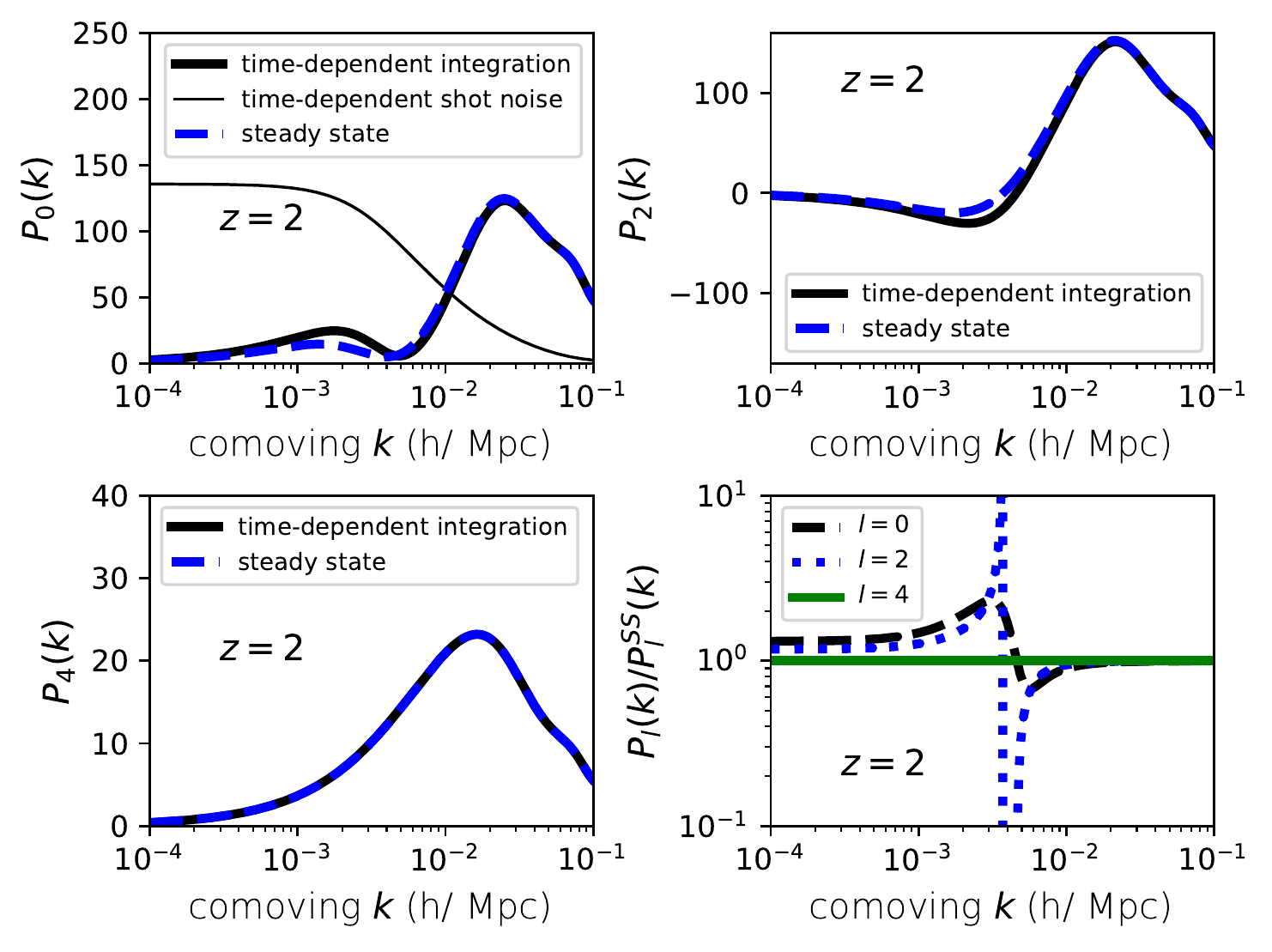}}
\caption{Legendre components of comoving redshift-space power spectrum (in units $h^{-3}\,{\rm
    Mpc}^3$) of \HI\ \Lya\ flux fluctuations as a function of comoving
  wavenumber at $z=2$. The BOSS estimate for QSO bias is used, and the
  galaxy bias is set at $b_G=3$. The emissivity parameters are
  $\alpha_j=1.8$ and $\alpha_S=0.8$, and $\beta=1.2$ is used for the
  attenuation coefficient. The shot noise term from sources is not
  included. The panels show $P_0(k)$ (top left), $P_2(k)$, (top
  right), and $P_4(k)$ (bottom left).) Solid lines show the power
  spectra for the time-dependent calculation; dashed lines show the
  steady-state results. The light solid line in the top left panel
  shows the shot noise contribution with $\tau_Q = 100$~Myr for the
  time-dependent calculation. The bottom right panel compares the
  time-dependent and steady-state calculations. Whilst the
  calculations for $P_0(k)$ and $P_2(k)$ agree well at high and low
  wavenumbers, they disagree for a range of intermediate
  wavenumbers. The calculations for $P_4(k)$ agree exactly for all
  wavenumbers, as the photoionization rate fluctuations do not
  contribute.
}
\label{fig:Pkl_TDVSS}
\end{figure}

The angular dependence introduced by redshift space distortions on the
\Lya\ flux power spectrum may be decomposed into its Legendre
components
\begin{equation}
P_{\it l}(k,z) = \frac{2{\it
    l}+1}{2}\int_{-1}^1\,d\mu_k\,P_\alpha(k,\mu_k,z)L_{\it l}(\mu_k),
\label{eq:Plkz}
\end{equation}
\citep{2013JCAP...03..024K}, where $L_{\it l}(\mu_k)$ is a Legendre
polynomial of order ${\it l}$. In the linear density approximation
(and assuming a flat sky) only the ${\it l}=0, 2$ and 4 components are
non-vanishing. The steady-state estimates for $P_0(k)$, without the shot noise contribution, and $P_2(k)$ disagree with the time-dependent calculation over
$0.001<k<0.01\,h\,{\rm Mpc^{-1}}$, depending on redshift, as shown in
Fig.~\ref{fig:Pkl_TDVSS}. Good agreement is found for $k>0.02\,h\,{\rm
  Mpc^{-1}}$, which includes the regions for the BAO peaks. Shot noise
dominates the $l=0$ component at wavenumbers $k<0.01\,h\,{\rm
  Mpc^{-1}}$ for the model shown, with $\tau_Q=100$~Myr and
$\tau_G=100$~Myr. (Shot noise does not contribute to the other components.) Both the time-dependent and the steady-state
estimates agree for $P_4(k)$ for all wavenumbers, as the $l=4$
component does not depend on the photoionization rate fluctuations.

\subsection{\Lya\ flux redshift-space correlation function}
\label{subsec:LyaXI}

\begin{figure}
\scalebox{0.55}{\includegraphics{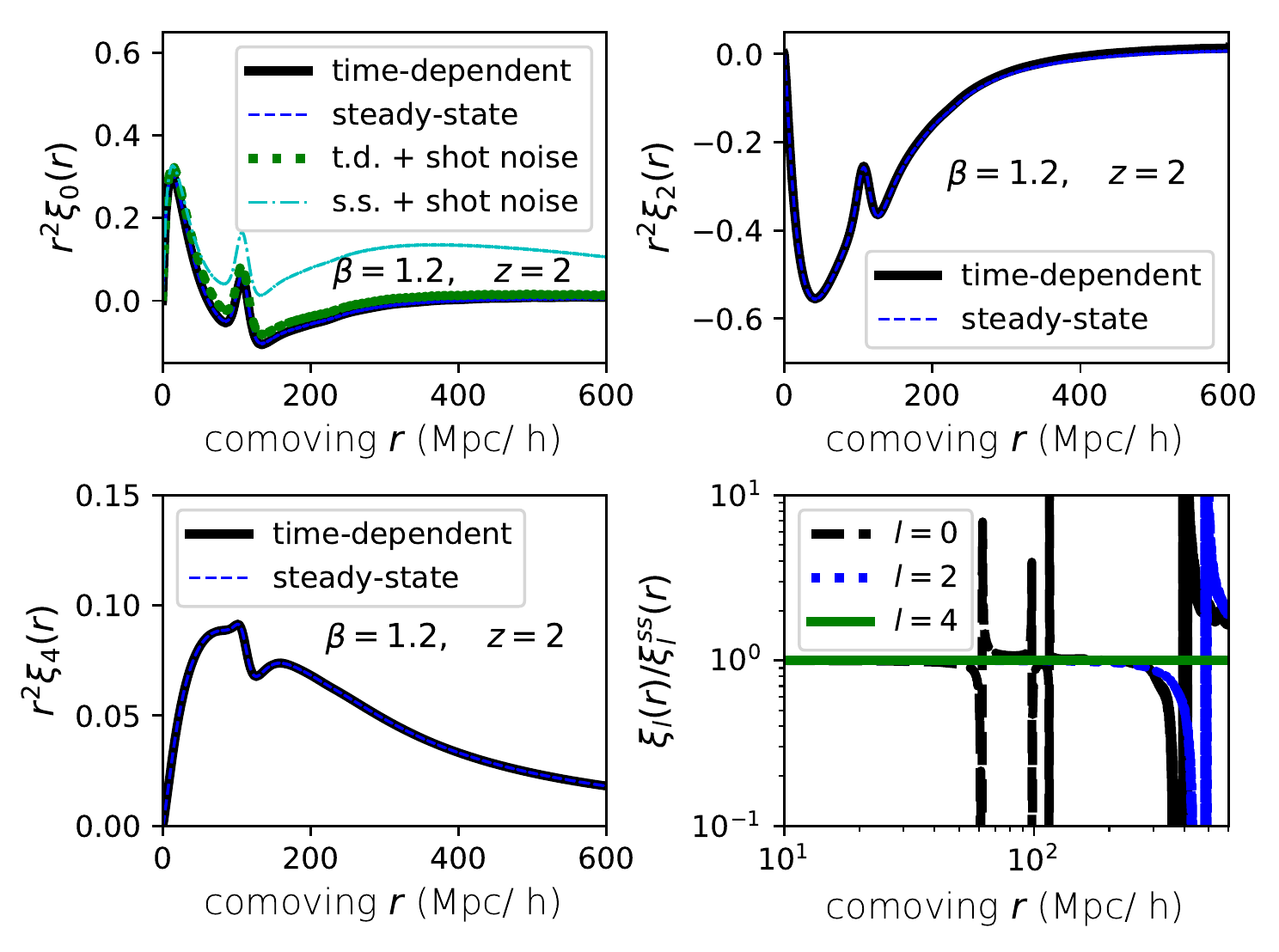}}
\caption{Legendre components of \HI\ \Lya\ flux redshift-space spatial
  correlation function as a function of separation at $z=2$. The BOSS
  estimate for QSO bias is used, and the galaxy bias is set at
  $b_G=3$. QSO and galaxy lifetimes of $\tau_Q=100$~Myr and
  $\tau_G=100$~Myr are assumed for the time-dependent calculation. The
  emissivity parameters are $\alpha_j=1.8$ and $\alpha_S=0.8$, and
  $\beta=1.2$ is used for the attenuation coefficient. The panels show
  $\xi_0(r)$ (top left), $\xi_2(r)$, (top right), and $\xi_4(r)$
  (bottom left). Solid lines show the correlation function for the
  time-dependent calculation; dashed lines show the steady-state
  results, both without the shot noise contribution. The dotted line
  shows the time-dependent solution for $\xi_0(r)$ including shot
  noise, and the dot-dashed line shows the steady state solution
  including shot noise. The bottom right panel compares the
  time-dependent and steady-state calculations (both without shot
  noise).
}
\label{fig:Xil_TDVSS_z2}
\end{figure}

\begin{figure}
\scalebox{0.55}{\includegraphics{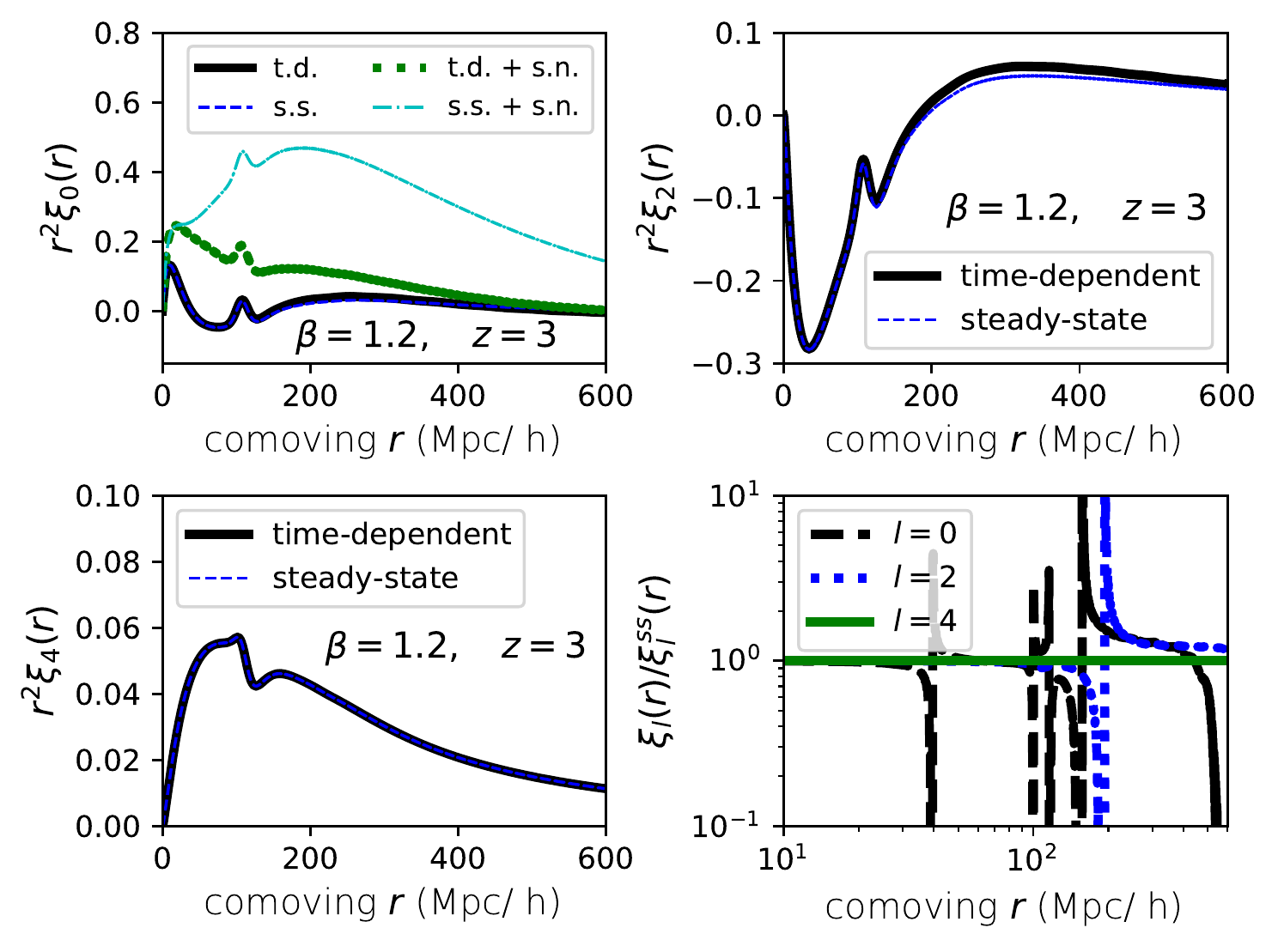}}
\caption{Legendre components of \HI\ \Lya\ flux redshift-space spatial
  correlation function as a function of separation as in Fig.~\ref{fig:Xil_TDVSS_z2}, at $z=3$. The steady-state estimate of $\xi_0(r)$ falls short of the
  time-dependent calculation for separations over a broad range at large separations.
}
\label{fig:Xil_TDVSS_z3}
\end{figure}

\begin{figure}
\scalebox{0.55}{\includegraphics{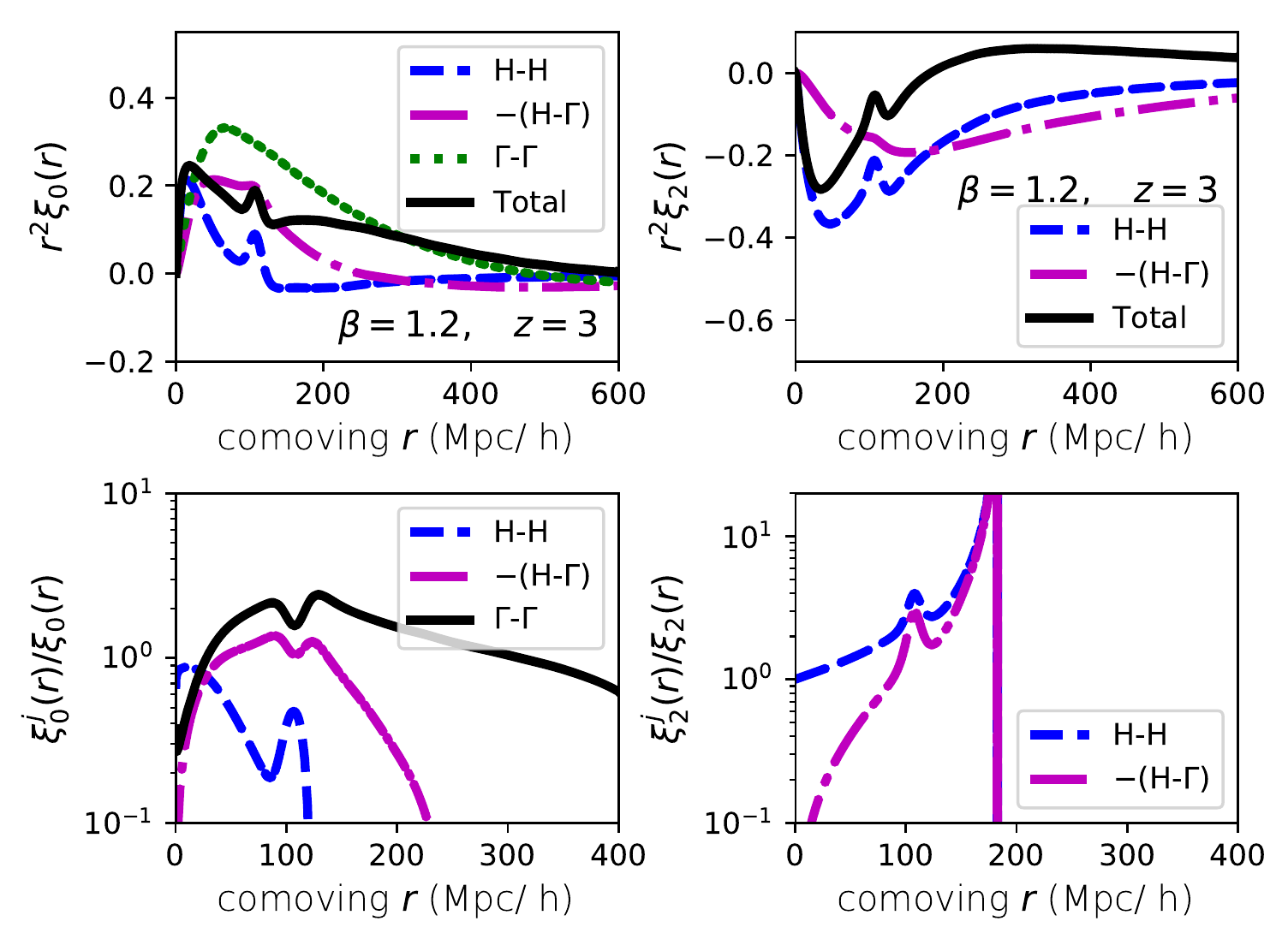}}
\caption{Break-down of the contributions to the $l=0$ and $l=2$ Legendre components of the \HI\ \Lya\ flux redshift-space spatial correlation
  function as a function of separation at $z=3$. The BOSS estimate for QSO bias is used, and the galaxy
  bias is set at $b_G=3$. The emissivity parameters are $\alpha_j=1.8$
  and $\alpha_S=0.8$, and $\beta=1.2$ is used for the attenuation
  coefficient. Shot noise is included (affecting only the $l=0$ component). Results are shown for QSO and
  galaxy lifetimes $\tau_Q=100$~Myr and $\tau_G=100$~Myr. The various components contribute comparable amounts for comoving separations $r>50\,h^{-1}\,{\rm Mpc}$. In particular, the shape of the BAO peak near $r\lsim110\,h^{-1}\,{\rm Mpc}$ is sensitive to the photoionization rate fluctuations, although the position of the peak is little affected.
}
\label{fig:Xil_TDV_components}
\end{figure}

The Legendre components of the redshift-space correlation function
corresponding to the power spectrum components are given by
\begin{equation}
\xi_{\it l}(r,z) = \frac{i^{\it
    l}}{2\pi^2}\int_0^\infty\,dk\,k^2j_{\it l}(kr)P_{\it l}(k,z).
\label{eq:xirz}
\end{equation}
The angular dependence may be recovered through
\begin{equation}
\xi(r,\mu,z)=\sum_{{\it l}=0}^2 L_{\it l}(\mu)\xi_{\it l}(r,z),
\label{eq:ximurz}
\end{equation}
where $\mu=\hat{\bm n} \cdot {\bm r}/ r$.

The Legendre components of the spatial correlation function of the
\HI\ \Lya\ flux are shown in Figs.~\ref{fig:Xil_TDVSS_z2} and
\ref{fig:Xil_TDVSS_z3} at $z=2$ and 3. Except for slight offsets in
the positions of the zero-crossings, the time-dependent and
steady-state calculations of $\xi_0(r)$, without the shot noise
contribution, and $\xi_2(r)$ agree well for separations
$r<200h^{-1}\,{\rm Mpc}$, while discrepancies arise at larger
separations. The range of discrepancy increases at the higher
redshift. As expected, no discrepancy is found for $\xi_4(r)$. The BAO
peak, prominent at $r\lsim110h^{-1}\,{\rm Mpc}$, is accurately
recovered by the steady-state calculation, with any shift in its comoving position compared with the time-dependent calculation smaller than $0.2\,h^{-1}\,{\rm Mpc}$ over $2<z<3$.

Adding in the shot noise can substantially alter the spatial correlations. Whilst the shot noise contributes little to the time-dependent solution at $z=2$, it is a
major contributor at $z=3$ for the model shown, with QSO and galaxy lifetimes of $\tau_Q=100$~Myr and $\tau_G=100$~Myr. The shot noise in the steady state solutions, corresponding to the infinite QSO lifetime limit, is much higher, dominating most of the signal. As for the redshift-space power spectrum, the correlation function provides a means of constraining the lifetimes of the sources.

The separate contributions to the $l=0$ and $l=2$ Legendre components of the spatial correlation function at $z=3$ are shown in Fig.~\ref{fig:Xil_TDV_components}. (We choose $z=3$ rather than $z=2$, at which the correlation function has multiple zero-crossings, for clarity of presentation.) While the gas density fluctuations dominate for comoving separations $r<10\,h^{-1}\,{\rm Mpc}$, by $r>50\,h^{-1}\,{\rm Mpc}$ the contributions from all the components are comparable. The structure near the BAO peak in particular is determined by all the components, with the photoionization rate shot noise term substantially boosting the peak value. Nonetheless, the positions of the peaks in the $l=0$ and 2 components are only slightly shifted compared with the gas density fluctuation contribution alone, increasing by about $\Delta r\simeq0.5\,h^{-1}\,{\rm Mpc}$ (comoving) at $z=2$, and by $\Delta r\simeq1\,h^{-1}\,{\rm Mpc}$ at $z=3$, when photoionization rate fluctuations are included, confirming the small values reported based on the steady-state approximation \citep{2014MNRAS.442..187G,2014PhRvD..89h3010P}. Because the scale of the BAO peak is fairly insensitive to $\Omega_m$, even a small change in the peak position may substantially bias estimates of $\Omega_m$ based on measurements of the \Lya\ flux redshift-space power spectrum, so that careful modelling of the effects of the photoionization background fluctuations is required for precision estimates.

\begin{figure}
\scalebox{0.55}{\includegraphics{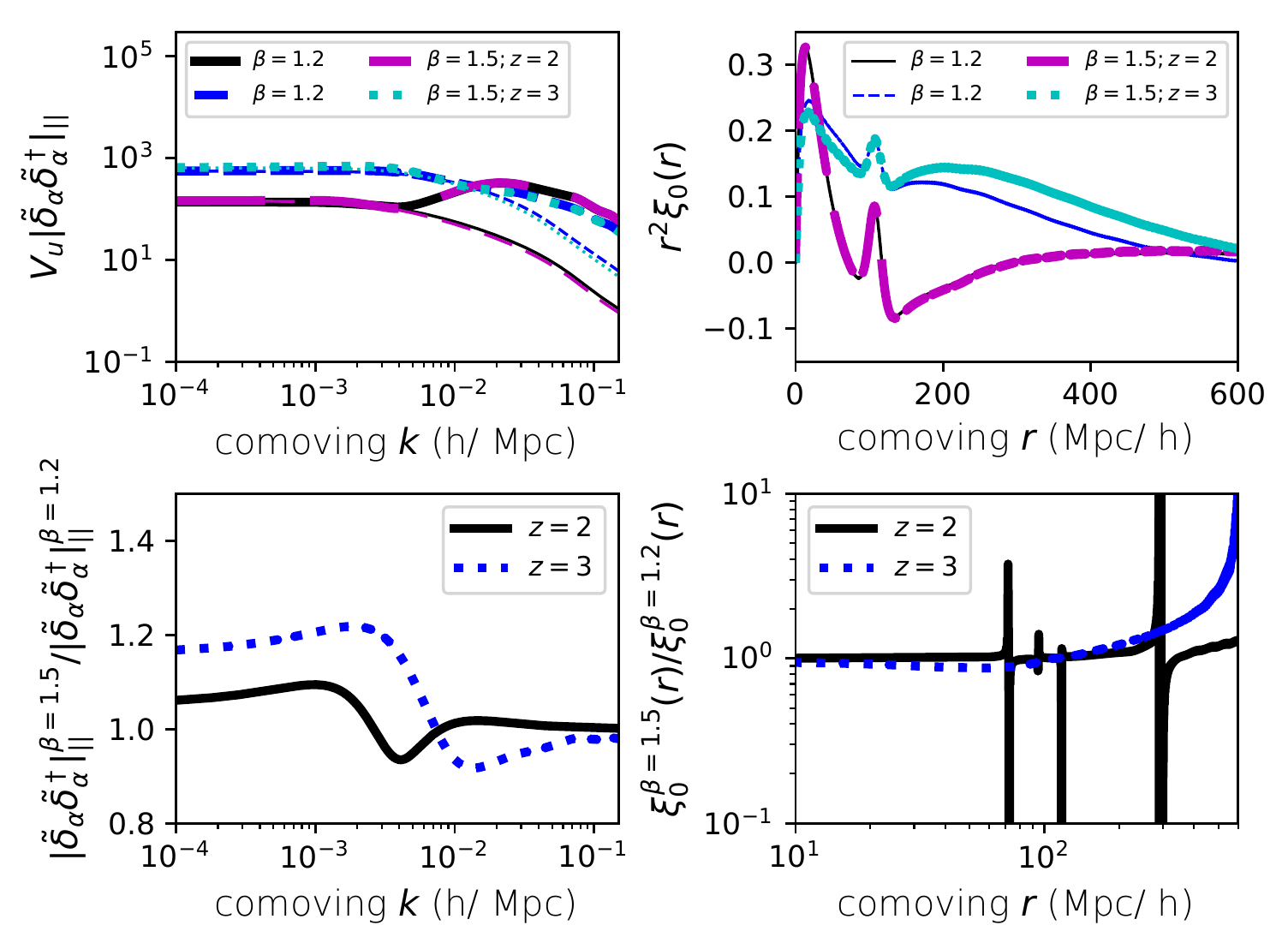}}
\caption{Comparison between attenuation coefficient cases $\beta=1.2$
  and 1.5 for the time-dependent calculation, at $z=2$ and 3. The BOSS
  estimate for QSO bias is used, and the galaxy bias is set at
  $b_G=3$. The emissivity parameters are $\alpha_j=1.8$ and
  $\alpha_S=0.8$. The left panels show the line-of-sight component of
  the redshift-space \Lya\ flux comoving power spectra (in units
  $h^{-3}\,{\rm Mpc}^3$) and their ratio (including shot noise) for
  $\beta=1.5$ to 1.2. The thick lines show the full power spectrum
  including shot noise, the thin lines in the upper panel show the
  shot noise contributions. The right panels show $\xi_0(r)$
  (including shot noise) and their ratios.
}
\label{fig:Pk_Xi0_beta_comp}
\end{figure}

As an alternative attenuation model, we also consider a column density
distribution with $\beta=1.5$, weighting the attenuation more towards
low \HI\ column density systems. Little difference is found in the
mean free paths and photoionization rates between the models, as shown
in Fig.~\ref{fig:UVbg}, although the required emissivity is somewhat
lower for the $\beta=1.5$ model. The resulting line-of-sight component
of the \Lya\ flux redshift-space power spectra and Legendre $l=0$
spatial correlation function components are shown in
Fig.~\ref{fig:Pk_Xi0_beta_comp}. In spite of the similarities in the
photoionization rates between the models, substantial differences are
found in the \Lya\ flux power spectra. Whilst at high wavenumbers the
power spectrum is little affected, it is substantially boosted for
$\beta=1.5$ compared with $\beta=1.2$ for (comoving) $k<0.01\,h\,{\rm
  Mpc^{-1}}$, by as much as a factor of 1.5--2, dependent on redshift.\footnote{Choosing $\beta=1.9$ instead of $\beta=1.2$, whilst renormalizing the emissivity to keep $\Gamma$ fixed at $z=3$, we find $\vert\tilde\delta_\Gamma/\tilde\delta_j\vert^2$ for $k\rightarrow0$ increases by 70 percent at $z=2$, a factor of 3 at $z=3$ and a factor of 6 at $z=4$ in the time-dependent computation. Very similar enhancements are found for the steady-state computation. An attenuation model treating the IGM absorption as arising predominantly from a diffuse component corresponds to taking $\beta\rightarrow2$, which will over-estimate the magnitude of the photoionization rate fluctuations at low wavenumbers.} The positions and depths of the shot-noise limited dips are almost
unaffected, suggesting the use of the dips for estimating the
lifetimes of QSOs is fairly robust against uncertainties in the spectral
shape of the attenuation coefficient. Except for slight offsets in the
positions of the zero crossings, $\xi_0(r)$ is little affected at
$z=3$. By $z=2$, the correlation functions are nearly identical. The
position of the BAO peak is essentially unaltered ($\Delta r < 0.2\,h^{-1}\,{\rm Mpc}$), between the two models.

\section{Frequency dependent solutions}
\label{sec:fullfreq}

The approach in this manuscript is to average the radiative transfer equation over frequency before solving it perturbatively.  Here we briefly comment on the full frequency-dependent solutions. Analogous to our Eq.~(\ref{eq:dfnd}), the general solution to the radiative transfer equation can be written as
\begin{equation}
 \widetilde{\delta_{I_\nu}}(\bar t) = \int_{\bar t_i}^{\bar t} d \bar t' G_{\nu'}(\bar t,  \bar t' ) b_{\chi_{\nu'}, \Gamma}(\bar t' ) \chi_{\nu'}(\bar t' ) \left[\tilde \delta_{S, {\nu'}}(\bar t' ) -\tilde \delta_\Gamma(\bar t' ) \right], 
 \label{eq:fullsoln}
\end{equation}
where $\nu'\equiv \nu \,[a(\bar t)/a(\bar t')]$ and all quantities are defined in same manner as Eqs.~(\ref{eq:favg})-(\ref{eq:chi}) except we do not average over frequency (keeping $\nu$ subscripts to indicate non-averaged quantities).\footnote{As with our solution to the frequency averaged equation, this solution to the frequency dependent does ignore the term from spatial fluctuations in the spectral index of the ionizing background as including this term requires solving for multiple frequencies rather than the photoionization weighting of frequencies done here.} For example, $\chi_\nu \equiv c\langle \alpha_\nu \rangle/H$.  Eq.~(\ref{eq:fullsoln}) makes the simplifying assumption that the dimensionless frequency-dependent attenuation coefficient $\chi_\nu$ traces fluctuations in $\Gamma$ with bias $b_{\chi_\nu, \Gamma}$ (rather than tracing the more general bias expansion in $\delta_{I_\nu}$).   The Green's function for the frequency-dependent solution is defined as\footnote{Note that we could have just written this off the bat from our knowledge of \citet{1996ApJ...461...20H}-like models, which solve
$$
\langle I_\nu \rangle = \int_{\bar t_i}^{\bar t} d\bar t' G_{\nu'}(\bar t, \bar t')\Big|_{k=0} \langle j_{\nu'} \rangle(\bar t'),
$$
where $ \langle j_{\nu'} \rangle$ is the average emission coefficient of all sources.
The frequency independent solution solves the same equation as these models (except for $k\neq 0$) when working with total fluctuations.}  
\begin{equation}
G_\nu(\bar t, \bar t' ) =   \frac{\langle I_\nu (\bar t') \rangle}{\langle I_\nu ( \bar t) \rangle}\left( \frac{a'}{a} \right)^{3} \exp \left\{ \int_{\bar t'}^{\bar t} d\bar t'' \left[i  \frac{{\bm \kappa}(\bar t'')\cdot \nhat}{a(\bar t'')} - \chi_\nu(\bar t'') \right] \right \}.
\end{equation}
The photoionization rate fluctuations can then be calculated by integrating $\langle  f \rangle^{-1} \sigma_\nu/[h_p \nu] \langle I_\nu \rangle   \widetilde{\delta_{I_\nu}}$ over frequency and angle and, then, solving for $\delta_\Gamma$.  This results in similar matrix equations to the ones we solved in \S~\ref{subsec:meth} except that to evaluate each matrix element requires an additional integral over frequency (which may be possible to evaluate analytically by breaking up the integrand into terms with different power-law dependences).  In addition, if one has a model for the bias as a function of column density (such as in \citealt{2018JCAP...04..026I}) and how $N_{\rm HI}$ responds to a change in $\Gamma$ (such as in \citealt{sanderbeckinprep}), this can be used to calculate  $b_{\chi_\nu, \delta}$ and $b_{\chi_\nu, \Gamma}$ from the equation for the effective opacity coefficient Eq.~(\ref{eq:attLLScolden}).  Once a solution for $\delta_\Gamma$ is obtained,  Eq.~(\ref{eq:fullsoln}) may be used to solve for any $\widetilde{\delta_{I_\nu}}$.

However, rather than solve the general time-dependent equations, here we consider the differences with the frequency dependent equations for the steady state limit.  The steady-state solution for the photoionization rate overdensity for the case with full frequency dependence is given by
\begin{equation}
\tilde\delta_{\Gamma, {\rm SS}}= \frac{1}{\langle f \rangle} \int_{\nu_L}^\infty \frac{\dnu}{h_P \nu} \sigma_\nu \langle I_\nu \rangle \frac{\phi_\nu (\chi_\nu +3 +\alpha) \tilde\delta_{j}
- b_{\chi_\nu, \delta}\chi_\nu\tilde\delta}{{\frac{\kappa}{a}\Bigl[{\rm
    atan}{\left(\frac{a^{-1} \kappa}{\phi_\nu (\chi_\nu +3+\alpha)}\right)\Bigr]^{-1}}}
+b_{\chi_\nu, \Gamma}\chi_\nu},
\label{eq:dGisofullSSnu}
\end{equation}
where we have assumed the source overdensity $\delta_{j}$ is frequency independent as would be expected if the sources have a single spectral index, $\alpha \equiv -d \log  \langle I_\nu \rangle/d\nu$, and note that our homogenous solution is $c {\langle j_\nu \rangle}/{\langle I_\nu \rangle} = \phi_\nu (\chi_\nu +3+\alpha) H$, in analogy to Eq.~(\ref{eq:fbg}).

For $\kappa \ll 1$, the steady-state frequency-dependent expression goes to
\begin{equation}
\tilde\delta_{\Gamma, {\rm SS}}= \frac{1}{\langle f \rangle} \int_{\nu_L}^\infty \frac{\dnu}{h_P \nu} \sigma_\nu \langle I_\nu \rangle \frac{  \tilde\delta_{j}
- b_{\chi_\nu, \delta} {\cal R}_\nu \tilde\delta }{1 + b_{\chi_\nu, \Gamma}{\cal R}_\nu},
\label{eq:dGisofullSSnukappall1}
\end{equation}
where ${\cal R}_\nu \equiv \chi_\nu/[\phi_\nu (\chi_\nu +3+\alpha)]$.  In the limit where the fluctuations in $\tilde \delta_j$ dominate over the fluctuations that owe to opacity, $\tilde\delta_{\Gamma, {\rm SS}} \rightarrow \tilde\delta_{j}$ as expected.

For the opposite limit of $\kappa \gg 1$, the steady-state frequency-dependent expression goes to
\begin{equation}
\tilde\delta_{\Gamma, {\rm SS}}= \frac{c}{\langle f \rangle H} \int_{\nu_L}^\infty \frac{\dnu}{h_P \nu} \sigma_\nu  \frac{\langle j_\nu \rangle \tilde\delta_{j}
- b_{\chi_\nu, \delta} \langle I_\nu \rangle \chi_\nu\tilde\delta}{{\frac{2 \kappa }{\pi a }}
+b_{\chi_\nu, \Gamma}\chi_\nu}
\label{eq:dGisofullSSnukappagg1}
\end{equation}
This expression shows that the total fluctuations at high wavenumbers are set by the spectrum of the sources, again considering the case $b_{\chi_\nu, \delta} = b_{\chi_\nu, \Gamma} = 0$  which is expected since the structures that affect the highest wavenumbers are the source proximity regions that experience little attenuation.  This again matches the $\kappa \gg 1$ frequency-averaged result. To see this, note that the $\sigma_\nu$-weighted frequency integral over $\langle j_\nu \rangle $ in Eq.~(\ref{eq:dGisofullSSnukappagg1}) is equal to $\pi a /[2 k \langle f \rangle] \langle j \rangle $ as is the solution to the frequency-averaged equation (Eq.~(\ref{eq:dGkSSO2}) noting the homogeneous solution Eq.~(\ref{eq:phi})). Thus, it is only at intermediate wavenumbers where the wavelength is comparable to the effective photon mean free path that differences are expected between the frequency-averaged solutions and the frequency-dependent solutions presented in this section.  At $z=2$ and $z=3$ in our model, we find numerically only sub-percent differences at intermediate wavenumbers (both with the biases set to zero and with $b_{\chi_\nu, \delta} =1$ and $b_{\chi_\nu, \Gamma} = 1-\beta$).  In conclusion, the solution for $\tilde\delta_{\Gamma, {\rm SS}}$ from the frequency-averaged equation is quite accurate.

\section{Discussion and conclusions}
\label{sec:conclusions}

Fluctuations in the \Lya\ forest flux as detected in background QSOs or galaxies depend on fluctuations in the gas density, temperature and photoionization rate. Motivated by the capacity to measure the 3D redshift-space power spectrum and correlation function of the \Lya\ forest made possible by multiple line-of-sight QSO and bright galaxy surveys, theoretical models of the expected signatures have recently been developed. Our approach differs from previous studies in two regards. Firstly, we consider an intergalactic attenuation model more closely tied to observations, taking care to match previous models of the unperturbed background radiation field. In particular, we do not decompose the opacity into an optically thin and optically thick component, as in \citet{2014PhRvD..89h3010P}, rather following HM12-like models that more smoothly interpolate between these two regimes and that use well-constrained empirical inputs for the IGM opacity. Our approach avoids artificial boosts in the large-scale power from specifying a large optically thin component. Secondly, previous studies only considered steady state solutions for the ionizing background. As the UV photoionizing radiation field depends on the contributions from sources distributed over cosmologically significant distances and times, both being attenuated by an evolving intervening IGM and produced by sources with finite lifetimes, the UV photoionizing radiation field is intrinsically time-dependent both in the evolution of its mean value and in its fluctuations. The finite lifetime of sources is especially important in the shot noise contribution to the signal, which may dominate the power spectrum over a wide range of wavenumbers and provides a substantial contribution to the flux correlation function. We have developed a formalism for time-dependent fluctuations in the radiation background taking into account these effects.
 
Whilst at high wavenumbers, the time-dependent fluctuations in the photoionization rate agree with an improved steady-state approximation (once corrected for evolution in the mean radiation background), deviations are found at lower wavenumbers. These arise from two factors, the evolution of the radiation background itself and of the source populations, and from a substantial reduction in the shot noise estimate by a factor proportional to the source lifetime. The neglect of the finite lifetime of the sources may overestimate the shot noise contribution by an order of magnitude or more on scales exceeding the attenuation length.

Application of the time-dependent solution to a $z=2-4$ UV
photoionizing background produced by QSOs and galaxies reveals an
increase in the non-shot noise contribution to the photoionization
rate power spectrum compared with the steady-state estimate by up to
30 percent at intermediate comoving wavenumbers
($0.001<k<0.01\,h\,{\rm Mpc^{-1}}$), and an asymptotic offset at low
wavenumbers of typically 10 percent. Much larger differences are found
in the shot noise contribution because of the finite lifetime of the
sources. We provide a general formalism for time-dependent shot noise,
which depends on the birthrate function of the sources. We then
specialize to the approximation that the birthrate is given by the
ratio of the source luminosity function to source lifetime, which we
take to be independent of source luminosity. For wavelengths short
compared with the distance light travels over the lifetime of the
sources, each source may be regarded as eternal and contributes fully
to the shot noise, as in the steady-state approximation. Both the
time-dependent and steady-state values for the shot noise agree in
this limit with $P_\Gamma \sim k^{-2}$.  For longer wavelengths, however,
sources expire before their emitted light can traverse the full
wavelength. Although the shot noise at small wavenumbers has a white
noise (flat) spectrum in both the time-dependent and steady-state
solutions, we show that the finite lifetime of the sources reduces the shot noise power compared with the
(infinite lifetime) steady-state solution on scales exceeding a total
effective mean free path $\lambda_*$ by a factor proportional to
$c\tau_S/\lambda_*$, where $\tau_S$ is the source lifetime. The reduction in the shot noise power reflects the larger number of sources that have contributed to the radiation field at a given time than is present at that time, as the photons continue to survive after the sources fade. Whilst
shot noise in the QSO counts dominates the photoionization rate power
spectrum at low wavenumbers in the steady-state approximation, in the
time-dependent calculation the large-scale contribution from shot noise is reduced to the point of being comparable to the power from clustering for QSO lifetimes of
$10^8\,{\rm yrs}$ at $z\sim 2$. In comparison to QSOs, the shot noise contribution
of galaxies is found to be negligible.

The photoionization rate power spectrum also depends on the power
spectrum of the sources, and so on their bias factors. This dependence
may provide a means of constraining the nature of the galaxies that contribute to the photoionizing radiation background in addition to QSOs. For expected bias factors of $b_G\simeq1-3$ the photoionization rate power spectrum is only weakly affected by the galaxy bias, as the dominant source of fluctuations is QSOs and is primarily sensitive to the fractional contribution of QSOs to the background.

The redshift-space power spectrum of \HI\ fluctuations, as would be measurable through fluctuations in the \Lya\ forest flux, is composed of three contributions, one arising from density fluctuations alone, one from photoionization rate fluctuations alone, and a cross-term (see Eq.~[\ref{eq:Pkaa}]). At low wavenumbers (comoving $k\lsim0.01\,h\,{\rm Mpc^{-1}}$), the terms depending on the photoionization rate fluctuations dominate, while at higher wavenumbers density fluctuations dominate. Because of the negative bias between gas density fluctuations and the measured \Lya\ flux, partial cancellation occurs on intermediate wavenumbers, producing a dip in the \Lya\ flux power spectrum. The depth of the dip is limited by the QSO shot noise, lending itself as a means for constraining the mean lifetime of QSOs. Accounting for the time-dependence of the shot noise is required for source lifetimes shorter than $10^9$~yr.

The time-dependent calculation is also essential for computing the non-shotnoise contribution to the \Lya\ flux redshift-space power spectrum at intermediate wavelengths. Whilst the non-shotnoise contribution is well estimated in the steady state limit at high wavenumbers ($k>0.02\,h\,{\rm Mpc^{-1}}$), more than order of magnitude deviations are found in the line-of-sight component of the flux redshift space power spectrum between the time-independent and steady-state estimates at intermediate wavenumbers.  The two estimates converge asymptotically at low wavenumbers at $z=3$, but an offset by as much as a factor of 2 remains at $z=2$.

In contrast, the non-shotnoise contribution to the redshift-space spatial correlations in the \Lya\ flux are remarkably robust against time-dependent effects. We decompose the spatial correlation function into its Legendre $l=0$, 2 and 4 components.
Except for slight shifts in the zero-crossings, the time-dependent and steady-state estimates are nearly identical for $l=0$ and 2 for comoving separations $r<200\,h^{-1}\,{\rm Mpc}$ at $z=2$ and $r<120\,h^{-1}\,{\rm Mpc}$ at $z=3$. (Photoionization rate fluctuations do not contribute to the $l=4$ term.) Relative discrepancies by factors of order unity or larger occur at larger separations, and these increase with redshift for $z>2$. 

The shot noise contribution to the $l=0$ component is highly redshift dependent.
(Shot noise does not contribute to the higher orders.) At small comoving separations ($r<10\,h^{-1}\,{\rm Mpc}$), the correlations are dominated by density fluctuations. At wider separations ($r>50\,h^{-1}\,{\rm Mpc}$), however, the density and photoionization fluctuations contribute comparable amounts, as does the cross term. At $z=2$, the shot noise contribution only slightly modifies the spatial correlations, but by $z=3$ it becomes the dominant contributor over a range of separations, including the BAO scale. Whilst the photoionization rate fluctuations, including the shot noise term, modify the shape and height of the BAO peak, they have only a small effect on its position, shifting it to wider separations by around $0.5-1$ percent at $z=2-3$. Because the position of the peak is weakly sensitive to $\Omega_m$, this magnitude shift may substantially bias cosmological parameter determinations and so should be taken into account for precision estimates.

Both the \Lya\ flux redshift-space power spectrum and spatial correlation function
are somewhat sensitive to the absorption properties of the IGM. Tilting the
\HI\ column density of \Lya\ absorbers from $\beta=1.2$ to 1.5,
increasing the weight of diffuse absorbers to the total absorption,
boosts the \Lya\ forest flux power spectrum by as much as 20 percent at intermediate wavenumbers. (Simulations suggest this steeper $\beta$ might better approximate the response of attenuation to the ionizing background.)  Except for slight shifts in the positions of the zero-crossings, the $l=0$ flux spatial correlation function is unaffected for comoving separations $r<200\,h^{-1}\,{\rm Mpc}$ at $z=2$ and $r<140\,h^{-1}\,{\rm Mpc}$ at $z=3$. The photoionization fluctuation corrections to the \Lya\ flux redshift-space spatial correlation function thus appear robust against uncertainties in IGM attenuation, permitting cosmological parameters to be probed through the BAO feature when the corrections are included. The correlation function for larger spatial separations may be useful for inferring physical properties of the IGM and the sources of the UV background which photoionizes it.

In spite of the differences in the flux redshift-space power spectrum introduced by time-dependent effects, the spatial correlation function smooths over the differences except at large separations. For estimates of the spatial correlation function, we find that direct integration of the time-dependent equations may be by-passed using asymptotic forms, provided here, for the flux power spectrum at low and high wavenumbers. Either patching the two together at an intermediate wavenumber or using a Lorentzian interpolation model are sufficient for recovering the flux spatial correlation function on measurable scales to better than 10 percent accuracy in the non-shotnoise component and better than 75 percent accuracy in the shotnoise component at $z=2.0-2.5$. This may provide a means of quickly honing in on the range of astrophysical parameters affecting the spatial correlation function. Full integration of the time-dependent equation for the shotnoise component may be preferable.

Whilst we primarily concentrated on solving the frequency-averaged radiative transfer equation, we commented briefly on frequency-dependent effects, showing that our same methods can be generalized to this limit. We demonstrated the frequency-dependent solution in the steady state limit matches the solution for frequency-averaged quantities at low and high wavenumbers. Through numerical solutions, we found the two agreed to sub-percent accuracy at intermediate wavenumbers.

There are additions to UVB fluctuation models beyond those considered here that may merit additional investigations. Our study concentrated on $z=2-4$, where the bulk of Ly$\alpha$ forest observations lie; however, time-dependent effects in the UVB will be even more pronounced at lower redshifts.  UVB fluctuations at these redshifts could be relevant for certain galaxy surveys \citep[as we explore in][]{sanderbeckinprep} and low-redshift Ly$\alpha$ forest observations \citep{2018arXiv180805605K}. We have not considered more complex source light curves such as long term variability. We also have not included source beaming in our calculations (but see \citealt{2017MNRAS.472.2643S}).  Our formalism can be generalized to include such effects. Finally, time-dependent effects generate angular anisotropy in observed correlations. Cross correlations with other large-scale structure tracers, such as with quasars \citep{2017A&A...608A.130D}, could result in a distinctive dipolar anisotropy that constrains quasar lifetimes. 

\section*{Acknowledgements}
We thank M. White for discussions, and A. Pontzen and the referee S. Gontcho A Gontcho for helpful comments that improved the clarity of the presentation. AM acknowledges support from the UK Science and Technology Facilities Council. MM acknowledges support from United States NSF award AST 1614439, NASA ATP award NNX17AH68G, and the Alfred P. Sloan foundation.

\bibliographystyle{mn2e-eprint}
\bibliography{ms}

\begin{thebibliography}{}

\bibitem[\protect\citeauthoryear{{Ahn}, {Alexandroff}, {Allende Prieto},
  {Anderson}, {Anderton}, {Andrews}, {Aubourg}, {Bailey}, {Balbinot}, {Barnes}
  \& et al.}{{Ahn} et~al.}{2012}]{2012ApJS..203...21A}
{Ahn} C.~P.,  {Alexandroff} R.,  {Allende Prieto} C.,  {Anderson} S.~F.,
  {Anderton} T.,  {Andrews} B.~H.,  {Aubourg} {\'E}.,  {Bailey} S.,  {Balbinot}
  E.,  {Barnes} R.,    et al. 2012, \apjs, 203, 21

\bibitem[\protect\citeauthoryear{{Bautista}, {Busca}, {Guy}, {Rich},
  {Blomqvist}, {du Mas des Bourboux}, {Pieri} \& {Font-Ribera}}{{Bautista}
  et~al.}{2017}]{2017A&A...603A..12B}
{Bautista} J.~E.,  {Busca} N.~G.,  {Guy} J.,  {Rich} J.,  {Blomqvist} M.,  {du
  Mas des Bourboux} H.,  {Pieri} M.~M.,    {Font-Ribera} A.,  2017, \aap, 603,
  A12

\bibitem[\protect\citeauthoryear{{Bielby}, {Hill}, {Shanks}, {Crighton},
  {Infante}, {Bornancini}, {Francke}, {H{\'e}raudeau}, {Lambas}, {Metcalfe} \&
  et al.}{{Bielby} et~al.}{2013}]{2013MNRAS.430..425B}
{Bielby} R.,  {Hill} M.~D.,  {Shanks} T.,  {Crighton} N.~H.~M.,  {Infante} L.,
  {Bornancini} C.~G.,  {Francke} H.,  {H{\'e}raudeau} P.,  {Lambas} D.~G.,
  {Metcalfe} N.,    et al. 2013, \mnras, 430, 425

\bibitem[\protect\citeauthoryear{{Bolton}, {Becker}, {Haehnelt} \&
  {Viel}}{{Bolton} et~al.}{2014}]{2014MNRAS.438.2499B}
{Bolton} J.~S.,  {Becker} G.~D.,  {Haehnelt} M.~G.,    {Viel} M.,  2014,
  \mnras, 438, 2499

\bibitem[\protect\citeauthoryear{{Bouwens}, {Illingworth}, {Oesch}, {Trenti},
  {Labb{\'e}}, {Bradley}, {Carollo}, {van Dokkum} \& {Gonzalez}}{{Bouwens}
  et~al.}{2015}]{2015ApJ...803...34B}
{Bouwens} R.~J.,  {Illingworth} G.~D.,  {Oesch} P.~A.,  {Trenti} M.,
  {Labb{\'e}} I.,  {Bradley} L.,  {Carollo} M.,  {van Dokkum} P.~G.,
  {Gonzalez} V.,  2015, \apj, 803, 34

\bibitem[\protect\citeauthoryear{{Busca}, {Delubac}, {Rich}, {Bailey},
  {Font-Ribera}, {Kirkby}, {Le Goff}, {Pieri} \& {Slosar}}{{Busca}
  et~al.}{2013}]{2013A&A...552A..96B}
{Busca} N.~G.,  {Delubac} T.,  {Rich} J.,  {Bailey} S.,  {Font-Ribera} A.,
  {Kirkby} D.,  {Le Goff} J.~M.,  {Pieri} M.~M.,    {Slosar} A.,  2013, \aap,
  552, A96

\bibitem[\protect\citeauthoryear{{Croft}}{{Croft}}{2004}]{2004ApJ...610..642C}
{Croft} R.~A.~C.,  2004, \apj, 610, 642

\bibitem[\protect\citeauthoryear{{D'Aloisio}, {McQuinn}, {Davies} \&
  {Furlanetto}}{{D'Aloisio} et~al.}{2018}]{2018MNRAS.473..560D}
{D'Aloisio} A.,  {McQuinn} M.,  {Davies} F.~B.,    {Furlanetto} S.~R.,  2018,
  \mnras, 473, 560

\bibitem[\protect\citeauthoryear{{du Mas des Bourboux}, {Le Goff}, {Blomqvist},
  {Busca}, {Guy}, {Rich}, {Y{\`e}che}, {Bautista}, {Burtin}, {Dawson} \& et
  al.}{{du Mas des Bourboux} et~al.}{2017}]{2017A&A...608A.130D}
{du Mas des Bourboux} H.,  {Le Goff} J.-M.,  {Blomqvist} M.,  {Busca} N.~G.,
  {Guy} J.,  {Rich} J.,  {Y{\`e}che} C.,  {Bautista} J.~E.,  {Burtin} {\'E}.,
  {Dawson} K.~S.,    et al. 2017, \aap, 608, A130

\bibitem[\protect\citeauthoryear{{Faucher-Gigu{\`e}re}, {Lidz}, {Zaldarriaga}
  \& {Hernquist}}{{Faucher-Gigu{\`e}re} et~al.}{2009}]{2009ApJ...703.1416F}
{Faucher-Gigu{\`e}re} C.-A.,  {Lidz} A.,  {Zaldarriaga} M.,    {Hernquist} L.,
  2009, \apj, 703, 1416

\bibitem[\protect\citeauthoryear{{Garzilli}, {Bolton}, {Kim}, {Leach} \&
  {Viel}}{{Garzilli} et~al.}{2012}]{2012MNRAS.424.1723G}
{Garzilli} A.,  {Bolton} J.~S.,  {Kim} T.-S.,  {Leach} S.,    {Viel} M.,  2012,
  \mnras, 424, 1723

\bibitem[\protect\citeauthoryear{{Gontcho A Gontcho}, {Miralda-Escud{\'e}} \&
  {Busca}}{{Gontcho A Gontcho} et~al.}{2014}]{2014MNRAS.442..187G}
{Gontcho A Gontcho} S.,  {Miralda-Escud{\'e}} J.,    {Busca} N.~G.,  2014,
  \mnras, 442, 187

\bibitem[\protect\citeauthoryear{{Haardt} \& {Madau}}{{Haardt} \&
  {Madau}}{1996}]{1996ApJ...461...20H}
{Haardt} F.,  {Madau} P.,  1996, \apj, 461, 20

\bibitem[\protect\citeauthoryear{{Haardt} \& {Madau}}{{Haardt} \&
  {Madau}}{2012}]{2012ApJ...746..125H}
{Haardt} F.,  {Madau} P.,  2012, \apj, 746, 125

\bibitem[\protect\citeauthoryear{{Hopkins}, {Richards} \&
  {Hernquist}}{{Hopkins} et~al.}{2007}]{2007ApJ...654..731H}
{Hopkins} P.~F.,  {Richards} G.~T.,    {Hernquist} L.,  2007, \apj, 654, 731

\bibitem[\protect\citeauthoryear{{Ir{\v s}i{\v c}} \& {McQuinn}}{{Ir{\v s}i{\v
  c}} \& {McQuinn}}{2018}]{2018JCAP...04..026I}
{Ir{\v s}i{\v c}} V.,  {McQuinn} M.,  2018, Journal of Cosmology \&
  Astroparticle Physics, 4, 026, 1801.02671

\bibitem[\protect\citeauthoryear{{Kaiser}}{{Kaiser}}{1987}]{1987MNRAS.227....1K}
{Kaiser} N.,  1987, \mnras, 227, 1

\bibitem[\protect\citeauthoryear{{Khaire}, {Walther}, {Hennawi}, {O{\~n}orbe},
  {Luki{\'c}}, {Prochaska}, {Tripp}, {Burchett} \& {Rodriguez}}{{Khaire}
  et~al.}{2018}]{2018arXiv180805605K}
{Khaire} V.,  {Walther} M.,  {Hennawi} J.~F.,  {O{\~n}orbe} J.,  {Luki{\'c}}
  Z.,  {Prochaska} J.~X.,  {Tripp} T.~M.,  {Burchett} J.~N.,    {Rodriguez} C.,
   2018, ArXiv e-prints, 1808.05605

\bibitem[\protect\citeauthoryear{{Kirkby}, {Margala}, {Slosar}, {Bailey},
  {Busca}, {Delubac}, {Rich}, {Bautista} \& {Blomqvist}}{{Kirkby}
  et~al.}{2013}]{2013JCAP...03..024K}
{Kirkby} D.,  {Margala} D.,  {Slosar} A.,  {Bailey} S.,  {Busca} N.~G.,
  {Delubac} T.,  {Rich} J.,  {Bautista} J.~E.,    {Blomqvist} M.,  2013,
  Journal of Cosmology and Astro-Particle Physics, 2013, 024, 1301.3456

\bibitem[\protect\citeauthoryear{{Laurent}, {Eftekharzadeh}, {Le Goff},
  {Myers}, {Burtin}, {White}, {Ross}, {Tinker} \& {Tojeiro}}{{Laurent}
  et~al.}{2017}]{2017JCAP...07..017L}
{Laurent} P.,  {Eftekharzadeh} S.,  {Le Goff} J.-M.,  {Myers} A.,  {Burtin} E.,
   {White} M.,  {Ross} A.~J.,  {Tinker} J.,    {Tojeiro} R.,  2017, Journal of
  Cosmology and Astro-Particle Physics, 2017, 017

\bibitem[\protect\citeauthoryear{{Lee}, {Bailey}, {Bartsch}, {Carithers},
  {Dawson}, {Kirkby}, {Lundgren} \& {Margala}}{{Lee}
  et~al.}{2013}]{2013AJ....145...69L}
{Lee} K.-G.,  {Bailey} S.,  {Bartsch} L.~E.,  {Carithers} W.,  {Dawson} K.~S.,
  {Kirkby} D.,  {Lundgren} B.,    {Margala} D.,  2013, \aj, 145, 69

\bibitem[\protect\citeauthoryear{{Lee}, {Krolewski}, {White}, {Schlegel},
  {Nugent}, {Hennawi}, {M{\"u}ller} \& {Pan}}{{Lee}
  et~al.}{2017}]{2017arXiv171002894L}
{Lee} K.-G.,  {Krolewski} A.,  {White} M.,  {Schlegel} D.,  {Nugent} P.~E.,
  {Hennawi} J.~F.,  {M{\"u}ller} T.,    {Pan} R.,  2017, ArXiv e-prints,
  1710.02894

\bibitem[\protect\citeauthoryear{{McQuinn}, {Hernquist}, {Lidz} \&
  {Zaldarriaga}}{{McQuinn} et~al.}{2011}]{2011MNRAS.415..977M}
{McQuinn} M.,  {Hernquist} L.,  {Lidz} A.,    {Zaldarriaga} M.,  2011, \mnras,
  415, 977

\bibitem[\protect\citeauthoryear{{McQuinn}, {Oh} \&
  {Faucher-Gigu{\`e}re}}{{McQuinn} et~al.}{2011}]{2011ApJ...743...82M}
{McQuinn} M.,  {Oh} S.~P.,    {Faucher-Gigu{\`e}re} C.-A.,  2011, \apj, 743, 82

\bibitem[\protect\citeauthoryear{{Meiksin}}{{Meiksin}}{2005}]{2005MNRAS.356..596M}
{Meiksin} A.,  2005, \mnras, 356, 596

\bibitem[\protect\citeauthoryear{{Meiksin} \& {Tittley}}{{Meiksin} \&
  {Tittley}}{2012}]{2012MNRAS.423....7M}
{Meiksin} A.,  {Tittley} E.~R.,  2012, \mnras, 423, 7

\bibitem[\protect\citeauthoryear{{Meiksin} \& {White}}{{Meiksin} \&
  {White}}{2004}]{2004MNRAS.350.1107M}
{Meiksin} A.,  {White} M.,  2004, \mnras, 350, 1107

\bibitem[\protect\citeauthoryear{{Meiksin}}{{Meiksin}}{2009}]{2009RvMP...81.1405M}
{Meiksin} A.~A.,  2009, Reviews of Modern Physics, 81, 1405

\bibitem[\protect\citeauthoryear{{P{\^a}ris}, {Petitjean}, {Ross}, {Myers},
  {Aubourg}, {Streblyanska}, {Bailey} \& {Armengaud}}{{P{\^a}ris}
  et~al.}{2017}]{2017A&A...597A..79P}
{P{\^a}ris} I.,  {Petitjean} P.,  {Ross} N.~P.,  {Myers} A.~D.,  {Aubourg}
  {\'E}.,  {Streblyanska} A.,  {Bailey} S.,    {Armengaud} {\'E}.,  2017, \aap,
  597, A79

\bibitem[\protect\citeauthoryear{{Planck Collaboration}, {Ade}, {Aghanim},
  {Arnaud}, {Ashdown}, {Aumont}, {Baccigalupi}, {Banday}, {Barreiro},
  {Bartlett} \& et al.}{{Planck Collaboration}
  et~al.}{2016}]{2016A&A...594A..13P}
{Planck Collaboration} {Ade} P.~A.~R.,  {Aghanim} N.,  {Arnaud} M.,  {Ashdown}
  M.,  {Aumont} J.,  {Baccigalupi} C.,  {Banday} A.~J.,  {Barreiro} R.~B.,
  {Bartlett} J.~G.,    et al. 2016, \aap, 594, A13

\bibitem[\protect\citeauthoryear{{Pontzen}}{{Pontzen}}{2014}]{2014PhRvD..89h3010P}
{Pontzen} A.,  2014, Phys. Rev. D, 89, 083010

\bibitem[\protect\citeauthoryear{{Pontzen}, {Bird}, {Peiris} \&
  {Verde}}{{Pontzen} et~al.}{2014}]{2014ApJ...792L..34P}
{Pontzen} A.,  {Bird} S.,  {Peiris} H.,    {Verde} L.,  2014, \apj, 792, L34

\bibitem[\protect\citeauthoryear{{Prochaska}, {Madau}, {O'Meara} \&
  {Fumagalli}}{{Prochaska} et~al.}{2014}]{2014MNRAS.438..476P}
{Prochaska} J.~X.,  {Madau} P.,  {O'Meara} J.~M.,    {Fumagalli} M.,  2014,
  \mnras, 438, 476

\bibitem[\protect\citeauthoryear{{Rorai}, {Carswell}, {Haehnelt}, {Becker},
  {Bolton} \& {Murphy}}{{Rorai} et~al.}{2018}]{2018MNRAS.474.2871R}
{Rorai} A.,  {Carswell} R.~F.,  {Haehnelt} M.~G.,  {Becker} G.~D.,  {Bolton}
  J.~S.,    {Murphy} M.~T.,  2018, \mnras, 474, 2871

\bibitem[\protect\citeauthoryear{{Rorai}, {Hennawi}, {O{\~n}orbe}, {White},
  {Prochaska}, {Kulkarni}, {Walther}, {Luki{\'c}} \& {Lee}}{{Rorai}
  et~al.}{2017}]{2017Sci...356..418R}
{Rorai} A.,  {Hennawi} J.~F.,  {O{\~n}orbe} J.,  {White} M.,  {Prochaska}
  J.~X.,  {Kulkarni} G.,  {Walther} M.,  {Luki{\'c}} Z.,    {Lee} K.-G.,  2017,
  Science, 356, 418

\bibitem[\protect\citeauthoryear{{Suarez} \& {Pontzen}}{{Suarez} \&
  {Pontzen}}{2017}]{2017MNRAS.472.2643S}
{Suarez} T.,  {Pontzen} A.,  2017, \mnras, 472, 2643

\bibitem[\protect\citeauthoryear{{Tittley} \& {Meiksin}}{{Tittley} \&
  {Meiksin}}{2007}]{2007MNRAS.380.1369T}
{Tittley} E.~R.,  {Meiksin} A.,  2007, \mnras, 380, 1369

\bibitem[\protect\citeauthoryear{{Upton Sanderbeck}, {Ir\v s i \v c}, {McQuinn}
  \& {Meiksin}}{{Upton Sanderbeck} et~al.}{2018}]{sanderbeckinprep}
{Upton Sanderbeck} P.~R.,  {Ir\v s i \v c} {McQuinn} M.,    {Meiksin} A.,
  2018, in preparation

\bibitem[\protect\citeauthoryear{{Worseck}, {Prochaska}, {O'Meara}, {Becker},
  {Ellison}, {Lopez}, {Meiksin}, {M{\'e}nard}, {Murphy} \&
  {Fumagalli}}{{Worseck} et~al.}{2014}]{2014MNRAS.445.1745W}
{Worseck} G.,  {Prochaska} J.~X.,  {O'Meara} J.~M.,  {Becker} G.~D.,  {Ellison}
  S.~L.,  {Lopez} S.,  {Meiksin} A.,  {M{\'e}nard} B.,  {Murphy} M.~T.,
  {Fumagalli} M.,  2014, \mnras, 445, 1745

\bibitem[\protect\citeauthoryear{{Zuo}}{{Zuo}}{1992}]{1992MNRAS.258...45Z}
{Zuo} L.,  1992, \mnras, 258, 45

\end{thebibliography}

\appendix 

\section{Perturbations of frequency-integrated radiative transfer equation}
\label{sec:PFIRT_ap}

\subsection{Frequency-integrated radiative transfer equation}
\label{subsec:FIRT_ap}

The radiative transfer equation for $I_\nu$, Eq.~(\ref{eq:RTInu}), simplifies for $\sigma_\nu=\sigma_L(\nu/\nu_L)^{s}\theta(\nu-\nu_L)$, where $\theta(x)$ is a step-function and $\nu_L$ is the frequency at the Lyman edge, when $s=3$:

\begin{eqnarray}
\int_0^\infty\,d\nu\,\nu\frac{\partial
  I_\nu(r,\nhat,t)}{\partial\nu}\frac{\sigma_\nu}{\nu}&=&\left[\sigma_\nu I_\nu\right]_0^\infty -
\int_0^\infty\,d\nu\,I_\nu\frac{d(\sigma_\nu)}{d\nu}\nonumber\\
&=&
-\frac{s\sigma_L}{\nu_L}\int_{\nu_L}^\infty\,d\nu\,I_\nu\left(\frac{\nu}{\nu_L}\right)^{-s-1}\nonumber\\&&+\int_0^\infty I_\nu\sigma_L\left(\frac{\nu}{\nu_L}\right)^{-s}\delta_D(\nu-\nu_L)\nonumber\\
&=&
-\frac{s\sigma_L}{\nu_L}\int_{\nu_L}^\infty\,d\nu\,I_\nu\left(\frac{\nu}{\nu_L}\right)^{-s-1}\nonumber\\&&+I_L\sigma_L
\end{eqnarray}
where $[\sigma_\nu I_\nu]_0^\infty=0$, noting $d\theta(\nu-\nu_L)/d\nu=\delta_D(\nu-\nu_L)$ and defining $I_L=I_{\nu_L}$.
Then
\begin{eqnarray}
\int_0^\infty\,&d\nu&\,\frac{\dot a}{a}\left[3I_\nu(r,\nhat,t) - \nu\frac{\partial I_\nu(r,\nhat,t)}{\partial\nu}\right]\frac{\sigma_\nu}{h_{\rm P}\nu}\nonumber\\&=&
\frac{(3-s)\sigma_L}{h_{\rm P}\nu_L}\frac{\dot a}{a}\int_{\nu_L}^\infty\,d\nu\,I_\nu\left(\frac{\nu}{\nu_L}\right)^{-s-1}+\frac{\dot a}{a}\frac{I_L\sigma_L}{h_{\rm P}}\nonumber\\
&=&\frac{\dot a}{a}\frac{I_L\sigma_L}{h_{\rm P}}
\end{eqnarray}
for any $I_\nu$ when $s=3$.

Defining  
\begin{equation} 
f=\int_0^\infty\,d\nu\,(I_\nu/h_{\rm P}\nu)\sigma_\nu,\quad j=\int_0^\infty\,d\nu\,(j_\nu/h_{\rm P}\nu)\sigma_\nu  
\end{equation} 
and  
\begin{equation}
\alpha_{\rm eff}=\frac{\int_0^\infty\,d\nu\,\frac{I_\nu}{h_{\rm P}\nu}\alpha_\nu\sigma_\nu}{\int_0^\infty\,d\nu\,\frac{I_\nu}{h_{\rm P}\nu}\sigma_\nu},
\label{eq:ap_aeff}
\end{equation} 
the integrated form of Eq.~(\ref{eq:RTInu}) becomes, for $s=3$,

\begin{equation}
\frac{1}{c}\dot f + \frac{1}{c}\frac{\dot 
  a}{a}\frac{I_L\sigma_L}{h_{\rm P}}+\nhat\cdot{\bm\nabla}f=-\alpha_{\rm 
  eff}f+j. 
\label{eq:ap_f}
\end{equation}

\subsection{Linear perturbations}
\label{subsec:LPFIRT_ap}

Consider linear planewave perturbations of the form
$$
f=\langle f\rangle + \Sigma_k{\tilde\delta f}(k)e^{-i{\bm k}\cdot{\bm x}},
$$
for comoving wavevector ${\bm k}$ and comoving coordinate ${\bm x}={\bm r}/a$. Noting ${\bm\nabla}=a^{-1}{\bm\nabla_x}$ gives

\begin{equation}
\frac{1}{c}\dot{\tilde\delta}_f + \frac{1}{c}\frac{\dot
  a}{a}\frac{\sigma_L\langle I_L\rangle}{h_{\rm P}\langle f\rangle}\tilde\delta_{I_L}-ia^{-1}{\bm k}\cdot\nhat\tilde\delta_f=-\alpha_{\rm
  eff}\tilde\delta_f-\tilde\delta\alpha_{\rm eff}+\frac{\langle j\rangle}{\langle
  f\rangle}\tilde\delta_j,
\label{eq:df2}
\end{equation}
where $\tilde\delta_f={\tilde\delta f}/\langle f\rangle$, $\tilde\delta_{I_L}={\tilde\delta I_L}/\langle I_L\rangle$ and $\tilde\delta_j={\tilde\delta j}/\langle j\rangle$. Writing $I_\nu$ as $I_\nu=I_Lg(\nu/\nu_L)$, $\log I_\nu=\log I_L + \log g(\nu/\nu_L)$ and $\tilde\delta_{I_\nu}=\tilde\delta_{I_L}+\tilde\delta_g$. The spectral fluctuation $\tilde\delta_g$ could arise from a fluctuation in the source spectra or evolution, or from a fluctuation in the attenuation if the attenuation is significant. Supposing the sources don't change character (same evolution and spectra, only the numbers change), and supposing the effect of attenuation is negligible, we may take $\tilde\delta_{I_L}=\tilde\delta_f$. A full frequency treatment is needed to test this assumption.

We define bias parameters $b_{\chi,\delta}$ and $b_{\chi,\Gamma}$ through $\tilde\delta\alpha_{\rm eff}=(b_{\chi,\delta}\tilde\delta+b_{\chi,\Gamma}\tilde\delta_\Gamma)\alpha_{\rm eff}$, where  $\tilde\delta$ is the fractional baryon density fluctuation and $\tilde\delta_\Gamma=\tilde\delta\Gamma/\Gamma$. A dependence on temperature fluctuations is absorbed into the density fluctuations, assuming an equation of state between gas temperature and density (see main text). We then obtain
\begin{eqnarray}
\frac{1}{c}\dot{\tilde\delta}_f &+& \left[\frac{\langle j\rangle}{\langle f\rangle} - ia^{-1}{\bm k}\cdot\nhat \right]\tilde\delta_f\\&=& \frac{\langle j\rangle}{\langle
  f\rangle}\tilde\delta_j -\left(b_{\chi,\delta}\tilde\delta + b_{\chi, \Gamma}\tilde\delta_\Gamma\right)\langle\alpha_{\rm eff}\rangle\nonumber.
\label{eq:dfHIatten_ap}
\end{eqnarray}

The corresponding Green's function is
$$ 
G(t,t') = \exp\left\{c\int_{t'}^t\,dt''\,\left[i\frac{{\bm k}\cdot{\nhat}}{a(t'')}  -\frac{\langle j\rangle}{\langle f\rangle}\right]\right\}.  
$$ 
The general solution is then, for $\tilde\delta_f(t)=0$ when $t<t_i$,
\begin{eqnarray} 
  \tilde\delta_f(t)&=&c\int_{t_i}^t\,dt'\,G(t,t') b_{\chi, \Gamma}(t')\langle\alpha_{\rm eff}(t')\rangle\nonumber\\
  &&\times\left[\tilde\delta_S(t') - \tilde\delta_\Gamma(t')\right],
\end{eqnarray}
where
\begin{equation}
  \tilde\delta_S(t)=q(t)\tilde\delta_j(t)-\frac{b_{\chi, \delta}(t)}{b_{\chi, \Gamma}(t)}\tilde\delta,
\end{equation}
with $q(t) = \langle j(t)\rangle/[b_{\chi, \Gamma}(t)\langle\alpha_{\rm eff}(t)\rangle\langle f(t)\rangle]$, is a source term. Here, $\langle\dots\rangle$ denotes a spatial average.

The photoionization rate is given by $\Gamma = \int\,d^2\nhat\,\langle f\rangle$, so that, for isotropic sources, $\tilde\delta_\Gamma(t)$ is given by the implicit equation
\begin{eqnarray}
\tilde\delta_\Gamma(t)&=&\frac{\langle f\rangle\int d^2\nhat\,\tilde\delta_f}{\int d^2\nhat\,\langle f\rangle}=\frac{1}{4\pi}\int d^2\nhat\,\tilde\delta_f\nonumber\\
&=&\frac{2\pi}{4\pi}\int_{-1}^{1}d\mu\nonumber\\
&\times&c\int_{t_i}^t\,dt'\,\left[\tilde\delta_S(t')-\tilde\delta_\Gamma(t')\right]\nonumber\\
&\times&e^{c\int_{t'}^t dt''\,\left[ia^{-1}(t'')k\mu-\frac{\langle j(t'')\rangle}{\langle f(t'')\rangle}\right]}\nonumber\\
&&\times\left[b_{\chi, \Gamma}(t')\langle\alpha_{\rm eff}(t')\rangle\right]\nonumber\\
&=&c\int_{t_i}^t\,dt'\,j_0[ck\eta(t,t')]\left[\tilde\delta_S(t')-\tilde\delta_\Gamma(t')\right]\nonumber\\
&&\times e^{-c\int_{t'}^t dt''\,\frac{\langle j(t'')\rangle}{\langle f(t'')\rangle}}\nonumber\\
&&\times\left[b_{\chi, \Gamma}(t')\langle\alpha_{\rm eff}(t')\rangle\right],
\label{eq:dGisofull_ap}
\end{eqnarray}
where $j_0(x)=\sin(x)/x$ and
$\eta(t,t')=\int_{t'}^t\,dt''a^{-1}(t'')$. (For an Einstein-deSitter cosmology,
$\eta(t,t')=3a(t)^{-1}(t-t^{2/3}t'^{1/3})=[2/(H(t)a(t))][1-(a(t^\prime)/a(t))^{1/2}]$,
and $ck\eta=(2\kappa(a)/a)[1-(a'/a)^{1/2}]$, where $\kappa(a)=ck/H(a)$.)

It is helpful to non-dimensionalize Eq.~(\ref{eq:dGisofull_ap}). We
define $d{\bar t'}=H(t')dt'$ and $\bar\eta({\bar
  t}, {\bar t'})=H(t)\eta(t, t')$, the dimensionless wavenumber
$\kappa(t)=ck/ H(t)$, so that $ck\eta(t,t')=\kappa(t){\bar\eta}({\bar
  t}, {\bar t'}) $, and introduce the dimensionless attenuation coefficient
\begin{equation}
  \chi=\frac{c}{H}\langle\alpha_{\rm eff}\rangle
\end{equation} 
and the dimensionless ratio
\begin{equation}
\zeta=\frac{\langle I_L\rangle\sigma_L}{h_{\rm P}\langle f\rangle}.
\end{equation} 

\noindent Eq.~(\ref{eq:dGisofull_ap}) may then be cast in the form

\begin{eqnarray}
  \tilde\delta_\Gamma(\bar t)&=&\int_{\bar t_i}^{\bar t}\,d\bar t'\,j_0[\kappa(\bar t)\bar \eta(\bar t,\bar t')]b_{\chi, \Gamma}(\bar t')\chi(\bar t')\left[\tilde\delta_S(\bar t')-\tilde\delta_\Gamma(\bar t')\right]\nonumber\\
&&  \times e^{-\int_{\bar  t'}^{\bar t} d\bar t''\,\phi(\bar t'')\left[\chi(\bar t'')+\zeta(\bar t'')\right]}.
\label{eq:dGisofull_nondim_ap}
\end{eqnarray}

In the special case $b_{\chi, \Gamma}\langle\alpha_{\rm eff}\rangle=0$, such as for negligible attenuation, Eq.~(\ref{eq:dGisofull_nondim_ap}) simplifies to the closed-form expression
\begin{eqnarray}
  \tilde\delta_\Gamma(\bar t) &=& \int_{\bar t_i}^{\bar t}\,d\bar t'\,j_0[\kappa(\bar t)\bar\eta(\bar t,\bar t')]\left[\chi(\bar t')+\zeta(\bar t')\right]\tilde\delta_S(\bar t')\nonumber\\
&&\times e^{-\int_{\bar t'}^{\bar t} d\bar t''\,\phi(\bar t'')\left[\chi(\bar t'')+\zeta(\bar t'')\right]},
\label{eq:dGisofullexpl_ap}
\end{eqnarray} 
where now
\begin{equation}
\tilde\delta_S(\bar t)=\phi(\bar t)\tilde\delta_j(\bar t)-b_{\chi, \delta}\tilde\delta\frac{\chi(\bar t)}{\chi(\bar t)+\zeta(\bar t)},
\end{equation}
and
$$
\phi=\frac{c\langle j\rangle/H}{(\chi+\zeta)\langle
  f\rangle}.
$$

\subsection{Asymptotic limits}
\label{subsec:asymps_ap}

 In the limit
$\kappa(t)\bar\eta(\bar t,\bar t_i)\ll1$, it follows from Eq.~(\ref{eq:dGisofull_nondim_ap}) that
\begin{eqnarray}
\frac{d{\tilde\delta}_\Gamma(\bar t)}{d\bar t}&=&
b_{\chi, \Gamma}(\bar t)\chi(\bar t)\left[\tilde\delta_S(\bar t)-\tilde\delta_\Gamma(\bar t)\right]\nonumber\\
&&-\phi(\bar t)\left[\chi(\bar t)+\zeta(\bar t)\right]\tilde\delta_\Gamma(\bar t)
\end{eqnarray}
so that
\begin{eqnarray}
  \frac{d{\tilde\delta}_\Gamma(\bar t)}{d\bar t}&+&
  \left[\left(\phi(\bar t)+b_{\chi, \Gamma}(\bar t)\right)\chi(\bar t)+\phi(\bar t)\zeta(\bar t)\right]\tilde\delta_\Gamma(\bar t)\nonumber\\ &=& b_{\chi, \Gamma}(\bar t)\chi(\bar t)\tilde\delta_S(\bar t),
\label{eq:dGammaODE}
\end{eqnarray}
which is exactly solvable by quadrature:
\begin{eqnarray}
  \tilde\delta_\Gamma(\bar t) &=&
  \int_{\bar t_i}^{\bar t}\,d\bar t'\,b_{\chi, \Gamma}(\bar t')\chi(\bar t')\tilde\delta_S(\bar t')\nonumber\\
  &\times&e^{-\int_{\bar t'}^{\bar t} d\bar t''\,\left[(\phi(\bar t'')+b_{\chi, \Gamma}(\bar t''))\chi(\bar t'')+\phi(\bar t'')\zeta(\bar t'')\right]}.
\label{eq:dGammakq_ap}
\end{eqnarray}

For $\kappa\gg1$, an asymptotic series may be developed. The case of time-independent $b_{\chi, \Gamma}$, $\chi$, $\zeta$ and $\phi$ is particularly straightforward. Making the substitution $u=(2\kappa(a)/a)(1-x'^{1/2})$ into Eq.~(\ref{eq:dGisofull_nondim_ap}), where $x'=a(\bar t')/a(\bar t)$,  then gives (for an Einstein-deSitter cosmology)

\begin{eqnarray}
  \tilde\delta_{\Gamma}(a)=b_{\chi, \Gamma}\chi\frac{a}{\kappa(a)}\int_0^{u_i}\,&du&\,j_0(u)\left(1-\frac{au}{2\kappa(a)}\right)^{-1+2\phi(\chi+\zeta)}  \nonumber\\
&\times&\left[\tilde\delta_S(u)-\tilde\delta_\Gamma(u)\right],
\label{eq:dGisofullu_ap}
\end{eqnarray}
where $u_i=(2\kappa(a)/a)(1-x_i^{1/2})$ for $x_i=a(\bar t_i)/a(\bar t)$.

This may be further simplified if in the source term $b_{\chi, \delta}=0$ and we adopt the form $b_j\sim (1+z)^{\alpha_b}$. Then, in the Einstein-deSitter limit, the source term evolves like

\begin{equation}
\tilde\delta_S(x')=\tilde\delta_S(a)x'^{1- \alpha_b}=\tilde\delta_S(a)\left(1-\frac{au}{2\kappa(a)}\right)^{2(1-\alpha_b)}.
\label{eq:deltaSevol_ap}
\end{equation}

It is helpful to introduce the rescaled radiation fluctuation
$Z(u)=\tilde\delta_\Gamma(u)/\tilde\delta_S(u)$. Then
Eq.~(\ref{eq:dGisofullu_ap}) gives

\begin{eqnarray} 
Z_0\equiv Z(u=0)=b_{\chi, \Gamma}\chi\frac{a}{\kappa(a)}&\int_0^{u_i}&\,du\,\left(1-\frac{au}{2\kappa(a)}\right)^{2\gamma_H}\nonumber\\
&\times&j_0(u)\left[1-Z(u)\right],
\label{eq:dGisofullZ_ap}
\end{eqnarray} 
where $\gamma_H=\phi(\chi+\zeta)-\alpha_b+1/2$.

This gives a programme for solving for $Z_0$ for
$\kappa\gg1$. Noting that $Z(u)$ is of order $\kappa^{-1}$, $Z_0$ may
be approximated iteratively as

\begin{equation}
Z_0^{(0)}=b_{\chi, \Gamma}\chi\frac{a}{\kappa(a)}\int_0^{u_i}\,du\,\left(1-\frac{au}{2\kappa(a)}\right)^{2\gamma_H}j_0(u),
\label{eq:dGisofullZ0_ap}
\end{equation}
\begin{equation}
Z_0^{(1)}=-b_{\chi, \Gamma}\chi\frac{a}{\kappa(a)}\int_0^{u_i}\,du\,\left(1-\frac{au}{2\kappa(a)}\right)^{2\gamma_H}j_0(u)Z^{(0)}(u),
\label{eq:dGisofullZ1_ap}
\end{equation}
with $Z_0=Z_0^{(0)}+Z_0^{(1)}$. This may be repeated indefinitely,
taking care to include all terms to the appropriate order in $1/\kappa$
from all previous levels.

The expansion in $1/\kappa(a)$ may be effected by solving
Eq.~(\ref{eq:dGisofullZ0_ap}) using contour integration. Define
$z=x+iy=re^{i\alpha}$. Since, for $\gamma_H>0$, the only
pole is at $z=0$, Cauchy's theorem may be used to deform the contour
into three contiguous parts:\ (i)\ a quadrant arc $C_\rho$ enclosing the pole above it
($y\ge0$) with radius $r=\rho$ and $\alpha$ running from 0 to
$\pi/2$, then $\rho\rightarrow0$ taken, (ii)\ the straight line $C_I$ along
the pure imaginary axis $z=iy$ with $y$ running from $\rho$ to $u_i$,
then $\rho\rightarrow0$ taken, and a quadrant arc $C_R$ of radius $R=u_i$
with $\alpha$ running from $\alpha=\pi/2$ to 0. This gives
\begin{eqnarray}
Z_0^{(0)}&=&b_{\chi, \Gamma}\chi\frac{a}{\kappa(a)}\lim_{\rho \to 0}
Im\int_{z=\rho}^{u_i}\frac{dz}{z}e^{iz}\left(1-\frac{az}{2\kappa(a)}\right)^{2\gamma_H}\nonumber\\
&=&b_{\chi, \Gamma}\chi\frac{a}{\kappa(a)}\lim_{\rho \to 0}
Im\int_{C_\rho+C_I+C_R}\frac{dz}{z}e^{iz}\left(1-\frac{az}{2\kappa(a)}\right)^{2\gamma_H}.
\label{eq:Z00_ap}
\end{eqnarray}

We perform each integral in turn:

\noindent (i)\ $C_\rho$
\begin{eqnarray}
  Im &\int_{C_\rho}&\frac{dz}{z}e^{iz}\left(1-\frac{az}{2\kappa(a)}\right)^{2\gamma_H}\\&=&Im\quad i\int_0^{\pi/2}\,
                                                                                     d\alpha\,
                                                                                     e^{i\rho\cos\alpha}e^{-\rho\sin\alpha}\left(1-\frac{1}{2}\frac{a}{\kappa(a)}\rho  
                                                                                     e^{i\alpha}\right)^{2\gamma_H}\nonumber\\
                                                                                 &=&\frac{\pi}{2}\nonumber
\label{eq:Crho_ap}
\end{eqnarray} 
as $\rho\rightarrow0$.

\noindent (ii)\ $C_I$
\begin{eqnarray}
Im
  &\int_{C_I}&\frac{dz}{z}e^{iz}\left(1-\frac{az}{2\kappa(a)}\right)^{2\gamma_H}\nonumber\\&=&Im\int_\rho^{u_i}\,\frac{dy}{y}e^{-y}\left(1-\frac{1}{2}i\frac{a}{\kappa(a)}y\right)^{2\gamma_H}\nonumber\\
&\simeq&Im\int_\rho^{u_i}\,\frac{dy}{y}e^{-y}\left(1-i\gamma_H\frac{a}{\kappa(a)}y\right)\nonumber\\
&\rightarrow&-\gamma_H\frac{a}{\kappa(a)}\int_0^\infty\,dy\,e^{-y}=-\gamma_H\frac{a}{\kappa(a)},
\label{eq:CI_ap}
\end{eqnarray}
where the quantity in brackets was expanded to first order in $y$
because $a/\kappa(a)\ll1$ and $e^{-y}$ cuts off exponentially with
$y$, the upper limit $u_i\rightarrow\infty$ was taken with only an
exponentially small error and the limit $\rho\rightarrow0$ was taken in the imaginary part.

\noindent (iii)\ $C_R$
\begin{eqnarray}
  Im &\int_{C_R}&\frac{dz}{z}e^{iz}\left(1-\frac{az}{2\kappa(a)}\right)^{2\gamma_H}\\&=&-Im\quad i\int_0^{\pi/2}\,
                                                                                     d\alpha\,
                                                                                     e^{iR\cos\alpha}e^{-R\sin\alpha}\left(1-\frac{1}{2}\frac{a}{\kappa(a)}R  
                                                                                     e^{i\alpha}\right)^{2\gamma_H}.\nonumber
\label{eq:CR1_ap}
\end{eqnarray} 
Noting $aR/2\kappa(a)=1-x_i^{1/2}$ (since $R=u_i$), the term in parentheses may be re-expressed as $[1-2(1-x_i^{1/2})\cos\alpha+(1-x_i^{1/2})^2]^{\gamma_H}e^{2i\gamma_H\theta}$, where $\theta=-{\rm atan}[(1-x_i^{1/2})\sin\alpha/(1-(1-x_i^{1/2})\cos\alpha)]$. Since $e^{-R\sin\alpha}$ cuts off exponentially with increasing $\alpha$, to obtain the leading order behaviour, we make the small angle approximations $\sin\alpha\sim\alpha$ and $\cos\alpha\sim1$. This then gives $\theta\sim-(1-x_i^{1/2})\alpha/x_i^{1/2}$ and $e^{2i\gamma_H\theta}\sim1+2i\gamma_H\theta$. Then
\begin{eqnarray}
  Im &\int_{C_R}&\frac{dz}{z}e^{iz}\left(1-\frac{az}{2\kappa(a)}\right)^{2\gamma_H}\nonumber\\
  &\simeq&-Im\quad ie^{iR}x_i^{\gamma_H}\int_0^{\pi/2}\,d\alpha\,e^{-R\alpha}\left(1-2i\gamma_H\frac{1-x_i^{1/2}}{x_i^{1/2}}\alpha\right)\nonumber\\
  &=&-x_i^{\gamma_H}\Biggl[\cos(R)\int_0^{\pi/2}\,d\alpha\,e^{-R\alpha}\nonumber\\
    &&+\sin(R)2\gamma_H\frac{1-x_i^{1/2}}{x_i^{1/2}}\int_0^{\pi/2}\,d\alpha\,\alpha e^{-R\alpha}\Biggr]\nonumber\\
  &\sim& -x_i^{\gamma_H}\frac{\cos\left[\frac{2\kappa(a)}{a}(1-x_i^{1/2})\right]}{\frac{2\kappa(a)}{a}(1-x_i^{1/2})},
\label{eq:CR2_ap}
\end{eqnarray}
with only an exponentially small error, noting that the second integral is sub-dominant compared with the first by a factor $1/\kappa(a)$, and substituting in $R=u_i$.

Combining the integrals gives

\begin{equation}
Z_0^{(0)}\sim b_{\chi, \Gamma}\chi\frac{a}{\kappa(a)}\left(\frac{\pi}{2}-\gamma_H\frac{a}{\kappa(a)}-x_i^{\gamma_H}\frac{\cos u_i}{u_i}\right).
\label{eq:Z00O2_ap}
\end{equation}

To obtain the complete expansion to order $1/\kappa^2(a)$, the leading order part of $Z_0^{(0)}$ must be inserted into Eq.~(\ref{eq:dGisofullZ1_ap}) in the form $(\pi/2)b_{\chi, \Gamma}\chi a'/\kappa(a')=(\pi/2)b_{\chi, \Gamma}\chi [1-au/(2\kappa(a))]^{-1}a/\kappa(a)$, where $\Omega_ma'^{-3}\gg\Omega_v$ was assumed to relate $\kappa(a')$ to $\kappa(a)$ for fixed comoving $k$. To lowest order, this gives

\begin{equation}
Z_0^{(1)}\sim-\left(\frac{\pi}{2}b_{\chi, \Gamma}\chi\frac{a}{\kappa(a)}\right)^2.
\label{eq:Z01O2_ap}
\end{equation}
Combining this with $Z_0^{(0)}$ gives to order $1/\kappa^2(a)$

\begin{eqnarray}
  \tilde\delta_\Gamma(\kappa\gg1)&\sim&\frac{\pi}{2}b_{\chi, \Gamma}\chi\frac{a}{\kappa(a)}\tilde\delta_S(a)\Biggl\{1-\frac{a}{\kappa(a)}\Biggl[\frac{\pi}{2}b_{\chi, \Gamma}\chi\nonumber\\
    &&+\frac{2}{\pi}\left(\phi(\chi+\zeta)-\alpha_b+\frac{1}{2}\right)\\
&&+\frac{1}{\pi}\frac{x_i^{\gamma_H}}{1-x_i^{1/2}}\cos\left[\frac{2\kappa(a)}{a}(1-x_i^{1/2})\right]\Biggr]\Biggr\}.\nonumber
\label{eq:dGkO2_ap}
\end{eqnarray}

\subsection{Steady state limit}
\label{subsec:SSlim_ap}

From Eq.~(\ref{eq:dfHIatten_ap}), setting $\dot{\tilde\delta}_f = 0$ gives
\begin{equation}
\tilde\delta_{f, {\rm SS}} = \frac{\frac{c\langle j\rangle}{H\langle
  f\rangle}\tilde\delta_j -b_{\chi, \Gamma} \chi\tilde\delta_\Gamma  - b_{\chi, \delta}\chi\tilde\delta}{-ia^{-1}{\bm \kappa}\cdot\nhat + \phi(\chi + \zeta)}.
\label{eq:dfHIattenSS_ap}
\end{equation}
Then the perturbation in the photoionization rate becomes
\begin{eqnarray}
\tilde\delta_{\Gamma, {\rm SS}}&=&\frac{\langle f\rangle\int d^2\nhat\,\tilde\delta_{f, {\rm SS}}}{\int d^2\nhat\,\langle f\rangle}=\frac{1}{4\pi}\int d^2\nhat\,\tilde\delta_{f, {\rm SS}}\\
&=&\frac{\phi(\chi+\zeta)\tilde\delta_j - b_{\chi, \delta}\chi\tilde\delta}{\frac{\kappa}{a}\left[{\rm atan}{\left(\frac{\kappa}{a\phi(\chi+\zeta)}\right)}\right]^{-1}
+b_{\chi, \Gamma}\chi},\nonumber
\label{eq:dGisofullSS_ap}
\end{eqnarray}
noting $(c/H)\langle j\rangle/\langle f\rangle=\phi(\chi+\zeta)$.

\subsection{Shot noise}
\label{subsec:sn_ap}

To include shot noise, the source emissivity is perturbed, including an evolving luminosity function and evolving source luminosity. The luminosity function is derived from a source creation rate function $\Psi({\bm x},L,t)$:\ the number of sources created in a volume $d^3x$ in the time interval $(t,t+dt)$ with luminosity track between $L$ and $LdL$ is $d^3x\;dt\;dL\Psi({\bm x},L,t)$. The notation is not ideal, as $L(t-t_i)$ is a function. For simplicity, each $L$ will be taken to correspond to a unique luminosity that is on only for a time interval $\tau_S(L)$. Then the luminosity function is
$$
\Phi(L,t)LdL = \int_0^T\;dt^\prime\;\langle\Psi({\bm x},L,t^\prime)\rangle_x L(t-t^\prime)dL,
$$
where $\langle\dots\rangle_x$ denotes a spatial average and $T>t$. Here, $L(t-t^\prime)=0$ for $t<t^\prime$. For the simplest case of a time-independent $\Psi$, $\Phi(L) = \langle\Psi({\bm x},L)\rangle\;\tau_S(L)$.

Consider the contribution from $N$ sources created during a time $T$, turning on at times $t_i$ at comoving positions ${\bm x}_i$, with luminosities $L_i(t-t_i)$. The emissivity is
$$
\epsilon^{(N)}({\bm x},t)=\frac{1}{a(t)^3}\sum_{i=1}^N\;L_i(t-t_i)\delta^3({\bm x}-{\bm x}_i),
$$
corresponding to a perturbed emissivity Fourier component in a volume $V_u$
\begin{eqnarray}
\tilde\delta_\epsilon^{(N)}({\bm k},t)&=&\frac{1}{\epsilon_{bg}(t)V_u}\int_{V_u}\;d^3x\;\epsilon^{(N)}({\bm
                         x},t)e^{i{\bm k}\cdot{\bm x}}\nonumber\\
&=&\frac{1}{\epsilon_{bg}(t)V_u}\sum_{i=1}^N\;L_i(t-t_i)e^{i{\bm k}\cdot{\bm x}_i},
\end{eqnarray}
where the background emissivity $\epsilon_{bg}(t)$ is
$$
\epsilon_{bg}(t) = \int_0^T dt^\prime\; \int_0^\infty dL\langle\Psi({\bm x},L,t^\prime)\rangle_u\;L(t-t^\prime),
$$
for $\Psi$ averaged over $V_u$:
$$
\langle\Psi({\bm x},L,t)\rangle_u = \frac{1}{V_u}\int_{V_u}\;d^3x\;\Psi({\bm x},L,t).
$$

Consider
$$
\tilde\delta_\epsilon^{(N)}({\bm k},t)\tilde\delta_\epsilon^{(N)*}({\bm k^\prime},t^\prime).
$$
To construct the correlation matrix of the emissivity fluctuations, this quantity is ensemble averaged over the positions and creation times of the sources, and over the Poisson process for the number of sources created, with mean value
$$
\bar N = V_u\int_0^Tdt\;\int_0^LdL\;\langle\Psi({\bm x},L,t)\rangle_u.
$$
The probability that a source is created in a volume element $d^3x$ at position ${\bm x}$ within a luminosity interval $(L,L+dL)$ and time interval $(t,t+dt)$ is
$$
p({\bm x},L,t)d^3xdLdt = {\bar N}^{-1}\Psi({\bm x},L,t)d^3xdLdt.
$$
Then, denoting an ensemble average by $\langle\dots\rangle$ and defining $\bar\Psi(L,t)=\langle\Psi({\bf x},L,t)\rangle$,
\begin{eqnarray}
&\epsilon_{bg}(t)&\epsilon_{bg}(t^\prime)V_u^2\langle\tilde\delta_\epsilon^{(N)}({\bm
  k},t)\tilde\delta_\epsilon^{(N)*}({\bm
  k^\prime},t^\prime)\rangle\nonumber\\
&=& \frac{1}{{\bar N}^N}\langle\prod_{j=1}^N\int_0^T
    dt_j\int_0^\infty\, dL_j\int_{V_u}d^3x_j\Psi({\bm
                        x}_j,L_j,t_j)\nonumber\\
&&\times\tilde\delta_\epsilon^{(N)}({\bm k},t)\tilde\delta_\epsilon^{(N)*}({\bm k^\prime},t^\prime)\rangle\nonumber\\
&=&{\bar N}^{-1}\sum_{i=1}^N\int_0^Tdt_i\;\int_0^\infty dL_i\;L_i(t-t_i)L_i(t^\prime-t_i)\nonumber\\
&&\times\int_{V_u}d^3x\;\bar\Psi(L_i,t_i)\;e^{i({\bm k}-{\bm k^\prime})\cdot{\bm x}_i}\nonumber\\
&&+{\bar N}^{-2}\sum_{i=1}^N\sum_{i^\prime\ne
   i}^N\int_0^Tdt_i\;\int_0^\infty dL_i\;L_i(t-t_i)\nonumber\\
&&\times\int_0^Tdt_{i^\prime}\;\int_0^\infty dL_{i^\prime}\;L_{i^\prime}(t^\prime-t_{i^\prime})\nonumber\\
&&\times\int_{V_u}d^3{\bm x}_i\int_{V_u}d^3{\bm
   x}_{i^\prime}\langle\Psi({\bm x}_i,L_i,t_i)\Psi({\bm
   x}_{i^\prime},L_{i^\prime},t_{i^\prime})\rangle\nonumber\\
&&\times e^{i({\bm k}\cdot{\bm x}_i-{\bm k^\prime}\cdot{\bm x}_{i^\prime})}\nonumber\\
&=&\frac{N}{\bar N}\int_0^Tdt^{\prime\prime}\;\int_0^\infty
    dL\;L(t-t^{\prime\prime})L(t^\prime-t^{\prime\prime}){\bar\Psi}(L,t^{\prime\prime})\nonumber\\
&&\times\int_{V_u}d^3x\;e^{i({\bm k}-{\bm k^\prime})\cdot{\bm x}}\nonumber\\
&&+\frac{N(N-1)}{{\bar N}^2}\left[\int_0^T
    dt^{\prime\prime}\int_0^\infty
    dL\;L(t-t^{\prime\prime})\right]\nonumber\\
&&\times\left[\int_0^T dt^{\prime\prime}\int_0^\infty dL\;L(t^\prime-t^{\prime\prime})\right]\nonumber\\
&&\times\int_{V_u}d^3{\bm x}\int_{V_u}d^3{\bm
         x}^\prime\langle\Psi({\bm x},L,t)\Psi({\bm
         x}^\prime,L^\prime,t^\prime)\rangle\nonumber\\
&&\times e^{i({\bm k}\cdot{\bm x}-{\bm k^\prime}\cdot{\bm x}^\prime)}.
\end{eqnarray}
The last term depends on the spatial correlation function $\xi$:
\begin{eqnarray}
\langle\Psi({\bm x},L,t)\Psi({\bm
  x}^\prime,L^\prime,t^\prime)\rangle&=&{\bar\Psi}(L,t){\bar\Psi}(L^\prime,t^\prime)\\
&\times&\left[1+\xi(\vert{\bm x}-{\bm x}^\prime\vert,t,t^\prime)\right].\nonumber
\end{eqnarray}
Note the time dependence in $\xi$. In the linear perturbation limit, the time-dependent correlation function may be expressed as
$$
\xi(\vert{\bm x}_i-{\bm x}_{i^\prime}\vert,t_i,t_{i^\prime})=D(t)D(t^\prime)b(t)b(t^\prime)\xi_{\rm init}(\vert{\bm x}_i-{\bm x}_{i^\prime}\vert),
$$
where $D(t)$ is the linear perturbation growth factor since some initial time when the matter spatial correlation function was $\xi_{\rm init}(\vert{\bm x}_i-{\bm x}_{i^\prime}\vert)$, corresponding to an initial matter power spectrum $P_{\rm init}(k)=\int_{V_u}d^3x\;\xi_{\rm init}(\vert {\bm x}\vert)e^{i{\bm k}\cdot{\bm x}}$, and $b(t)$ is the (time-dependent) bias factor for the sources.

Allowing for periodic boundary conditions, the emissivity correlations vanish for ${\bm k}\ne{\bm k^\prime}$. Then for ${\bm k}={\bm k^\prime}\ne 0$, taking the Poisson average over $N$, noting the Poisson average of $[N(N-1)]$ is ${\bar N}^2$, assuming the volume $V_u$ is sufficiently large that the spatial and ensemble averages are the same ($\langle\dots\rangle_u=\langle\dots\rangle$), and noting the definition of $\epsilon_{bg}$, the emissivity fluctuation power spectrum is
\begin{eqnarray}
P_\epsilon(k,t,t^\prime)&=&V_u\langle\tilde\delta_\epsilon({\bm
                            k},t)\tilde\delta^*_\epsilon({\bm
                            k},t^\prime)\rangle\nonumber\\
&=&\frac{1}{n_{\rm eff}(t,t^\prime)}+D(t)D(t^\prime)b(t)b(t^\prime)P_{\rm init}(k),
\label{eq:powem_ap}
\end{eqnarray}
where
\begin{eqnarray}
\frac{1}{n_{\rm
  eff}(t,t^\prime)}&=&\left[\epsilon_{bg}(t)\epsilon_{bg}(t^\prime)\right]^{-1}\\
&&\times\int_0^\infty dt^{\prime\prime}\;\int_0^\infty dL\;L(t-t^{\prime\prime})L(t^\prime-t^{\prime\prime}){\bar\Psi}(L,t^{\prime\prime})\nonumber\\
&=&\left[\epsilon_{bg}(t)\epsilon_{bg}(t^\prime)\right]^{-1}\nonumber\\
&&\times\int_0^\infty dL\;\int_0^\infty dt^{\prime\prime}\;L(t-t^{\prime\prime})L(t^\prime-t^{\prime\prime})\nonumber\\
&&\times\Phi(L,t^{\prime\prime})\tau_S(L)^{-1},\nonumber
\end{eqnarray}
for the simple evolution model $\bar\Psi(L,t)=\Phi(L,t)/\tau_S(L)$. If
$\tau_S(L)=\tau_S$ for all $L$ and the luminosity function evolves
slowly, so that $\tau_S\vert\dot\Phi\vert\ll\Phi$, $\Phi(L)$ may be approximated
as $\bar\Phi(L,t,t^\prime)=(1/2)[\Phi(L,t)+\Phi(L,t^\prime)]$, and the
expression for $n_{\rm eff}$ simplifies to

\begin{eqnarray}
\frac{1}{n_{\rm
  eff}(t,t^\prime)}&=&\frac{\int_0^\infty\;dL\;L^2\bar\Phi(L,t,t^\prime)}{[\int_0^\infty\;dL\;L\Phi(L,t)][\int_0^\infty\;dL\;L\Phi(L,t^\prime)]}\nonumber\\
&&\times{\rm Max}\left[0,\left(1-\frac{\vert t-t^\prime\vert}{\tau_S}\right)\right].
\label{eq:neffttp_ap}
\end{eqnarray}

For a mixed population of sources, such as QSOs and galaxies, the shot
noise terms add (eg, for populations \lq a\rq and \lq b\rq),

\begin{eqnarray}
\frac{1}{n_{\rm eff}(t,t^\prime)}&=&\frac{1}{[\int_0^\infty\;dL\;L\Phi(L,t)][\int_0^\infty\;dL\;L\Phi(L,t^\prime)]}\\
&\times&\Biggl\{\int_0^\infty\;dL\;L^2\bar\Phi_a(L,t,t^\prime)
{\rm Max}\left[0,\left(1-\frac{\vert t-t^\prime\vert}{\tau_{S,
         a}}\right)\right]\nonumber\\
& +&
\int_0^\infty\;dL\;L^2\bar\Phi_b(L,t,t^\prime){\rm Max}\left[0,\left(1-\frac{\vert t-t^\prime\vert}{\tau_{S, b}}\right)\right]\Biggr\},\nonumber
\label{eq:neffttpab_ap}
\end{eqnarray}
where $\Phi=\Phi_a+\Phi_b$. The contributions to the power spectrum of the background radiation field are weighted by the contribution of each population to the mean background emissivity,
\begin{equation}
\epsilon_{bg}^{(i)}(t)= \int\;dt^\prime\;\int_0^\infty\;dL\; L(t-t^\prime)\bar\Psi_i(L,t^\prime)\simeq\int_0^\infty\;dL\; L\Phi_i(L,t).
\end{equation}
Then
\begin{eqnarray}
P_\epsilon(k,t,t^\prime)&=&\frac{1}{n_{\rm eff}(t,t^\prime)}+\frac{1}{\epsilon_{bg}(t)\epsilon_{bg}(t^\prime)}\left[\epsilon_{bg}^{(a)}(t)\epsilon_{bg}^{(a)}(t^\prime)P_{aa}(k,t,t^\prime)\right.\nonumber\\
&&\left.+\epsilon_{bg}^{(a)}(t)\epsilon_{bg}^{(b)}(t^\prime)P_{ab}(k,t,t^\prime)\right.\nonumber\\
&&\left.+\epsilon_{bg}^{(b)}(t)\epsilon_{bg}^{(a)}(t^\prime)P_{ba}(k,t,t^\prime)\right.\nonumber\\
&&\left.+\epsilon_{bg}^{(b)}(t)\epsilon_{bg}^{(b)}(t^\prime)P_{bb}(k,t,t^\prime)\right],
\label{eq:powemab_ap}
\end{eqnarray}
where $P_{ij}(k,t,t^\prime)=D(t)D(t^\prime)b_i(t)b_j(t^\prime)P_{\rm init}(k)$ and $\epsilon_{bg}=\epsilon_{bg}^{(a)}+\epsilon_{bg}^{(b)}$.

\subsection{Shot noise estimates}
\label{subsec:sn_est}

\begin{figure}
\scalebox{0.55}{\includegraphics{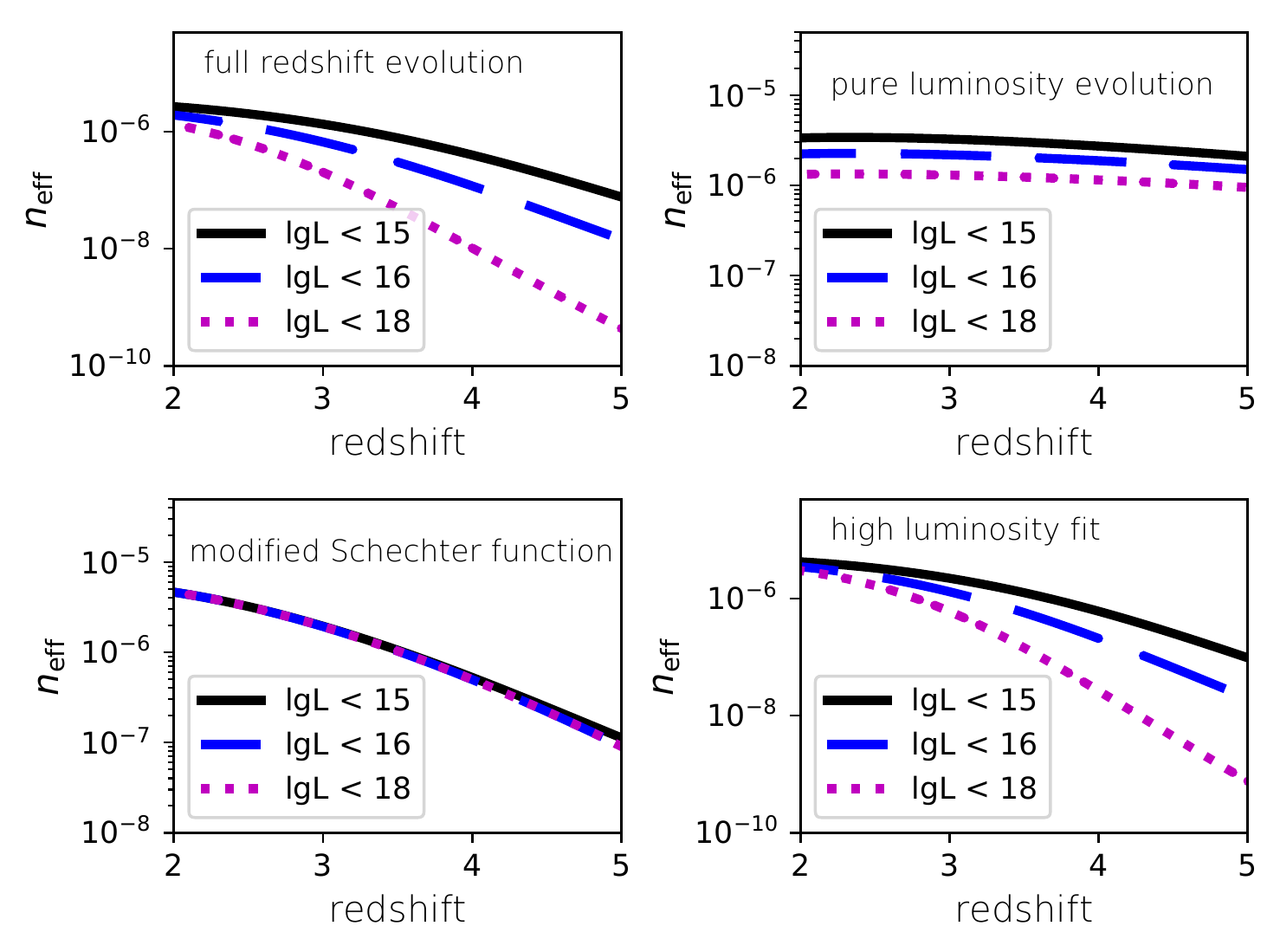}}
\caption{The evolution in the effective comoving number density
  $n_{\rm eff}$ (in units ${\rm Mpc}^{-3}$), of QSO sources. Four
  model fits for the QSO luminosity function are assumed:\ a full
  redshift evolution model ($z$-evol), a pure luminosity evolution
  model (PLE), a modified Schechter function model (mS) and a redshift
  evolution model fit to the high luminosity end. See
  \citet{2007ApJ...654..731H} for details. A source lifetime of
  $\tau_S=100$~Myr is adopted for the calculations.
}
\label{fig:Neff4QSOLFs}
\end{figure}

Fig.~\ref{fig:Neff4QSOLFs} shows the effective comoving number density
of sources computed at the Lyman edge, allowing for the obscuration
estimate of \citet{2007ApJ...654..731H} in the $B$-band (their
Eq.~4). A minimum bolometric QSO luminosity of $10^{10}L_\odot$ is
assumed. (The results are not very sensitive to the low luminosity
end.) A black hole mass of $10^{10}\,M_\odot$ corresponds to an
Eddington luminosity of about $0.3\times10^{15}\,L_\odot$. Higher
luminosity QSOs would suggest they are fed by super-Eddington
accretion, as may occur for non-spherically symmetric
accretion. Results are shown for upper bolometric luminosities of
$10^{15}\,L_\odot$, $10^{16}\,L_\odot$ and $10^{18}\,L_\odot$, noting
that values above $10^{16}\,L_\odot$ may correspond to rare transient
accretion phases. Each panel corresponds to a different model fit for the QSO
luminosity function:\ a full
redshift evolution model ($z$-evol), a pure luminosity evolution model
(PLE), a modified Schechter function model (mS) and a redshift
evolution model fit to the high luminosity end. See
\citet{2007ApJ...654..731H} for details.  Whilst the
      PLE and mS models are quite insensitive to the upper limit, the
      full $z$-fit and HL models are extremely sensitive to the upper
      limit, with the rare high luminosity QSOs driving $n_{\rm eff}$
      down. Fig.9 of \citet{2007ApJ...654..731H} for the full
      $z$-evolution model shows there should only be fewer than 30
      QSOs brighter than $10^{15}\,L_\odot$ in a comoving volume of
      1~Gpc$^3$ at $z=3$, and fewer than 1 by $z=6$, with the number
      density declining rapidly with luminosity. It is unclear what
      the appropriate upper limit is. The averages correspond to an
      ensemble average, as could be applied to an infinite
      universe. But if the actual universe has no sources within the
      horizon brighter than, say, $10^{16}\,L_\odot$, it makes little
      sense to estimate an ensemble average based on higher
      luminosities.

\begin{figure}
\scalebox{0.55}{\includegraphics{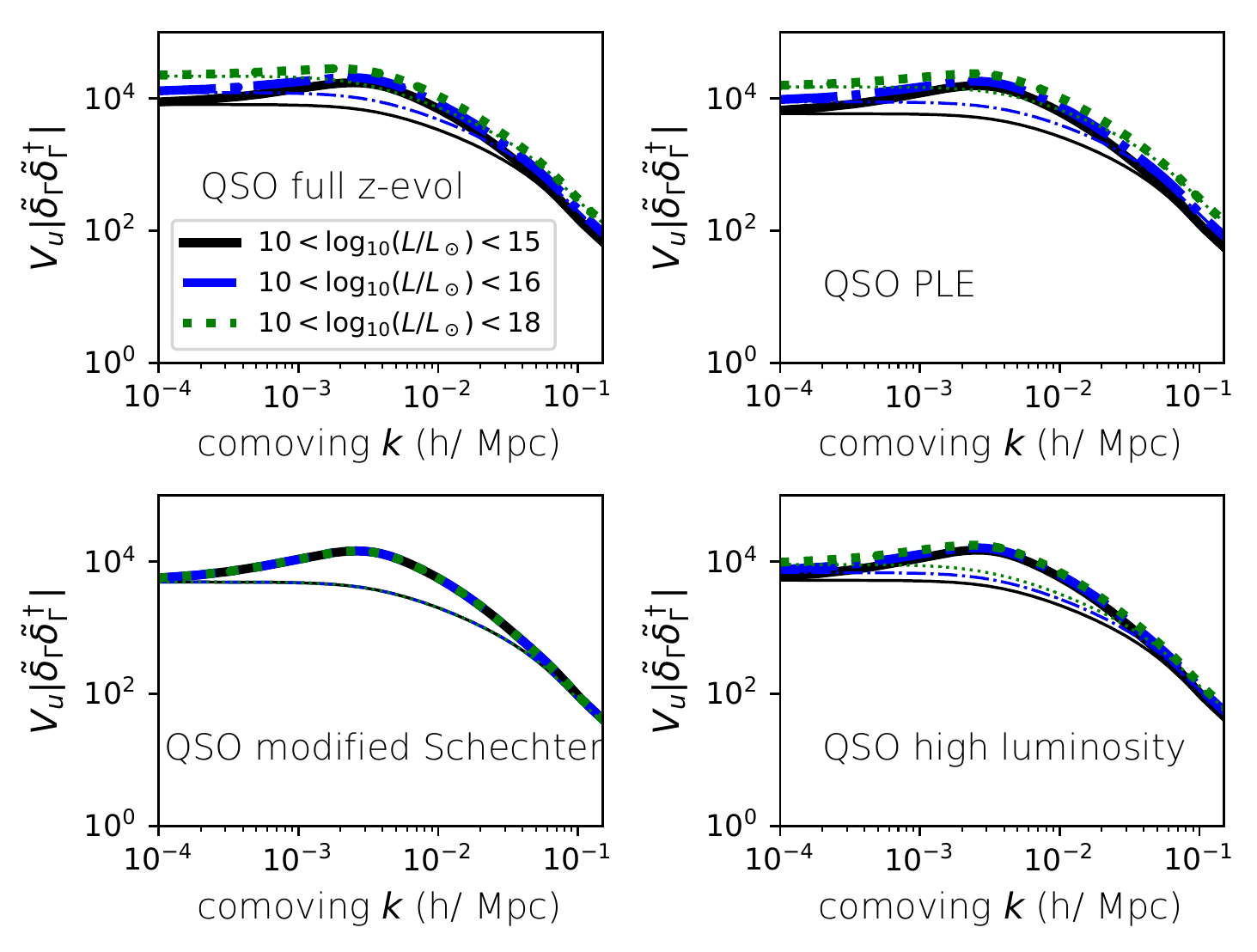}}
\caption{Comoving power spectrum (in units $h^{-3}\,{\rm Mpc}^3$), of
  fluctuations in the photoionization rate as a function of comoving
  wavenumber at $z=2$. The BOSS estimate for QSO bias is used, the
  galaxy bias is set at $b_G=3$, and galaxies contribute about half
  the flux. The emissivity parameters are $\alpha_j=1.8$ and
  $\alpha_S=0.8$, and $\beta=1.2$ is adopted for the attenuation
  coefficient. Results are shown for QSO and galaxy lifetimes of
  $\tau_Q=100$~Myr and $\tau_G=100$~Myr. Four model fits for the QSO
  luminosity function are assumed:\ a full redshift evolution model
  ($z$-evol), pure luminosity evolution (PLE), modified Schechter
  function (mS) and a redshift evolution model fit to the high
  luminosity end. Heavy lines show the full power spectrum; light
  lines show the shot noise contribution.
}
\label{fig:SolvedGammaCorr4QSOLFs}
\end{figure}

The effect of the upper QSO bolometric luminosity on the power
spectrum of the photoionization rate fluctuations is shown in
Fig.~\ref{fig:SolvedGammaCorr4QSOLFs} at $z=2$ for upper limits of
$10^{15}\,L_\odot$, $10^{16}\,L_\odot$ and $10^{18}\,L_\odot$. (The
lower limit is $10^{10}\,L_\odot$.) Shot noise is subdominant for
comoving wavenumbers $0.001\lsim k\lsim 0.01-0.02\,h\,{\rm Mpc}^{-1}$
(depending on luminosity function model), for an upper luminosity up
to $10^{16}\,L_\odot$. For an upper luminosity of $10^{18}\,L_\odot$,
shot noise dominates everywhere in the full redshift evolution and PLE
models.

\subsection{Interpolation formulas}
\label{subsec:interp}

\begin{figure}
\scalebox{0.55}{\includegraphics{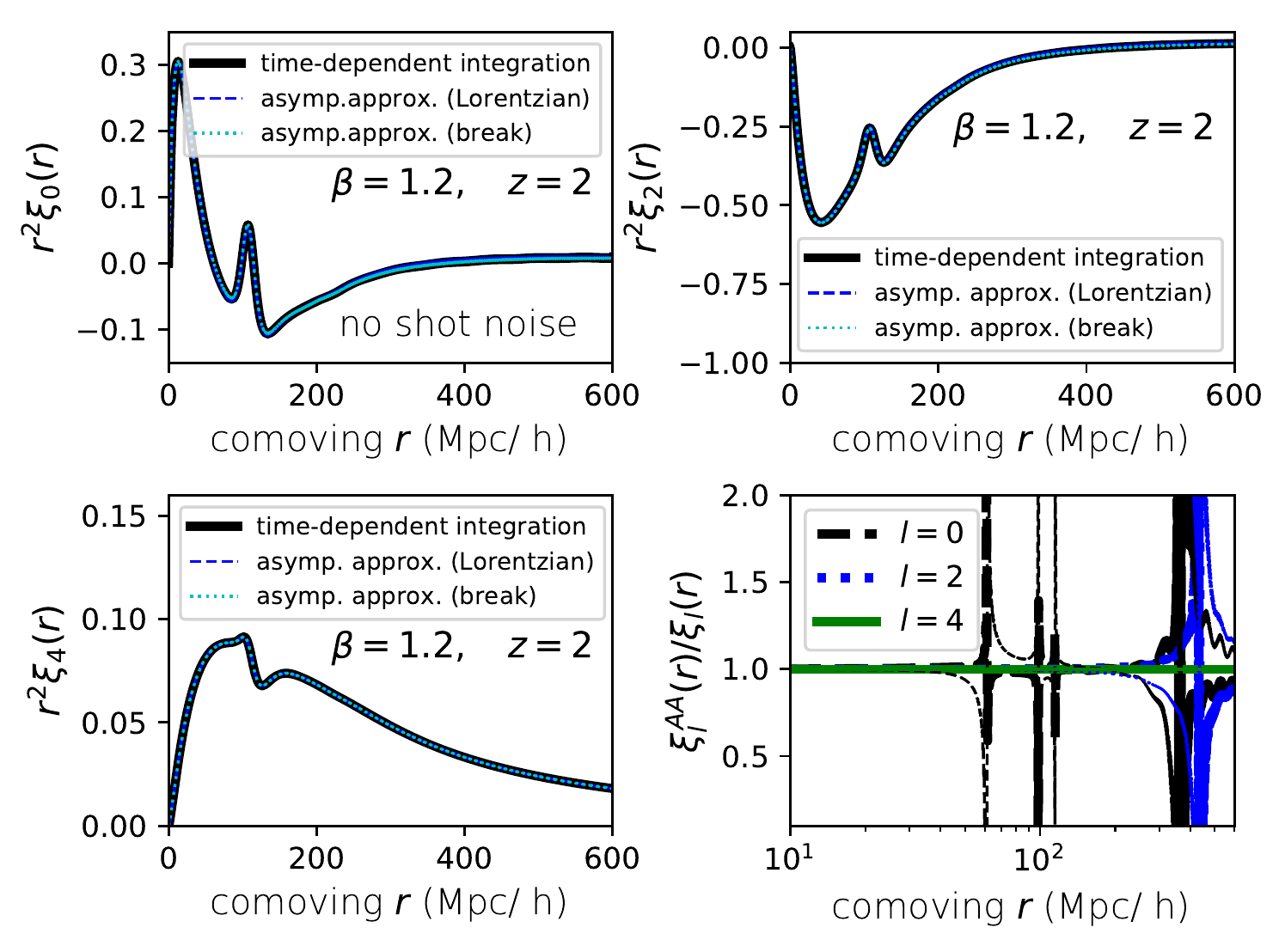}}
\caption{Comparison between the time-dependent calculation and
  approximate solutions for the Legendre components of the redshift space \HI\ \Lya\ flux
  spatial correlation function without shot noise, shown at $z=2$. The
  BOSS estimate for QSO bias is used, and the galaxy bias is set at
  $b_G=3$. The emissivity parameters are $\alpha_j=1.8$ and
  $\alpha_S=0.8$, and $\beta=1.2$ is used for the attenuation
  coefficient. Results for two approximations are shown, one modelling
  the photoionization rate power spectrum as a Lorentzian (blue dashed lines) and a second
  matching the asymptotic limiting values of the fluctuations at an
  intermediate wavenumber $k_{\rm match}=0.2\,h\,{\rm Mpc^{-1}}$ (cyan dotted lines) (see
  text). The panels show $\xi_0(r)$ (top left), $\xi_2(r)$, (top
  right), and $\xi_4(r)$ (bottom left). Solid lines show the
  correlation function for the full time-dependent calculation. The
  bottom right panel compares the asymptotic approximations with the
  full time-dependent calculation. The thick lines correspond to the
  approximation based on matching the asymptotic forms at $k_{\rm
    match}$, and the thin lines to the Lorentzian
  approximation. Except near the zero-crossings, the asymptotic
  approximations provide an accurate basis for computing the
  \Lya\ flux correlations.
}
\label{fig:Xil_TDVAA_nsn}
\end{figure}

\begin{figure}
\scalebox{0.55}{\includegraphics{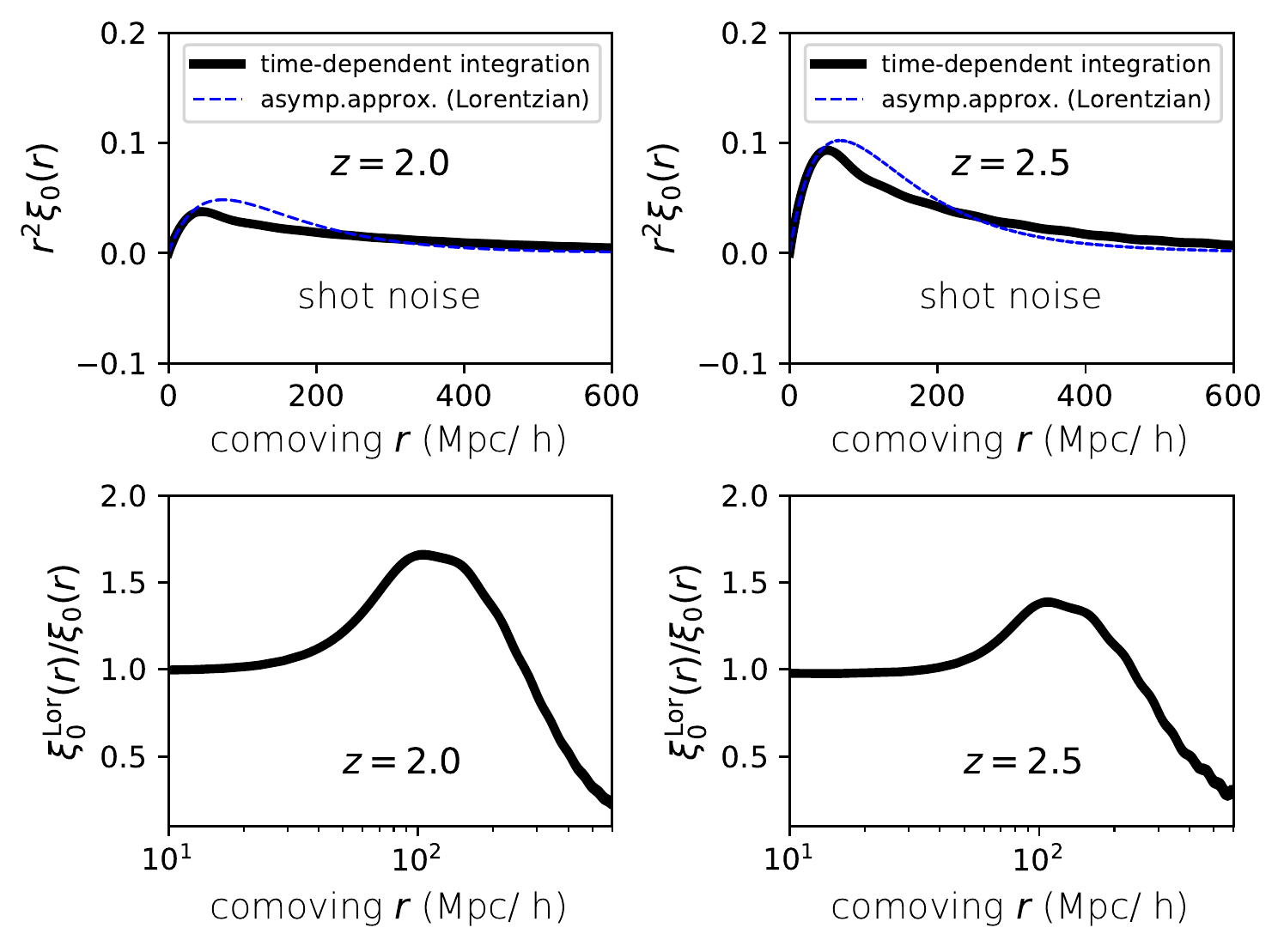}}
\caption{Comparison between the time-dependent calculation and
  Lorentzian approximation to the photoionization rate power spectrum
  for the shot noise contribution to the $l=0$ Legendre component of
  the redshift space \HI\ \Lya\ flux spatial correlation function,
  shown at $z=2$ and 2.5. The BOSS estimate for QSO bias is used, and
  the galaxy bias is set at $b_G=3$. The emissivity parameters are
  $\alpha_j=1.8$ and $\alpha_S=0.8$, and $\beta=1.2$ is used for the
  attenuation coefficient. The panels show $\xi_0(r)$ at $z=2.0$ (top
  left) and $z=2.5$ (top right). Solid lines show the correlation
  function for the full time-dependent calculation; dashed lines show
  the results using only the Lorentzian approximation. The bottom
  panels show the ratio of the correlation function using the
  Lorentzian approximation to the correlation function from the full
  time-dependent solution.
}
\label{fig:Xil_TDVAA_sn}
\end{figure}

As a substitute for using the full time-dependent machinery for
computing the \Lya\ flux correlations, the asymptotic approximations
to the photoionization rate power spectrum discussed in
Sec.~\ref{subsubsec:bgflucs:limits} may be used instead by matching to
the low and high $k$ limits. We consider two approximations, a
Lorentzian form for the power spectrum, Eq.~(\ref{eq:aplor}), and
matching the asymptotic forms at an intermediate
wavenumber. Fig.~\ref{fig:Xil_TDVAA_nsn} shows the correlation
function components without the shot noise contribution for the Lorentzian approximation and for matching
the asymptotic forms at $k_{\rm match}=0.2\,h\,{\rm Mpc^{-1}}$. Except
very near the zero-crossings, the Legendre components are
well-recovered, to better than 10 percent for $l=0$ and to a few
percent or better for $l=2$. Direct matching yields better agreement
at the zero-crossings, although the Lorentzian model does well
overall. The Lorentzian approximation for the shot noise contribution,
Eq.~(\ref{eq:apshotLor}), also recovers the full computation, but, as
shown in Fig.~\ref{fig:Xil_TDVAA_sn}, the match is not as good as for
the non-shotnoise contribution. We use $\alpha_n=2$, although the results are not very sensitive to this choice. For comoving separations
$r<300\,h^{-1}\,{\rm Mpc}$, agreement is unfortunately poorest near
the BAO peak, over-shooting the shot noise by 70 percent at $z=2$ and
40 percent at $z=2.5$. Such approximations may nonetheless save
considerable computational expense in winnowing parameter space for
modelling measurements of the \Lya\ flux correlation function. The
full time-dependent solution may be preferable for the shot noise
contribution when varying only the QSO and galaxy bias parameters
since the shot noise term does not depend on these, so that the
shot noise computation need be done only once. Final comparisons with
data should use the full time-dependent integrations for both the
non-shotnoise and shot noise contributions for precision work.

\label{lastpage}

\end{document}